\documentclass[11pt]{article}
\usepackage[subrefformat=parens,labelformat=parens]{subfig}
\usepackage{geometry}
\usepackage{setspace}
\usepackage{bm}
\usepackage{setspace}
\usepackage[utf8]{inputenc}
\usepackage{authblk}
\usepackage{algorithm}
\usepackage{algpseudocode}
\usepackage{color,soul}
\usepackage[colorlinks=TRUE,linkcolor=blue,citecolor=blue]{hyperref}
\usepackage{float}
\usepackage{amsmath} 
\usepackage{amsfonts}
\usepackage{amsthm}
\usepackage{amssymb}
\usepackage{graphicx,psfrag,epsf}
\usepackage{enumerate}
\usepackage{appendix}

\usepackage{natbib}
\usepackage{pdfpages}
\usepackage{placeins}
\usepackage{forest}
\usepackage{textcomp}
\usepackage[normalem]{ulem}
\usepackage{tabularx,booktabs}
\usepackage{comment}
\newtheorem{assumption}{Assumption}

\newcommand*{\T}{\top}

\newcommand*{\bD}{{\bm{D}}}

\newcommand{\xddots}{%
  \raise 4pt \hbox {.}
  \mkern 6mu
  \raise 1pt \hbox {.}
  \mkern 6mu
  \raise -2pt \hbox {.}
}

\newcommand{\commentGG}[1]{{\color{orange} \it [[#1]]}}
\usepackage[capitalise,noabbrev,compress]{cleveref}
\crefformat{equation}{#2\textup{(#1)}#3}
\crefformat{assumption}{Assumption #1}

\title{
Spatial vertical regression for spatial panel data: Evaluating the effect of the Florentine tramway's first line on commercial vitality
}
\author[1]{Giulio Grossi \footnote{Corresponding Author: giulio.grossi@unifi.it}}
\author[1]{Alessandra Mattei}
\author[2]{Georgia Papadogeorgou}
\affil[1]{Department of Statistics, Computer Science, Applications, University of Florence}
\affil[2]{Department of Statistics, University of Florida}
\date{}

\allowdisplaybreaks
\begin{document}

\maketitle
\setstretch{1.4}

\begin{abstract}
    Synthetic control methods are commonly used in panel data settings to evaluate the effect of an intervention. In many of these cases, the treated and control units correspond to spatial units such as regions or neighborhoods. Our approach addresses the challenge of understanding how an intervention applied at specific locations influences the surrounding area. Traditional synthetic control applications do not address how to define the effective area of impact, the extent of treatment propagation across space, and the variation of effects with distance from the treatment sites.
To address these challenges, we introduce Spatial Vertical Regression (SVR) within the Bayesian paradigm. This innovative approach allows us to accurately predict the outcomes in varying proximities to the treatment sites, while meticulously accounting for the spatial structure inherent in the data. Specifically, rooted on the vertical regression framework of the synthetic control method, SVR employs a Gaussian process to ensure that the imputation of missing potential outcomes for areas of different distance around the treatment sites is spatially coherent, reflecting the expectation that nearby areas experience similar outcomes and have similar relationships to control areas.
This approach is particularly pertinent to our study on the Florentine tramway's first line construction. We study its influence on the local commercial landscape, focusing on how business prevalence varies at different distances from the tram stops.
\end{abstract}


\maketitle

\newpage

\section{Introduction}

In causal inference, researchers often deal with panel data where one or a few units experience the treatment starting from a common time point, while the remaining units remain control throughout.
Treatment effects on the treated units are often evaluated using the Synthetic Control (SC) method \citep{abadie2010synthetic, abadie2021, abadie2022synthetic}. With this method, the impact of an intervention is evaluated as the difference between the observed value of the outcome and its counterfactual value, imputed using a weighted average of outcomes of the control units, termed the synthetic control. 
The popularity of SC methods is growing rapidly, and its field of application ranges from the social sciences, to ecological studies, to policy evaluation. 
\cite{athey2017state} refer to it as ``arguably the most important innovation in the policy evaluation literature in the last 15 years'' and nowadays (March 2026), the reference paper, \cite{abadie2010synthetic}, counts over 8,600 citations on Google Scholar. 

In several applied problems, data have a spatial structure. For instance, in \cite{abadie2003}, Basque terrorism may reasonably affect the GDP in nearby regions, or in \cite{lee2023policy}, a VAT tax on alcoholics can affect beer vending in nearby counties.
As \cite{abadie2021} discusses, SC methods may be particularly suitable to assess causal effects in settings where units are   spatial areas such as neighborhoods, cities, counties, or states.

In several policy evaluation studies, the treatment is applied at some {\it locations}, referred to as {\it treatment sites}, and may affect the outcome up to an unknown distance. For instance, as in our case study of the Florentine tramway, 
the construction of new infrastructure at specific locations in a city may affect the socio-economic development of surrounding areas up to some distance.
Currently, there is no guidance on how to deploy SC methods accounting for the spatial propagation of the effects when the treatment is applied to a given spatial location.
In practice, one can apply SC methods separately to areas of different size around the treatment site, essentially neglecting the spatial correlation across these areas, which in turn may lead to less precise inferential results. 
However, exploiting the common spatial patterns across units may allow for information sharing in the imputation of counterfactuals for treated units. This is particularly relevant in contexts with few time periods with observed data before the treatment occurs, where SC methods are commonly deployed.

In this work, we aim to fill this gap by bridging  spatial statistics and synthetic control methodology.
We focus on evaluating the effect of a treatment applied on a set of locations on a sequence of treated areas, which are defined as the geographical units of varying distances from the treatment sites.
Under the Rubin causal model \citep{rubin1974}, we propose a Bayesian approach for imputing the missing potential outcomes that accounts for the presence of spatial correlation among the treated units.  
We refer to this method as \textit{Spatial Vertical Regression} or SVR. 
The term vertical regression was first introduced by \cite{athey2021matrix} to describe synthetic control-based methods. 
Adopting this approach,
we express the missing control potential outcome for a treated unit as a linear combination of the observed outcomes for the control units, up to residual error terms.
The key insight is that the relationship of a control unit with the treated areas is expected to be spatially coherent, in that the weight of a control unit in imputing a treated area's missing potential outcome should be similar across contiguous treated areas.
SVR encodes this by employing a Gaussian process prior on the coefficients of each control unit with a spatial correlation matrix.
This specification expresses that the weights given to each control unit in imputing the missing control potential outcome for the treated units may be similar for treated units of similar distance from the treatment locations. 
Positioned within the Bayesian paradigm, SVR allows us to easily quantify the uncertainty of the estimates, a still debated issue when SC methods are used. From a substantive perspective, the proposed approach is able to provide information on the spread of the treatment effects over space, which may be crucial to policy makers.

We evaluate SVR's performance and compare it against alternative approaches through an extensive simulation study. 
The simulation results suggest that SVR 
leads to efficiency gains in comparison to the competitors when spatial structure is present, and performs similarly in its absence. By leveraging the spatial structure of the data, SVR performs well and exhibits good inferential properties even in scenarios with few pre-treatment time periods, when existing methods 
may work poorly or even be infeasible.

Our motivating context is the evaluation of the effect that the construction and operational phases of the Florentine tramway network's first line had on the surrounding neighborhood's commercial vitality. This application is relevant as the effect of an urban transport facility may spread in the surrounding areas \citep{guerra2012half}. We study the effect of the tramway for the served 
neighbourhood at different distances from the stops. 
We find that the construction and operation of the tramway line had small and mostly statistically insignificant effects on the neighborhood's commercial vitality. More specifically, we observe an increase in the number of stores close to the tramway stops during the construction phase, likely due to the anticipation of the tramway's operational phase. This effect is very localized, within a two-minute walk from the stops, and diminishes quickly at larger distances. Negative impacts due to traffic congestion caused by construction work are not present. Areas at farther distances experience a steep decline in the inauguration year of the tramway, followed by a large recovery that granted an overall increase at the end of the observation period.

The paper will follow this outline: Section~\ref{sec:literature} briefly reviews the literature on causal inference with  panel data, Section \ref{sec:application} describes the motivating application, Section \ref{sec:cauframe} introduces the causal estimands, and Section \ref{sec:met} introduces the proposed approach. 
Section \ref{sec:sim} presents the simulations studies while Section \ref{sec:res} presents our application results. Finally Section \ref{sec:con} concludes and discusses future research.



\section{Related work}\label{sec:literature}
%


In the last years, a growing number of studies has contributed to the literature on causal inference with panel data, where a number of units are followed over time, and a subset of them is exposed to a treatment starting at a specific time point. Recent work applies, extends, or provides theoretical advances to the SC method introduced initially by \cite{abadie2003} \cite[see][for methodological details]{abadie2010synthetic, abadie2015}. 
See \cite{abadie2021} and \cite{arkhangelsky2023causal} for extended reviews.

In panel data settings, estimating causal effects often relies on estimating counterfactual control outcomes for the treated units in the post-intervention period.
Vertical Regression (VR) methods are often employed towards this goal. The term ``vertical regression'' was first introduced by \cite{athey2021} to
describe methods
in which  treated units' potential outcomes under control are expressed as a linear combination of the outcomes of control units. 
The coefficients of this linear combination are estimated by regressing  treated units' outcomes  on control units' outcomes at each pre-treatment time point.
These estimated coefficients are then applied to the control units' outcomes in the post-treatment period to predict the treated units' missing potential outcomes under control.
The original SC method imputes the missing control potential outcomes for the treated units in the post-treatment period as convex linear combinations of the control units. Therefore, the SC method has been characterized as a constrained least squares estimator of a vertical regression model.

In turn, vertical regression methods can be viewed as a modification of the SC method that relax the convexity restrictions and add intercepts \citep{hsiao2012panel, doudchenko2016balancing,ferman2021synthetic}.

A number of additional modifications have been suggested to the original SC method, including estimators that combine SC methods with matching \citep{abadie2021penalized, kellogg2021combining} or difference-in-differences \citep{arkhangelsky2021synthetic}, and the augmented SC estimators which also use a linear regression of the treated periods outcomes on the pre-treatment outcomes for the control units \citep{ben2021, ben2022}.
A strand of the literature has opted for factor-model approaches that motivate synthetic-control type algorithms and matrix completion methods \citep{xu2017generalized, amjad2018robust, agarwal2020synthetic, athey2021} and approaches with nonlinear panel data models such as the changes-in-changes model introduced by \cite{athey2006identification} and the distributional SC method \citep{gunsilius2023distributional}.

In synthetic control (SC) settings, quantifying uncertainty around the counterfactual trajectory is non-trivial, as standard errors are not available off-the-shelf. 
A classic sampling-based perspective, where units are viewed as a sample from a larger population, is controversial when the study includes all the units of the population of interest, as it is quite common in SC settings. 
A design-based or model-based perspective are alternative approaches.
A design-based perspective attributes randomness to the treatment assignment mechanism, viewing the potential outcomes as fixed \cite[e.g.,][]{bottmer2024design, athey2021}. As \cite{bottmer2024design} noted, the design-based perspective underlies the placebo tests that are commonly used in synthetic controls analyses \cite[e.g.,][]{abadie2010synthetic}, which requires exchangeability either over units or over time.
Conformal inference procedures have also been proposed \cite [e.g.,][]{chernozhukov2021exact, cattaneo2021prediction}, which  rely on exchangeability of the residuals from some model over time,  and thus, are closely related to the placebo test methods. The model-based perspective views the potential outcomes as random variables. Under this perspective, Bayesian approaches provide a natural framework to handle the large number of
structurally missing potential outcomes with automatic uncertainty quantification, without relying on large sample approximations.
Model-based Bayesian inference has been widely used for causal inference with panel data in general, and in SC settings in particular  
\cite[e.g.,][]{tuomaala2019bayesian, kim2020bayesian, samartsidis2020bayesian, menchetti2020estimating, pang2022bayesian, goh2022synthetic, giudice2022inference, ben2023estimating, fernandez2025bayesian}. See also \cite{martinez2022bayesian} for a discussion on Bayesian and Frequentist inference for SCs.

In this paper, we use a model-based Bayesian approach to inference, which allows us to naturally incorporate prior substantive knowledge about the spatial data structure into the causal analysis. 
The key feature of the SVR approach we propose is the use of a Gaussian process prior distribution to borrow information across spatially adjacent units in a panel data setting. 
SVR has similarities with the methods proposed by  \cite{vega2023spatio}  and \cite{ben2023estimating}, both of which employ latent factor models for the potential outcomes.
In a similar vein to us, the authors of these manuscripts deal with the spatial structure of the data using prior distributions in a Bayesian framework. 
Specifically, \cite{vega2023spatio} specify spatially-informed priors on the unit-specific factor loadings, 
and \cite{ben2023estimating} use multitask Gaussian Processes, a non-parametric Bayesian approach,  to generalize  low-rank factor models with a prior that encourages smoothness in the underlying factors.  
In contrast, our work is set within the SC framework which specifies that the control potential outcomes for the treated units are a unit-specific linear combination of the control potential outcomes for the control units. We capture spatial correlation across the treated units by employing a Gaussian process prior distribution on the coefficients of each control unit in the vertical regression model for the treated areas.
On a different vein, \cite{fernandez2025bayesian} use Bayesian shrinkage priors  to penalize the inclusion of short-distance control units to avoid biased estimates due to spillover effects. 

Our work is also related to the literature on causal inference with spatially correlated data in panel data settings, which often arise in environmental studies, possibly generating spillover effects \cite[e.g.,][]{delgado2015difference, sanford2018geospatial, bardaka2018causal, antonelli2020heterogeneous, butts2021difference, hettinger2023estimation}
and the literature on SC methods with  multiple outcomes \cite[see, e.g.,][]{billmeier2013assessing, acemoglu2016value, cunningham2018decriminalizing, kasy2023employing, ben2023estimating, sun2023using, tian2023synthetic}.
In causal inference with spatial panel data, focus is usually on accounting for spatial correlation or spatial spillover effects, whereas we are interested in assessing how the treatment effects propagate across space. 
In the presence of multiple  outcomes, SC methods are often applied separately to each  outcome variable. Recently, \cite{ben2023estimating, sun2023using, tian2023synthetic} proposed methods for estimating effects on multiple outcomes simultaneously. 
These methods are based on low-rank factor models with shared latent factors across different outcomes, and it is not completely clear which aspect of the data informs the correlation among outcomes. 
In the presence of spatial outcomes, we exploit the external information about the spatial structure in a principled way. 

Our work can be also interpreted as a contribution to the literature on causal inference with interference. 
Because we formulate the causal inference problem by viewing each spatial entity as a unit of which some are the treatment sites, we can interpret treatment effect propagation across space in terms of spatial spillover effects. Then, SVR can be viewed as a method to estimate treatment effects with panel data under interference   \cite[see, e.g.,][for recent contributions to causal inference under interference with panel data]{cao2019, grossi2025direct, di2020inclusive, agarwal2022network}. Indeed, our motivating study is 
the evaluation of the effects of a new tramway line construction in Florence, previously analyzed by \cite{grossi2025direct} with the objective to disentangle direct and spillover effects. Here, we reformulate their research question
in terms of assessing causal effects on spatial units.

\FloatBarrier

\section{Florence's first tramway line}
\label{sec:application}

The first line of the Florentine transit network began an ambitious urban renovation, improving accessibility in one of Italy’s most traffic-congested cities. Although plans for an integrated tramway date back to the 1990s, construction began in 2006, culminating in an inaugural run in 2010. As of 2026, two lines are operational, and three more are in development, slated for completion by 2030.

Investments in transportation infrastructure, especially in light rail systems, spark debate in urban and transport economics (e.g., \citealp{cervero1993assessing}, \citealp{landis1995rail}, \citealp{baum2000effects}, \citealp{hess2007impact}, \citealp{pan2013impacts}, \citealp{papa2015accessibility}). Such projects aim to reduce travel time and congestion while often revitalizing public spaces and commercial areas. These benefits can be more pronounced where there is a well-connected transit network, as shown by \cite{mejia2012transportation} and \cite{credit2018transit}.

Research commonly focuses on the positive correlation between improved transport and real estate prices \citep{pagliara2011urban, yan2012impact}. Complementary to that literature, some studies focus on the effects of changes in transportation systems on commercial vitality and location choices \citep{ray2015open, nilsson2016measuring, neto2025paths}. Some work focuses specifically on the causal effects of the construction of the first line of the Florentine tramway. \cite{grossi2025direct} study its effect on the commercial environment whereas \cite{budiakivska2018please} find an overall positive effect on real estate prices, coherent with the literature.

We contribute to this literature by evaluating the impact of the tramway on commercial vitality during the construction and operational phases of the tramway line, and, importantly, how it diffuses spatially at different distances from the tram stops.
In an evaluation study, it is crucial to consider both the construction and operational phases, since reduced accessibility during construction may displace businesses, especially those near the site. Nevertheless, once the tramway becomes operational, nearby areas may enjoy better accessibility and surroundings \citep{credit2018transit, pogonyi2021metros}. 



\begin{figure}[!b]
    \centering
   \includegraphics[width=0.75\textwidth]{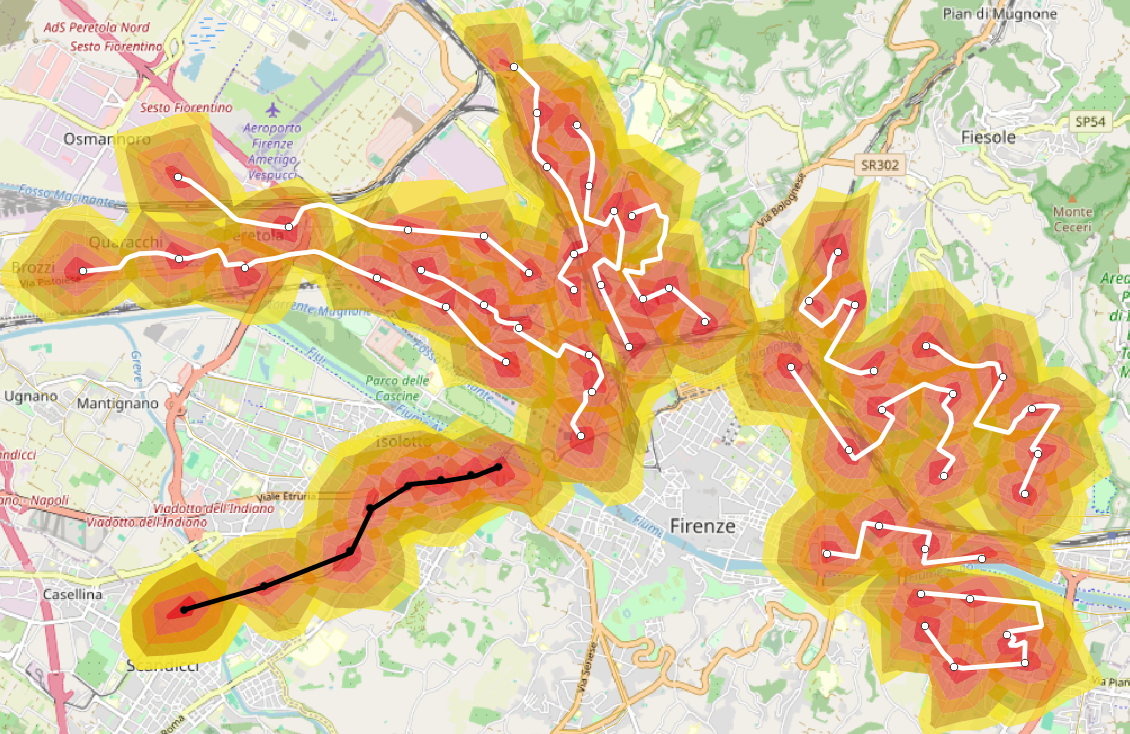}
    \caption{Catchment areas around the tramway stops representing areas within 120, 240, 360, 480 and 600 seconds of walking from a tramway stop, shown in red to yellow color. Black points and line represent the realized stops and tramway line, respectively, while white points and lines represent the hypothetical stops and tramway lines.}
    \label{fig:buffer}
\end{figure}

Figure \ref{fig:buffer} shows the study area: the black line and points indicate the first Florentine tramway line and stops, respectively. 
As we aim to assess how the treatment effect spreads across space, we define five catchment areas within 600 seconds walking distance from any tram stop, in two-minute increments.  
Walking distances are calculated using the API for isochrones from Openroute service and the \texttt{osrm} package in R \citep{osrm}.
We draw 11 hypothetical, alternative tramway lines and corresponding stops, shown in the figure in white color. These lines and stops will be the basis for acquiring a synthetic control. They are chosen either because they follow lines that have been constructed after our study period, or similar paths to the realized one \citep[see also][for more description on these alternative tramway lines]{grossi2025direct}. We define control catchment areas in a similar manner to the treated ones.

\begin{figure}[htbp]
    \centering
            \includegraphics[width=0.85\linewidth]{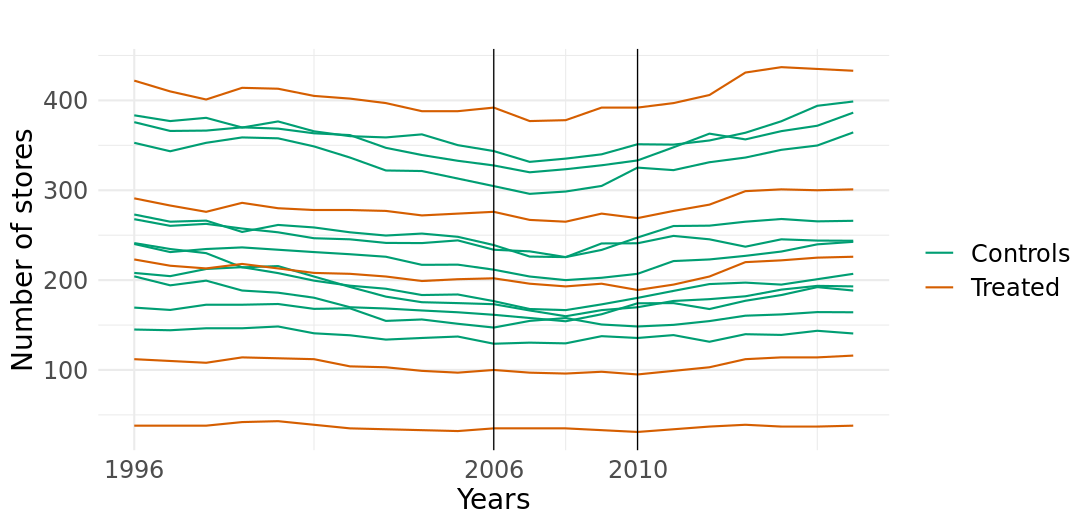}
    \caption{Number of Stores in Treated and Control Areas Across Time. 
    Treated units, shown in orange, correspond to five catchment areas within 120, 240, 360, 480 and 600 seconds of walking distance from any of the realized tramway stops.
    Control units, shown in light blue, correspond to average number of stores in their catchment areas, corresponding to the 11 hypothetical tramway lines.
    }
    \label{fig:florence_time}
\end{figure}

Our primary outcome to capture commercial vitality is the number of stores within each catchment area.
Store locations come from yearly data from ISTAT’s Statistical Archive of Active Firms (SAAF) for 1996--2016. Construction of the tramway began in early 2006 and ended in 2009. Therefore, the pre-treatment period includes only 10 time periods. 
The outcome of the five treated areas is defined as the number of stores within the corresponding catchment areas at the five different distances from the tram stops.
For the control areas, the outcome is defined as the average number of stores in the five catchment areas around the hypothetical tramway line.
Figure \ref{fig:florence_time} presents time series for treated and control units, showing limited variation in store counts and a common citywide trend, largely driven by macroeconomic factors. Vertical lines represent the beginning of the construction and operational phases of the tramway line.

\FloatBarrier

\section{Causal framework on spatial panel data}\label{sec:cauframe}

\subsection{The setup}

Consider a geography of interest denoted by $\Omega$ which we follow at time points $t = 1, 2, \dots, T$. Let $S = \{s_1, s_2, \dots, s_n\} \subset \Omega$ denote the locations on which an intervention occurred {\it simultaneously} at some intermediate time point $T_0+1$ where $1 \leq T_0 < T$. We refer to these locations as {\it treatment sites}. If $Z_t$ denotes the treatment level of the points in $S$ at time $t$, we have that
$$
Z_t = \begin{cases} 0, & \text{if } t \leq T_0 \\ 1, & \text{if } t > T_0 
\end{cases}.
$$
Therefore, we observe the region of interest under the treatment vector $\bm z_1 = (z_1, z_2, \dots, z_T)$ where
\begin{align}
\bm z_1 = ( \ \underbrace{0, 0, \dots, 0}_{T_0}, \ \underbrace{1, 1, \dots, 1}_{T - T_0} \ ) .
\label{eq:treatment}
\end{align}

Consider $N$ subsets of $\Omega$ representing $N$ different areas in the geography of interest, denoted by $A_i$, for $i = 1, 2, \dots, N$. Assume that the first $N_1$ units are defined as the areas within a given distance of the locations in $S$:
$$
A_i = \{\omega \in \Omega: d(\omega; S) < d_i \}, \quad i = 1, 2, \dots, N_1,
$$
where $d(\cdot ; \cdot)$ denotes some pre-specified distance function, $d(\omega; S)$ is the minimum distance of $\omega \in \Omega$ from the points in $S$, and $d_i > 0$ are pre-specified constants. Without loss of generality, assume that $d_1 < d_2 < \dots < d_{N_1}$. 
Therefore, the areas $A_1, A_2, \dots, A_{N_1}$ represent areas of increasing size around the treatment sites.
The remaining $N_0 = N - N_1$ areas are located at least $d_\infty$ away from all points in $S$ for some large value of $d_\infty$, implying that $A_i \cap A_j = \varnothing$ for $i \in \{1, 2, \dots, N_1\}$ and $j \in \{N_1 + 1, N_1 + 2, \dots, N\}$.

We measure an outcome on the geography of interest for each time period, denoted as $\mathcal Y_t = \{\mathcal Y_t(\omega), \ \omega \in \Omega \}$, and $t = 1, 2, \dots, T$. The outcome is compacted over the areas $A_i$ to represent an overall outcome in the area at time $t$, denoted as $Y_{it}$. For example, if $\mathcal Y_t$ is an integrable surface on $\Omega$, $Y_{it}$ might represent the total value over the set $A_i$, defined as $Y_{it} = \int_{A_i} \mathcal Y_t(\omega) \mathrm{d}\omega$. In the case where $\mathcal Y_t$ is binary over $\Omega$, the outcome $Y_{it}$ might instead represent the number of locations in $A_i$ for which $\mathcal Y_t$ is equal to 1 at time $t$.

In our study, $\Omega$ corresponds to the geographical area of the municipality of Florence, which we follow during the years 1996 to 2016. The set $S$ denotes the set of coordinates of the Florentine tramway stops, for which construction started in 2006. Therefore, we follow the area of Florence before the initiation of construction, and after construction has begun. The $N_1$ units correspond to areas surrounding the tramway stops that are within $d_i$ walking time distance from a tramway stop, whereas the remaining $N_0$ units correspond to areas that are far away from the locations of the tramway, as shown in \cref{fig:buffer}. Then, the spatial outcome $\mathcal Y_t$ is a binary indicator for each location denoting the presence or absence of a business at location $\omega$ at year $t$, and $Y_{it}$ denotes the number of businesses in area $A_i$ during the same year.

\subsection{Causal estimands and assumptions}
\label{subsec:estimands_assumptions}

We want to evaluate what would have happened had the points in $S$ not been treated at all. To do so, we consider a counterfactual treatment vector $\bm z_0$ where
\begin{align}
\bm z_0 = ( \ \underbrace{0, 0, \dots, 0}_{T_0}, \ \underbrace{0, 0, \dots, 0}_{T - T_0} \ ) .
\label{eq:control}
\end{align}
We assume that there exist potential outcome surfaces that correspond to the realized treatment vector in \cref{eq:treatment}, and the counterfactual treatment vector in \cref{eq:control}. Denote these potential outcome surfaces as $\mathcal Y_t^{\bm z_1}$ and $\mathcal Y_t^{\bm z_0}$, respectively. Then, the corresponding potential outcomes in the areas $A_i$ are denoted as $Y_{it}^{\bm z_1}$ and $Y_{it}^{\bm z_0}$, respectively. We assume that the observed outcome corresponds to the potential outcome under the observed treatment vector, $Y_{it} = Y_{it}^{\bm z_1}$.

We are interested in understanding the effect that the treatment initiation over locations $S$ at time $T_0+1$ had on the outcome of the areas near the treatment sites. Therefore, we are interested in studying
\begin{align}
\Delta_{it} = Y_{it}^{\bm z_1} - Y_{it}^{\bm z_0}, \quad \text{for } i = 1, 2, \dots, N_1, \text{ and } t = T_0 + 1, \ldots T.
\label{eq:effect}
\end{align}
We also define the intertemporal average:
\begin{eqnarray}\label{eq:ATE}
\Delta_{i} = \frac{1}{T-T_0}\sum_{t=T_{0}+1}^T\Delta_{it}
\end{eqnarray}    
to evaluate the overall effect throughout the treatment period for the area within distance $d_i$ from the treatment sites. 
The comparison of effects for areas at different distances 
can shed light on how the impact of the treatment emanates across space.

For units $i = 1, 2, \ldots, N_1$, the potential outcome  $Y_{it}^{\bm z_1}$ is observed at times after treatment initiation, $Y_{it}^{\bm z_1}= Y_{it}$, for $t = T_0 + 1, \ldots T$. Therefore, for these units and time periods, $\Delta_{it}= Y_{it}- Y_{it}^{\bm z_0}$, and only the control potential outcomes, $Y_{it}^{\bm z_0}$, for $t = T_0 + 1, \ldots T$ need to be estimated.
We make the following assumptions.

\begin{assumption}
The outcomes at time periods up to $T_0$ are unaltered by whether the treatment is initiated at $T_0$ or not, and $Y_{it}^{\bm z_1} = Y_{it}^{\bm z_0}$ for all $i = 1, 2, \dots, N$ and time periods $t = 1, 2, \dots, T_0 $.
\label{ass:nonanticipating}
\end{assumption}
\begin{assumption}
The potential outcomes for areas at least $d_\infty$ away from $S$ are unaffected by the treatment, and $Y_{it}^{\bm z_0} = Y_{it}^{\bm z_1}$ for $i = N_1 + 1, N_1 + 2, \dots, N$, and $t = 1, 2, \dots, T$.
\label{ass:controls}
\end{assumption}

\cref{ass:nonanticipating} formalizes that none of the areas responded to the treatment before the treatment was actually initiated, and that the observed outcomes before the intervention time $T_0+1$ are not affected by whether treatment will be initiated or not. This assumption is often referred to as the non-anticipation of the treatment assumption \citep[e.g.][]{bojinov2019time, menchetti2020estimating}. 

For that reason, we set $T_0$ based on the initiation of the line's construction phase, rather than based on its time of operation \citep{papadogeorgou2023evaluating}.
We believe that this choice renders \cref{ass:nonanticipating} plausible in our setting, as it is unlikely that the tramway line affected outcomes prior to its construction phase. The reason is that, even though the light rail project in Florence was announced in 2000, its implementation was marked by repeated failed tenders, prolonged legal disputes, and substantial public uncertainty regarding its feasibility and timing. Given this highly unstable and contested process, it is unlikely that private economic agents, in particular store owners, engaged in systematic anticipatory behaviour prior to the actual construction phase.

\cref{ass:controls} states that the $N_0$ areas that are far from the locations in $S$ are always unaffected by whether these locations receive the treatment or not. Therefore, under \cref{ass:controls}, the $N_0$ areas $A_i$, for $i = N_1 + 1, N_1 + 2, \dots, N$ are referred to as the control areas/units and the $N_1$ areas $A_i$, for $i=1, 2, \dots, N_1$, are referred to as the treated areas/units. Since the control areas are unaffected by the treatment initiation, they form a valid donor pool for imputing the missing control potential outcomes for the treated areas.

Let $\bm Y^1_t= (Y_{1t}, \dots, Y_{N_1t})$ and $\bm Y^0_t = (Y_{N_1+1t}, \dots, Y_{N_1+2t} \dots, Y_{N_1+N_0t})$ be the vectors representing the observed outcome for  the treated areas  and the control areas, respectively,  at time  $t$, $t \in \{1, \dots, T_0, \dots, T\}$.    

\section{Bayesian approach for Spatial Vertical Regression}\label{sec:met}

The goal is to impute the missing control potential outcomes, $ Y_{it}^{\bm z_0}$, for the treated units in the post-intervention time periods, $i=1, 2, \ldots, N_1$ and $t = T_0 + 1, T_0 + 2, \ldots, T$ using the following available information: pre-treatment period observed outcomes for the treated units, $\bm Y^1_t$, $t=1, 2, \ldots, T_0$, and pre- and post-treatment period outcomes for the control units, $\bm Y^0_t$, $t=1, 2, \ldots, T_0, T_0+1\ldots, T$. To do so, we borrow from the vertical regression framework.

\subsection{Formalization of vertical regression within the Bayesian paradigm}

Formally, we assume that the control potential outcomes for each treated unit are unit-specific linear combinations of the control potential outcomes of control units. Specifically, for $i=1, 2, \ldots, N_1$:
\begin{equation} \label{eq:model}
\begin{alignedat}{2}
 Y_{it}^{\bm z_0} &= \beta_0^{(i)} + \sum_{c=N_1+1}^{N_1+N_0} \beta_c^{(i)}
 Y_{ct}^{\bm z_0} + \epsilon_{it} 
 &\quad\quad&\raisebox{-\normalbaselineskip}[0pt][0pt]{for $t=1,\ldots, T_0,T_0+1, \ldots, T.$}
 \\
 &= \beta_0^{(i)} + \sum_{c=N_1+1}^{N_1+N_0} \beta_c^{(i)} Y_{ct}+ \epsilon_{it},
 \end{alignedat}
\end{equation}
The parameters of the linear combinations are the intercepts $\beta_0^{(1)}, \ldots, \beta_0^{(N_1)}$ and the weights for each control unit across the regressions of all treated units,
$\bm \beta_c \equiv \left(\beta_c^{(1)}, \ldots, \beta_c^{(N_1)}\right)$, $c=N_1+1, \ldots, N_1+N_0$. 
The model in \cref{eq:model} on the potential outcomes is the crucial assumption for estimation, as also employed in \cite{abadie2010synthetic} and \cite{athey2021matrix}. Therefore, we assume stable pattern between pre and post-treatment outcomes which implies that the conditional mean of the potential outcomes can be written in the form of \cref{eq:model} with fixed coefficients across time. We also assume the absence of idiosyncratic shocks in that the distribution of the error terms is unaltered across time, as in \cite{abadie2015}.

These coefficients are chosen using a Bayesian regression approach, where the pre-treatment outcomes of the treated units are regressed on the pre-treatment outcomes of the control units. Formally, for $i=1, 2, \ldots, N_1$,
\begin{equation} \label{eq:model_observed}
\begin{alignedat}{2}
 Y_{it} &= \beta_0^{(i)} + \sum_{c=N_1+1}^{N_1+N_0} \beta_c^{(i)}
 Y_{ct}^{\bm z_0} + \epsilon_{it}
  &\quad\quad&\raisebox{-\normalbaselineskip}[0pt][0pt]{for $t=1,\ldots, T_0$.}
 \\
 & = \beta_0^{(i)} + \sum_{c=N_1+1}^{N_1+N_0} \beta_c^{(i)} Y_{ct}+ \epsilon_{it},
\end{alignedat}
\end{equation}
In Equation \cref{eq:model_observed} the observed data in the pre-intervention period are used to estimate the model parameters for the potential outcomes model in \cref{eq:model}.
We specify that the $\epsilon_{it}$ are zero mean Normal error terms, possibly spatially correlated,
 $
\bm \epsilon_t = (\epsilon_{1t}, \ldots,  \epsilon_{N_1t})^{\T} \sim \text{MVN}_{N_1}(\bm 0_{N_1}, \Sigma_e),$ where $\text{MVN}_{N_1}(\bm 0_{N_1}, \Sigma_e)$ denotes a $N_1-$multivariate Normal distribution with vector mean zero and covariance matrix $\Sigma_e$. We discuss the choice of $\Sigma_e$ below.

When the units of interest are spatial areas, we propose to account for the spatial structure of the units by linking the estimation of the parameters in \cref{eq:model_observed} across treated areas. Doing so may provide valuable information for imputing the missing control outcomes, and thus, improve inference on the causal effects. In contrast, in the presence of multiple treated units, SC methods are generally applied to each treated unit separately, ignoring the correlation structure across the treated units and potentially incurring a loss in efficiency. 
This is particularly relevant in most settings where SC methods are applied, where the pre-intervention treatment period is relatively short, such as in our study of \cref{sec:application}.


\subsection{Spatially-informed coefficient vectors and residuals}

\subsubsection{Gaussian process specification on control unit's regression coefficients}

The key insight underlying our approach is that the treated units that are close will exhibit similar relationship with each control unit.
Thus, the coefficients of the linear combinations that define the control potential outcomes for the treated areas should be similar. 
Specifically, we expect that the coefficients of the linear combinations in Equation~\eqref{eq:model} are correlated and vary smoothly across treated units.

We embed these prior beliefs in the model using a Bayesian hierarchical approach with a Gaussian process prior on the coefficients $\bm \beta_c$. Formally, for $c=N_1+1, \ldots, N_1+N_0$,
\begin{equation}
\bm{\beta}_c \sim \mathcal{GP}(b_c \mathbf{1}_{N_1}, \Sigma_{\beta}),
\label{eq:GP_prior}
\end{equation}
where $b_c\mathbf{1}_{N_1}$ is the vector of means for the Gaussian process with $b_c\in \mathbb{R}$ a scalar parameter and $\mathbf{1}_{N_1}$ the $N_1-$dimensional vector with all elements equal to one.  
The coefficients' covariance matrix is set to 
$$
\Sigma_{\beta}=\sigma^2_\beta \mathcal{K}(D, \rho^2_\beta).
$$
Here, $D$ is the $N_1\times N_1$ matrix of the spatial distances among treated units, and $\mathcal{K}(D, \rho^2_\beta)$ denotes the quadratic exponential kernel. Then, the $(i,i')$ element of $\Sigma_{\beta}$ is \begin{equation}\label{eq:beta}
    \operatorname{Cov}(\beta^{(i)}_c,\beta^{(i')}_c) = \sigma^2_\beta \operatorname{exp}\left( - \frac{D_{ii'}^2}{2\rho_\beta^2}  \right).
\end{equation}
We specify the $(i,i')$ element of $D$ as $D_{ii'} = d_i - d_{i'}$, representing the difference of the distance of the treated areas from the treatment sites.  Alternative definitions of $D$ and kernel functions could be considered depending on the spatial relationships observed in the data. We opt for the quadratic exponential kernel because it naturally allows us to account for a-priori higher correlations among coefficients pertaining to adjacent treated units.

Imposing spatial correlation across the coefficients $\bm \beta_c$ aligns with the spatial structure of the data. We provide the following example as an illustration that the spatial correlation structure across units can affect the value of the coefficient for each control area across treated units. Suppose that for $t=1, 2, \ldots, T$,
$$
\left[ \begin{array}{c}
     \bm{Y}_t^{\bm z_0,1}\\
     \bm{Y}_t^{\bm z_0,0}
\end{array}\right]\sim \text{MVN}_N \left(\left[\begin{array}{c}
     \bm{\mu}^1\\
     \bm{\mu}^0
\end{array}\right]; \left[\begin{array}{cc}
     \Sigma_{11} & \Sigma_{10}\\
    \Sigma_{01} &  \Sigma_{00}
\end{array}\right]\right)
$$
where $\bm{Y}_t^{\bm z_0,1}$ and $\bm{Y}_t^{\bm z_0,0}$ denote the control potential outcomes for the treated units and the control units, and $\bm \mu^1, \bm \mu^0$ their means, respectively.
 Therefore, the matrices $\Sigma_{11}$,
$\Sigma_{00}$ and $\Sigma_{10}=\Sigma_{01}^\top$ are of dimension $N_1 \times N_1$, $N_0 \times N_0$, and  $N_1 \times N_0$ and they describe the spatial correlation structure across treated units, control units, and across treated and control units, respectively. 
Then, 
$$
 \bm{Y}_t^{\bm z_0,1}\mid 
    \bm{Y}_t^{\bm z_0,0} \sim MVN_{N_1} \left( 
     \bm{\mu}^{1|0}; \Sigma_e\right)
$$
where 
$
\bm{\mu}^{1|0} = \bm \mu_1 + \Sigma_{10} \Sigma_{00}^{-1} \left(\bm{Y}_t^0-\bm{\mu}^0\right)$,
and 
$\Sigma_e = \Sigma_{11}-\Sigma_{10} \Sigma_{00}^{-1}\Sigma_{01}$. Therefore $\bm \beta_c$ is the $c^{th}$ column of the $N_1 \times N_0$ matrix
$\Sigma_{10} \Sigma_{00}^{-1}$. By the block Cholesky decomposition, we have that
$$
\left[\begin{array}{cc}
     \Sigma_{11} & \Sigma_{01}^\top\\
    \Sigma_{01} &  \Sigma_{00}
\end{array}\right] =
\left[\begin{array}{cc}
     L_{11}& 0 \\
     L_{01}      & L_{00}
\end{array}\right] \left[\begin{array}{cc}
     L_{11}^{\T} & L_{01}^{\T}\\
     0      & L_{00}^{\T}
\end{array}\right]=
\left[\begin{array}{cc}
     L_{11}L_{11}^{\T}&  L_{11}L_{01}^{\T} \\
     L_{01} L_{11}^{\T}     & L_{00}  L_{00}^{\T}
\end{array}\right]
$$
where  $L_{11}$ is the Cholesky decomposition of the submatrix $ \Sigma_{11}$: $ \Sigma_{11}=L_{11}L_{11}^{\T}$, $L_{01}$ is the $N_0 \times N_1$ matrix equal to $\Sigma_{01}L_{11}^{-\T}$ and $L_{00}$ is the Cholesky decomposition of the matrix 
$\Sigma_{00} - L_{01}L_{01}^{\T}$.  Then,
$$
\Sigma_{10} \Sigma_{00}^{-1} = \Sigma_{01}^{\T} \Sigma_{00}^{-1} =
L_{11} L_{01}^{\T}  (L_{00} L_{00}^{\T})^{-1} =
L_{11} L_{01}^{\T}  L_{00}^{-\T} L_{00}^{-1}.
$$
Therefore, $\bm \beta_c$ is the $c^{th}$ column of the $N_1 \times N_0$ matrix $L_{11} L_{01}^{\T}  L_{00}^{-\T} L_{00}^{-1}$. This decomposition illustrates that the vector of regression coefficients corresponding to each control unit, $\bm \beta_c$, depends on the spatial correlation structure across treated units through $L_{11}$. 

The spatial dependence in the regression coefficients propagates to the imputation of the missing control potential outcomes for the treated units. If $\bm \theta$ denotes the full parameter vector,
then for $t =T_0+1, \ldots, T$, 
$$
\bm Y_t^{\bm z_0, 1} \mid \bm Y_1^{1}, \ldots, \bm Y_{T_0}^1,
\bm Y_1^{0}, \ldots, \bm Y_{T}^0, \bm \theta \sim \text{MVN}_{N_1}(\bm \mu_t, \Sigma_e)
$$
where $\bm Y_t^{\bm z_0, 1} = (Y_{1t}^{\bm z_0}, \ldots, Y_{N_1t}^{\bm z_0})^{\T}$. The $i^{th}$ element of the $N_1-$dimensional mean vector $\bm \mu_t$ is 
$$
\mu_{it} = \beta_0^{(i)} + \sum_{c=N_1+1}^{N_1+N_0} \beta_c^{(i)}
 Y_{ct}^{\bm z_0},
$$
illustrating that the spatial information entering the regression coefficients propagates to the imputed outcomes.

\subsubsection{Spatial dependence in outcome model residuals}

By specifying the Gaussian process prior on the regression coefficients in \cref{eq:GP_prior}, we specify that the treated units might form similar relationship with each control unit. In addition, we allow for spatial dependence in the control potential outcomes of the treated units. To do so, we return to the choice of $\Sigma_e$ in \cref{eq:model_observed}.
We specify that the $(i,i')$ element of the covariance matrix $\Sigma_e$ is
\begin{equation}\label{eq:cov_errors}
    \operatorname{Cov}(e_{it},e_{i't}) = \sigma^2_e \times \left (w\operatorname{exp}\left(-\frac{D_{ii'}^2}{2\rho_e^2}  \right) + (1-w)  \right),
\end{equation}
with $w \in [0,1]$. For $i=i'$ we have $\operatorname{Cov}(e_{it},e_{i't})=\operatorname{Var}(e_{it}) =\sigma^2_e$.  The overall variance, $\sigma^2_e$, scales the spatial correlation, which is a convex linear combination of two components. One component depends on the 
spatial distance of
units $i$ and $i'$, in an exponential quadratic form, where $\rho_e^2$ is a decay parameter that adjusts the rate at which the spatial correlation diminishes with  the distance.
The other component represents an idiosyncratic term according to which the error terms are independent and identically distributed.

It is useful for illustration to discuss the two extremes where $w \in \{0,1\}$. For $w=0$, the spatial correlation is completely captured by the outcomes of the control units and the error terms $e_{it}$ are $N(0, \sigma^2_e)$ independent  random variables. For $w=1$, the variance-covariance matrix of the error terms across treated units only depends on the spatial structure. Therefore, the errors are specified to be a mixture of a spatial and an independent component. 

The error terms are specified to be independent over time. Therefore, the temporal correlation structure of the potential outcomes is assumed to be captured by the temporal correlation structure of the control units. 
We can easily relax this assumption, but we maintain it for simplicity and also for comparison with original SC methods where it is often made.

\paragraph{}

The Bayesian regression approach we propose presents several advantages.
We embed information on the spatial correlation structure across units directly into the model. Through the Gaussian Process priors on the weights of the linear combinations defining the control potential outcomes for the treated units, we   inform the model that those weights are expected to vary smoothly across treated units whose distance from the treatment sites increases progressively. 
This reflects our belief that nearby treated units have similar relationships to control units, 
and the similarity diminishes as the distance of the treated units increases. Additionally, we allow for spatial correlation in the outcome model residuals.  
Last but not least, the Bayesian regression approach naturally provides a quantification of the uncertainty of the imputed control potential outcomes for the treated units, and thus, of the causal effects. This is an important  advantage.  The quantification of the uncertainty in causal panel studies, especially when SC methods are used,  is still an open issue and subject to debate \citep{shen2022same}.

The importance of accounting for dependence  across units, time and outcomes in causal panel-data studies has recently attracted the attention of the causal inference community \cite[e.g.,][]{ben2023estimating, sun2023using}. 
Our method presents some   similarities with the method recently proposed by \cite{ben2023estimating}. We both use Gaussian processes to account for dependence across units, but in a different way. \cite{ben2023estimating} assume that the  model  for the control outcomes for all units 
is  a unit-specific linear combination of   latent time-varying  variables shared across units. They specify a Gaussian process for each vector of  time-varying latent variables, where the kernel of the Gaussian process captures the correlation of the latent variables across time. 
Instead, we embed the spatial structure information in the Gaussian Process priors of the coefficients of the control units in the linear combinations which define the control potential outcomes for the treated units. The kernels of the Gaussian Process priors impose a structure across the unit-specific weights of the treated units that reflects the spatial correlation structure across them. 

\subsection{Choice of prior distributions and computation}
\label{subsec:priors}

First, we discuss priors that relate to the control units' coefficients. To encourage regularization in low-information settings such as data with a small number of pre-intervention time periods, we specify a Laplace prior on a control unit's overall importance across treated areas, 
$b_c \sim \hbox{Laplace}(\lambda_b)$.
We adopt prior distributions 
$\sigma^2_\beta  \sim \mathcal{N^+}(a_{\beta}, \eta_\beta)$
and
$\rho^2_{\beta} \sim  \hbox{Gamma}(a_{\rho_\beta}, \eta_{\rho_\beta}),$
where the Gamma distribution is specified using the shape and rate parameters, and $\mathcal{N^+}$ represents the Normal distribution truncated below at zero.

For the remaining parameters of the model, including the intercepts, 
the variance of the residual error terms,  
and the parameters 
governing the errors' correlation structure, we specify the  following priors: 
\begin{eqnarray*}
&     \beta_0^{(i)} \sim N(\mu_0, \sigma^2_0), \quad i=1, \ldots, N_1,& \\
    &
    \sigma^2_e \sim \mathcal{N^+}(a_e, \eta_e), &\\ 
    &\rho^2_e \sim  \hbox{Gamma}(a_{\rho_e}, \eta_{\rho_e}),&
     \\
    &w \sim \hbox{Beta}(a_w, b_w). &
    \end{eqnarray*}

Disparities in outcome magnitudes across units can affect the estimation of spatially correlated coefficients, leading the model to focus on absolute levels rather than on the underlying trends. We bypass this challenge by normalizing the data to correct for differences in the overall level and size across units. We normalize the time series of each treated and control units' outcomes with respect to the pre-treatment mean and standard deviation \citep[as suggested also in][]{abadie2010synthetic, abadie2021}. Specifically, we use
$$ \widetilde{Y}_{it}= \dfrac{Y_{it} - \overline{Y}_i}{ \sqrt{\frac{1}{T_0 -1} \sum_{t=1}^{T_0} (Y_{it} - \overline{Y}_i)^2 }}, $$
where $\overline{Y}_i=\sum_{t=1}^{T_0}Y_{it}/T_0$. In what follows, we assume that the outcomes $Y$ are normalized, and avoid using the notation $\widetilde Y$.  Along with standardizing the outcomes, we rescale the values $d_i$ to vary from 0 to 1.

Working with normalized data allows us to specify values for hyperparameters that are expected to perform well across a variety of settings (as also evident in our simulations in \cref{sec:sim}). Since employing diffuse, non-informative prior distributions can perform poorly \citep{gelman2008weakly}, we instead adopt weakly informative prior distributions. For normalized data, we set the hyperparameters of the prior distributions as follows:  $\lambda_b = 0.1$ $\mu_0=0$, $\sigma^2_0=1$, $a_\beta=0$, $\eta_\beta=0.35$, $a_{\rho_\beta}=0.5$, $\eta_{\rho_\beta}=2$, $a_e=0$, $\eta_e=0.5$, $a_{\rho_e}=0.5 $, $\eta_{\rho_e}=1.5$, and $a_w=2, \text{ } b_w =2$. These choices are adopted throughout our work.
A formal justification of these choices along with draws from the induced prior distributions for $\bm \beta_c$ and the residual spatial correlation of treated areas are given in Supplement \ref{supp_sec:prior_justification}.

The posterior distribution of the model parameters and the missing control potential outcomes for the treated units is derived using Hamiltonian Monte Carlo (HMC) implemented in Stan \citep{Carpenter2017stan} and its R interface \citep{Rstan}. 
The code developed to obtain the results presented in this paper is publicly available at the following GitHub repository: 
\href{https://github.com/Giulio-Grossi/SVR}{https://github.com/Giulio-Grossi/SVR}. 
This software enables the replication of all experiments described in the paper and it can be employed in any setting where a generic distance measure between units is defined. As a result, the proposed method is highly flexible and can be adapted to a variety of applications and analytical contexts. Furthermore, the repository includes the code for obtaining estimates and confidence intervals using the other methods evaluated in the simulations of \cref{sec:sim}.

\FloatBarrier
\section{Simulation Study}\label{sec:sim}
\subsection{Data generating process}


In this section, we present a simulation study designed to assess how spatial correlation among units affects the regression coefficients and to evaluate the performance of our method relative to the original SC approach and a standard vertical regression under various scenarios, including a benchmark scenario with no spatial structure.

We consider a causal panel study with $N_1=5$ treated areas and $N_0=10$ control areas observed  in  $T_0$ pre-treatment periods and $T-T_0$   post-treatment periods. We consider three lengths of the pre-treatment  period, $T_0=10,20,40$, combined with the following lengths of the post-treatment period: $T-T_0=5,10, T_0/2$, for a total of seven distinct scenarios (for $T_0=10$, $T-T_0=T_0/2=5$, and for $T_0=20$, $T-T_0=T_0/2=10$). 
Therefore, we consider simulation scenarios closely aligned with the empirical setting described in \cref{sec:application}. In particular, we consider a short panel ($T_0=10$) with a similar number of treated and control units as in our study, and allow for different levels of spatial correlation to reflect empirically plausible spatial dependence structures.

The control potential outcomes for the treated units are generated according to Equation~\eqref{eq:model}, with  the intercepts, $\beta_0^{(1)}, \ldots, \beta_0^{(N_1)}$,  the weights $\bm \beta_c = (\beta_c^{(1)}, \ldots, \beta_c^{(N_1)})^{\T}$, $c=N_1+1, \ldots, N_1+N_0$, the observed control potential outcomes for the control areas, $\bm Y_t^0$, $t=1,2, \ldots, T$ and the error terms  generated as follows.
The intercepts, $\beta_0^{(1)}, \ldots, \beta_0^{(N_1)}$, are generated independently from a standard Normal distribution.
The coefficients of a control unit in the regression for all treated areas, $\bm \beta_c = (\beta_c^{(1)}, \ldots, \beta_c^{(N_1)})^{\T}$ for $c=N_1+1, \ldots, N_1+N_0$, are generated from 
a multivariate Normal distribution. The mean vector is set to $m_c \bm 1_{N_1}$, where $m_c \sim N(0, 1)$. The covariance matrix is set to $\sigma^2_s\mathcal{K}(D, \rho_s^2)$, with element $(i,i')$ equal to $$
    \operatorname{Cov}(\beta^{(i)}_c,\beta^{(i')}_c) = \sigma^2_s\operatorname{exp}\left(- \frac{D_{ii'}^2}{2\rho_s^2}  \right).
$$
We set $\sigma^2_s=0.4$ and consider four values of the lengthscale parameter $\rho_s^2$: $0.001^2$, $0.2^2$, $0.4^2$, and $0.6^2$  corresponding to scenarios where the weights associated with the same control unit for different treated units are essentially independent, weakly correlated, mid-correlated, and strongly correlated, respectively.  
These scenarios mimic studies where accounting for the spatial correlation structure among the treated units in estimating the regression weights may be progressively more relevant, and the case of $\rho_s^2 = 0.001^2$ represents a scenario where our approach is not expected to provide any advantages. In Supplement \ref{supp_sec:illustration_simulation}, we include a visualization of $\bm \beta_c$ in our simulations under the different values of $\rho_s^2$.

We generate temporally-correlated outcomes for the control areas. Specifically, let $\bm Y_c^{\bm z_0} = \bm Y_c =(Y_{c1}, \ldots, Y_{cT})^{\T}$ represent the times-series under no intervention for the $c^{th}$ unit for $c=N_1+1, \ldots, N_1+N_0$, which we observe. We draw $\bm Y_c$ from a $T-$dimensional multivariate normal distribution. We specify mean vector $t_c \bm 1_T$ where $t_c \sim \mathcal{N}(0, 0.7^2)$. The covariance matrix is set to $\Sigma_{T\times T} = \sigma^2_c \mathcal{K}(D_T, \rho^2_c)+\eta_c \mathbf{I}_T$, where  $D_T$ is a $T\times T$ matrix of time differences and  $\mathbf{I}_T$ is the $T \times T$ identity matrix, so that
$$
\operatorname{Cov}(Y_{ct},Y_{ct'}) = \sigma^2_c\operatorname{exp}\left( - \frac{[D_T]_{tt'}}{2\rho_c^2}  \right) + \eta_c \mathrm{I}\{t = t'\}, $$
and $\mathrm{I}\{\cdot\}$ is the indicator function.
We set $\sigma^2_c=0.3^2$, $\rho^2_c=0.05^2$, $\eta_c=0.15^2$, for all $c=N_1+1, \ldots, N_1+N_0$.
In Supplement \ref{supp_sec:illustration_simulation}, we illustrate that the outcome time series for control units under this design has temporal correlation that resembles the data in \cref{fig:florence_time}.

Finally, we generate $
\bm \epsilon_t = (\epsilon_{1t}, \ldots,  \epsilon_{N_1t})^{\T} \sim \text{MVN}_{N_1}(\bm 0_{N_1}, \Sigma_e),$ where the covariance matrix is constructed according to Equation \eqref{eq:cov_errors}. We consider three scenarios for the error structure. In the first scenario we set $w=0$, which implies that the error terms are independent $N(0, \sigma^2_e)$ random variables. To ensure a realistic signal to noise ratio, we set $\sigma_e^2$ equal to the $40\%$ of the mean (over the treated units) of the variance of the linear combinations $\beta_0^{(i)} + \sum_c \beta_c^{(i)} Y_{ct}$, $t=1, \ldots, T$. 
In the other two scenarios we set $w=0.5$, and lengthscale parameter $\rho^2_e=0.2^2$. The two scenarios differ with respect to the value of $\sigma^2_e$, which we set equal to $40\%$ and $70\%$ of the mean of the variance 
defined as in the independent case, representing scenarios with medium and high noise, respectively.

These specifications result in $7 \times 4 \times 3=84$ different scenarios that vary the lengths of the pre- and post-treatment periods, the strength of the spatial correlation across the regression coefficients, and the residual error structure.
We consider such diverse scenarios in order to investigate the role of using information on the spatial structure of the data in different challenging settings.

\subsection{Alternative approaches}

We compare our approach (SVR) with five alternative existing methods: the original SC method, three vertical regression methods (two Frequentist and one Bayesian), and a Bayesian synthetic control method.
All the methods except SVR are implemented separately for each treated unit; thus, they do not account for information on the spatial correlation among treated units.

To implement the original SC method \citep{abadie2010synthetic}, we adopt Euclidean distance for the estimation of the weights.
%
Furthermore, we implement three vertical regression methods \citep{doudchenko2016balancing} using the simple Ordinary Least Squares estimator (OLS), the ridge regression estimator (SR; which stands for ``separate ridge'' since we implement the ridge regression estimator for each treated unit separately), and a Bayesian
Vertical Regression (BVR) approach.
Finally, we implement a Bayesian synthetic control method \cite[BSC, implemented in Stan; e.g.][]{kim2020bayesian, martinez2022bayesian} using a Bayesian simplex regression approach.
See Supplement \ref{supp_sec:prior_justification} for the choice of prior distributions.
For the Bayesian methods, we consider three chains of $10,000$ iterations each and a burn-in of 5,000. 
We assess the convergence using the potential scale-reduction statistic \citep{gelman1992inference}.

Since inference in the original SC method remains challenging and subject to debate, 
we opt for using the original synthetic control method only to obtain point estimates of 
the causal effects, without uncertainty quantification.
For OLS, we use the results obtained by \cite{shen2022same} to construct confidence intervals.
We use the ridge regression only to obtain point estimates of the missing control potential outcomes for the treated units and the causal effects, 
given that quantification of the uncertainty in this context is still an open issue.
The Bayesian approaches, including the proposed SVR and BSC, have the great advantage to naturally provide a quantification of the uncertainty, having as output the posterior predictive distribution of the missing control potential outcomes and the posterior distribution of the causal effects.

\subsection{Evaluation metrics}


We generate $L=200$ data sets for each scenario. We evaluate and compare the performance of each method using  the bias and the Mean Squared Error (MSE) of the imputed  missing control potential outcomes for the treated units averaged over the post-treatment periods and over the treated units. Specifically,
$$
\text{Bias} = \dfrac{1}{TN_1}\sum_{t=T_0+1}^T \sum_{i=1}^{N_1} \text{Bias}_{it} = 
\dfrac{1}{TN_1}\sum_{t=T_0+1}^T \sum_{i=1}^{N_1} \dfrac{1}{L} \sum_{\ell=1}^{L}\left[
\widehat Y_{it}^{\bm z_0, (\ell)} - Y_{it}^{\bm z_0, (\ell)}
\right]
$$
and
$$
\text{MSE} = \dfrac{1}{TN_1}\sum_{t=T_0+1}^T \sum_{i=1}^{N_1} \text{MSE}_{it} = 
\dfrac{1}{TN_1}\sum_{t=T_0+1}^T \sum_{i=1}^{N_1} \dfrac{1}{L} \sum_{\ell=1}^{L}\left[
\widehat Y_{it}^{\bm z_0, (\ell)} - Y_{it}^{\bm z_0, (\ell)}
\right]^2,
$$
where $\widehat Y_{it}^{\bm z_0, (\ell)}$ is the imputed value and $Y_{it}^{\bm z_0, (\ell)}$ is the true value of the control potential outcome for treated unit $i$ at time $t$ in the $\ell^{th}$ replication from the DGP.  
Average bias and MSE for the Bayesian methods are calculated using the posterior medians of the posterior predictive distributions of the missing control potential outcomes.

For the vertical regression OLS method, and the Bayesian methods (BVR,  BSC and  our SVR method)  we also calculate the average coverage probabilities  (ACP) of the 95\% confidence / credible  intervals: Let $C_{0.95}(Y_{it}^{\bm z_0, (\ell)})$ denote the 
95\% confidence / credible  intervals for $Y_{it}^{\bm z_0, (\ell)}$, then
$$
\text{ACP} = \dfrac{1}{TN_1}\sum_{t=T_0+1}^T \sum_{i=1}^{N_1} \text{CP}_{it} = 
\dfrac{1}{TN_1}\sum_{t=T_0+1}^T \sum_{i=1}^{N_1} \dfrac{1}{L} \sum_{\ell=1}^{L} \mathrm{I}\left\{Y_{it}^{\bm z_0, (\ell)} \in C_{0.95}(Y_{it}^{\bm z_0, (\ell)}) \right\}.
$$
We opt for these four methods as they are the ones with a clear quantification of the uncertainty.

\subsection{Simulation results}


We focus on the results in the $7 \times 3 = 21$ scenarios with mid-correlation across the control units' coefficients, that is,
$\rho^2_{s} = 0.4^2$. 
We discuss the remaining results briefly afterwards.

\begin{table}[!b]
\centering
\resizebox{0.75\textwidth}{!}{ 
\begin{tabular}{cccccccccc}
\hline \\[-8pt]
& \multicolumn{3}{c}{$T_0=10$} & \multicolumn{3}{c}{$T_0=20$} & \multicolumn{3}{c}{$T_0=40$} \\
& IID & Sp--40\% & Sp--70\% & IID & Sp--40\% & Sp--70\% & IID & Sp--40\% & Sp--70\%  \\
\cmidrule(lr){2-4} \cmidrule(lr){5-7} \cmidrule(lr){8-10}
& \multicolumn{9}{c}{$T-T_0 = 5$} \\
\cmidrule(lr){2-10}
\textbf{True} & 1.71 & 1.71 & 1.71 & 1.68 & 1.67 & 1.67 & 1.66 & 1.67 & 1.67\\
\addlinespace
SC & 0.05 & 0.08 & 0.08 & -0.03 & -0.02 & -0.03 & 0.06 & 0.06 & 0.05\\
SR & 0.07 & 0.08 & 0.08 & 0.00 & 0.00 & 0.01 & 0.00 & 0.00 & 0.00\\
OLS & -0.07 & -0.11 & -0.15 & -0.01 & -0.03 & -0.04 & -0.01 & -0.01 & -0.02\\
BVR & 0.01 & 0.10 & 0.10 & 0.00 & -0.02 & -0.02 & -0.01 & -0.01 & -0.02\\
BSC & 0.07 & 0.08 & 0.08 & -0.03 & -0.02 & -0.03 & 0.06 & 0.06 & 0.06\\
\textbf{SVR} & 0.03 & 0.08 & 0.10 & 0.01 & 0.00 & 0.00 & -0.01 & -0.01 & -0.01\\
\hline \\[-8pt]
& \multicolumn{9}{c}{$T-T_0 = 10$} \\
\cmidrule(lr){2-10}
\textbf{True} & 1.69 & 1.68 & 1.68 & 1.65 & 1.66 & 1.66 & 1.70 & 1.71 & 1.71\\
\addlinespace
SC & 0.06 & 0.09 & 0.09 & -0.01 & 0.00 & -0.01 & 0.05 & 0.05 & 0.05\\
SR & 0.07 & 0.08 & 0.09 & 0.02 & 0.01 & 0.01 & 0.00 & 0.00 & 0.00\\
OLS & -0.10 & -0.08 & -0.11 & -0.01 & -0.04 & -0.05 & -0.01 & -0.01 & -0.01\\
BVR & -0.01 & 0.06 & 0.05 & 0.00 & -0.03 & -0.03 & -0.01 & -0.01 & -0.01\\
BSC & 0.07 & 0.08 & 0.08 & 0.00 & 0.01 & 0.00 & 0.04 & 0.05 & 0.05\\
\textbf{SVR} & 0.02 & 0.06 & 0.07 & 0.01 & 0.01 & 0.01 & -0.01 & 0.00 & -0.01\\
\hline\\[-8pt]
& \multicolumn{9}{c}{$T-T_0 = 0.5 \times T_0$} \\
\cmidrule(lr){2-10}
\textbf{True} &  &  &  &  &  &  & 1.71 & 1.71 & 1.71\\
\addlinespace
SC &  &  &  &  &  &  &  0.06 & 0.06 & 0.06\\
SR & &  &  &  &  &  &  0.00 & 0.00 & 0.00\\
OLS & \multicolumn{3}{c}{same as $T-T_0 =5 $} &  \multicolumn{3}{c}{same as $T-T_0 =10$} &  0.00 & 0.00 & -0.01\\
BVR &  &  &  &  &  &  &  0.00 & 0.00 & -0.01\\
BSC & &  &  &  &  &  &  0.05 & 0.05 & 0.05\\
\textbf{SVR} &  &  &  &  &  &  &  0.00 & 0.00 & 0.00\\
\hline \\
\end{tabular}
}
\caption{Average post-treatment bias of Synthetic Control (SC); Separate Ridge (SR) regression; OLS  vertical regression; Bayesian Vertical Regression (BVR); Bayesian Synthetic Control (BSC) and Spatial Vertical Regression (SVR) in scenarios with spatially mid-correlated weights ($\rho^2_s = 0.4^2$). Columns correspond to the length of the pre-intervention period $T_0$ and the three cases of residual error structure, IID, Spatial with 40\% noise, and Spatial with 70\% noise. Rows correspond to the length of the post-intervention period and the different methods. Bias and outcome mean value are expressed in the same unit of measure. }
\label{tab:bias_04}
\end{table}

\begin{table}[!t]
\centering
\resizebox{0.75\textwidth}{!}{  
\begin{tabular}{cccccccccc}
\hline \\[-8pt]
& \multicolumn{3}{c}{$T_0=10$} & \multicolumn{3}{c}{$T_0=20$} & \multicolumn{3}{c}{$T_0=40$} \\
& IID & Sp--40\% & Sp--70\% & IID & Sp--40\% & Sp--70\% & IID & Sp--40\% & Sp--70\%  \\
\cmidrule(lr){2-4} \cmidrule(lr){5-7} \cmidrule(lr){8-10}
& \multicolumn{9}{c}{$T-T_0 = 5$} \\
\cmidrule(lr){2-10}
SC & 0.97 & 1.00 & 1.36 & 0.72 & 0.73 & 0.98 & 0.73 & 0.74 & 0.99\\
SR & 0.83 & 0.84 & 1.04 & 0.58 & 0.58 & 0.90 & 0.40 & 0.42 & 0.71\\
OLS & 12.65 & 16.22 & 28.36 & 1.00 & 1.05 & 1.83 & 0.44 & 0.47 & 0.82\\
BVR & 2.19 & 2.20 & 3.53 & 0.82 & 0.84 & 1.40 & 0.43 & 0.46 & 0.79\\
BSC & 0.96 & 0.97 & 1.21 & 0.80 & 0.80 & 1.02 & 0.81 & 0.84 & 1.08\\
\textbf{SVR} & 0.57 & 0.65 & 0.95 & 0.38 & 0.46 & 0.72 & 0.31 & 0.37 & 0.62\\
\hline \\[-8pt]
& \multicolumn{9}{c}{$T-T_0 = 10$} \\
\cmidrule(lr){2-10}
SC & 1.26 & 1.32 & 1.76 & 0.87 & 0.88 & 1.15 & 0.82 & 0.82 & 1.06\\
SR & 0.98 & 0.97 & 1.18 & 0.69 & 0.69 & 1.03 & 0.44 & 0.45 & 0.75\\
OLS & 20.85 & 23.60 & 41.22 & 1.27 & 1.33 & 2.34 & 0.49 & 0.51 & 0.89\\
BVR & 3.25 & 3.25 & 5.20 & 1.01 & 1.02 & 1.71 & 0.47 & 0.49 & 0.86\\
BSC & 1.14 & 1.14 & 1.39 & 0.92 & 0.93 & 1.16 & 0.91 & 0.92 & 1.14\\
\textbf{SVR} & 0.71 & 0.84 & 1.17 & 0.42 & 0.52 & 0.80 & 0.32 & 0.38 & 0.64\\
\hline \\[-8pt]
& \multicolumn{9}{c}{$T-T_0 = 0.5 \times T_0$} \\
\cmidrule(lr){2-10}
SC &  &  &  &  &  &  & 0.85 & 0.86 & 1.10\\
SR &   &  &  &  &  &  & 0.46 & 0.48 & 0.78\\
OLS &  \multicolumn{3}{c}{same as $T-T_0 = 5$} & \multicolumn{3}{c}{same as $T-T_0 = 10$} & 0.52 & 0.54 & 0.95\\
BVR &   &  &  &  &  &  & 0.50 & 0.52 & 0.90\\
BSC &   &  &  &  &  &  & 0.93 & 0.94 & 1.17\\
\textbf{SVR} &   &  &  &  &  &   & 0.33 & 0.40 & 0.66\\

\hline \\
\end{tabular}
}
\caption{Average post-treatment MSE of Synthetic Control (SC); Separate Ridge (SR) regression; OLS  vertical regression; Bayesian Vertical Regression (BVR); Bayesian Synthetic Control (BSC) and Spatial Vertical Regression (SVR) in scenarios with spatially mid-correlated weights ($\rho^2_s = 0.4^2$). Columns correspond to the length of the pre-intervention period $T_0$ and the three cases of residual error structure: IID, Spatial with 40\% noise, and Spatial with 70\% noise. Rows correspond to the length of the post-intervention period and the different methods.}
\label{tab:mse_04}
\end{table}

\begin{table}[!t]
\centering
\resizebox{0.75\textwidth}{!}{  
\begin{tabular}{cccccccccc}
\hline \\[-8pt]
& \multicolumn{3}{c}{$T_0=10$} & \multicolumn{3}{c}{$T_0=20$} & \multicolumn{3}{c}{$T_0=40$} \\
& IID & Sp--40\% & Sp--70\% & IID & Sp--40\% & Sp--70\% & IID & Sp--40\% & Sp--70\%  \\
\cmidrule(lr){2-4} \cmidrule(lr){5-7} \cmidrule(lr){8-10}
& \multicolumn{9}{c}{$T-T_0 = 5$} \\
\cmidrule(lr){2-10}
OLS & 0.00 & 0.00 & 0.00 & 0.84 & 0.83 & 0.83 & 0.74 & 0.74 & 0.74\\
BVR & 0.93 & 0.92 & 0.91 & 0.94 & 0.94 & 0.93 & 0.95 & 0.94 & 0.94\\
BSC & 0.80 & 0.79 & 0.81 & 0.88 & 0.87 & 0.88 & 0.90 & 0.90 & 0.90\\
\textbf{SVR} & 0.93 & 0.92 & 0.93 & 0.94 & 0.94 & 0.94 & 0.94 & 0.94 & 0.94\\
\hline \\[-8pt]
& \multicolumn{9}{c}{$T-T_0 = 10$} \\
\cmidrule(lr){2-10}
OLS & 0.00 & 0.00 & 0.00 & 0.85 & 0.84 & 0.84 & 0.77 & 0.77 & 0.77\\
BVR & 0.93 & 0.93 & 0.91 & 0.94 & 0.94 & 0.93 & 0.95 & 0.95 & 0.94\\
BSC & 0.78 & 0.78 & 0.80 & 0.86 & 0.86 & 0.87 & 0.88 & 0.89 & 0.90\\
\textbf{SVR} & 0.93 & 0.93 & 0.93 & 0.94 & 0.94 & 0.94 & 0.94 & 0.94 & 0.94\\
\hline \\[-8pt]
& \multicolumn{9}{c}{$T-T_0 = 0.5 \times T_0$} \\
\cmidrule(lr){2-10}
OLS &  &  &  &  &  &  & 0.78 & 0.77 & 0.77\\
BVR & \multicolumn{3}{c}{same as $T-T_0=5$}  & \multicolumn{3}{c}{same as $T-T_0=10$} & 0.97 & 0.96 & 0.95\\
BSC &  &  &  &  &  &  & 0.88 & 0.88 & 0.89\\
\textbf{SVR} &  &  &  &  &  &  & 0.94 & 0.94 & 0.94\\

\hline \\
\end{tabular}
}
\caption{Average post-treatment   coverage of  95\% confidence / posterior credible intervals for the imputed post-treatement control  potential outcomes for the treated units based on  OLS vertical regression; Bayesian Vertical Regression (BVR); Bayesian Synthetic Control (BSC) and Spatial Vertical Regression (SVR) in scenarios with spatially mid-correlated weights ($\rho^2_s = 0.4^2$). Columns correspond to the length of the pre-intervention period $T_0$ and the three cases of residual error structure: IID, Spatial with 40\% noise, and Spatial with 70\% noise. Rows correspond to the length of the post-intervention period and the different methods.}
\label{tab:cov_04}
\end{table}

Tables \ref{tab:bias_04} and \ref{tab:mse_04} show the average bias and MSE for SVR and the five alternative methods, respectively, and Table~\ref{tab:cov_04} shows ACP
for OLS vertical regression, BVR, BSC and our SVR method.
Unsurprisingly, all the methods perform well in terms of bias under all scenarios.
%
At the same time, SVR has the smallest MSE in all scenarios with mid spatial correlation ($\rho^2 = 0.4^2$),  illustrating that it leads to more efficient imputations of the missing control potential outcome in the presence of spatial correlation in the controls' coefficients. Notably, this result is confirmed for a short pre-intervention period ($T_0 = 10 $), which is very common in applied studies.

At the same time, SVR leads to straightforward and appropriate uncertainty quantification, as we can see in Table~\ref{tab:cov_04}. The ACP of the posterior 95\% credible intervals 
using SVR is very close to the nominal level of 95\% in all scenarios. BVR yields similar ACP results, but a higher MSE across all scenarios.
In contrast, 
OLS and BSC have coverage of the 
95\% intervals that is substantially lower than their nominal level.

\begin{figure}[!t]
    \centering
    \includegraphics[width=.9\linewidth]{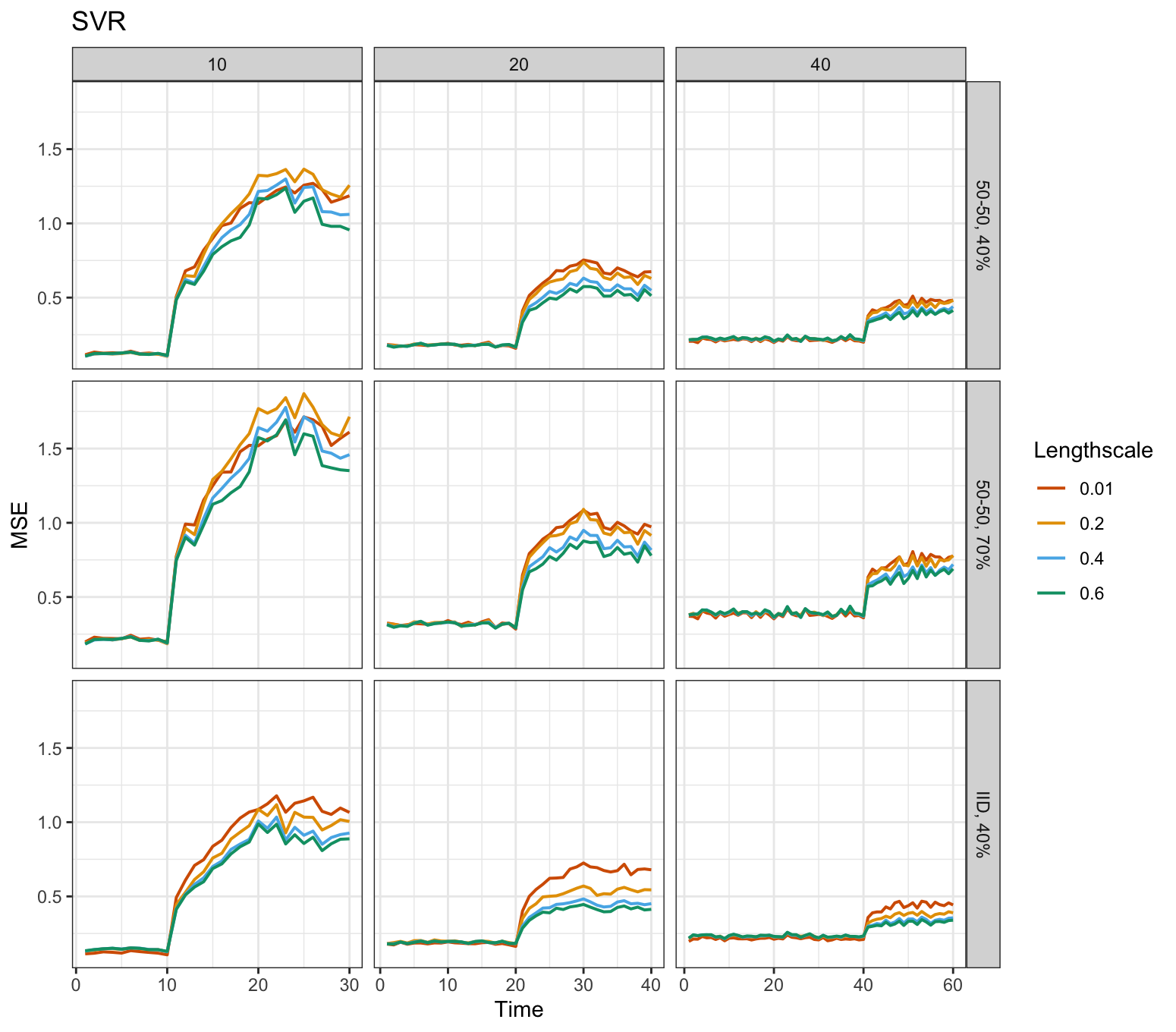}
    \caption{MSE for SVR estimates when $T_0 = 20$ by varying the spatial correlation $\rho^2_s$ (colors),  the length of the pre-treatment period (columns) and the error correlation structure (rows).
    }
    \label{fig:mse_smac_main}
\end{figure}

Additional simulation results are included in Supplement \ref{supp_sec:additional_simulations}, including results for the other levels of spatial correlation across the control units' coefficients, $\rho^2_s$. We summarize the results here. Conclusions remain overall similar, with SVR performance being stronger against alternatives for larger values of spatial correlation, as expected. SVR has lower MSE comparing with alternative methods, excluding the scenario with short pre-treatment, low spatial correlation across outcomes, and high error, where SR has marginally lower MSE. At the same time, its performance is comparable to alternative methods even in the absence of spatial correlation ($\rho^2_s = 0.001^2$), especially for larger values of $T_0$. Lastly, Figure~\ref{fig:mse_smac_main} reports the MSE for SVR estimates when $T_0 = 20$ varying the spatial correlation, $\rho^2_s$,  the length of the post-treatment period and
the error correlation structure. Evidently, the MSE of SVR decreases with higher level of spatial correlation under all the scenarios considered, illustrating that SVR's performance improves in the presence of spatial correlation across the coefficients. On the other hand, the alternative methods do not exploit the spatial structure to reduce MSE (see Figures \ref{fig:mse_bsc}, \ref{fig:mse_sc}, \ref{fig:mse_bvr} in Supplement \ref{supp_sec:additional_simulations}). 



Overall, our simulation results show that SVR performs well under different scenarios.
In scenarios where the presence of spatial correlation implies that weights defining the control potential outcomes vary smoothly across spatially correlated treated units, SVR outperforms the competitors, improving the precision of inferences in terms of MSE and coverage probability of  corresponding intervals. At the same time, it performs similarly to the competitors  even in scenarios where there is no spatial correlation in the regression coefficients across treated units. Lastly, SVR leads to good inferential results throughout all cases considered, a challenging task in the synthetic control literature.

\section{Estimating the effect of the Florentine tramway construction}
\label{sec:res}

%

We apply SVR to assess the effect of the first line of the Florentine tramway network on the commercial vitality of the surrounding areas, measured by the number of stores in the area.
First, we normalize the pre-intervention time series for each unit following the procedure discussed in \cref{subsec:priors}.
Because there is an evident temporal trend across all the time series (see \cref{fig:florence_time}), we de-trend the normalized outcomes for both the treated and control units. To do so, we use the control areas to calculate Florence's overall time trend, $ \text{Trend}_t = \frac{1}{N_0} \sum_{c= N_1+1}^N {Y}_{ct}$, and we subtract it from all units. Therefore, the outcomes used in the estimation are 
$$ Y_{it}^* = {Y}_{it} - \text{Trend}_t. $$
%
%
Reported results are on the original scale.
In 
Supplement \ref{supp_sec:additional_application} we include the time series and correlation plot for the treated and control outcomes after normalization and de-trending. 

\begin{figure}[!t]
    \centering
    \includegraphics[width=\linewidth]{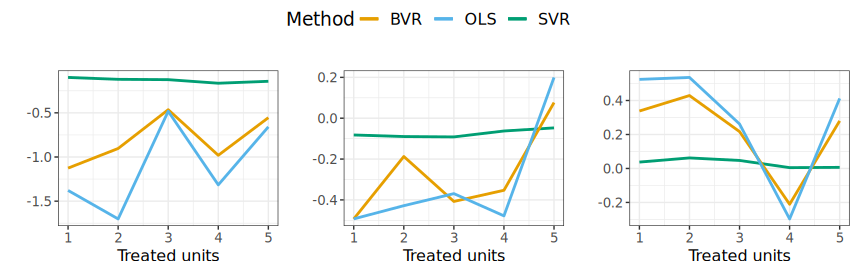}
    \caption{Estimated $\bm \beta_c$ coefficients for three randomly chosen control units from three different estimation methods. See \cref{fig:trace_all} in Supplement \ref{supp_sec:additional_application} for the figures corresponding to all control units.}
    \label{fig:smoothed_beta}

\vspace{10pt}
    \includegraphics[width=\textwidth]{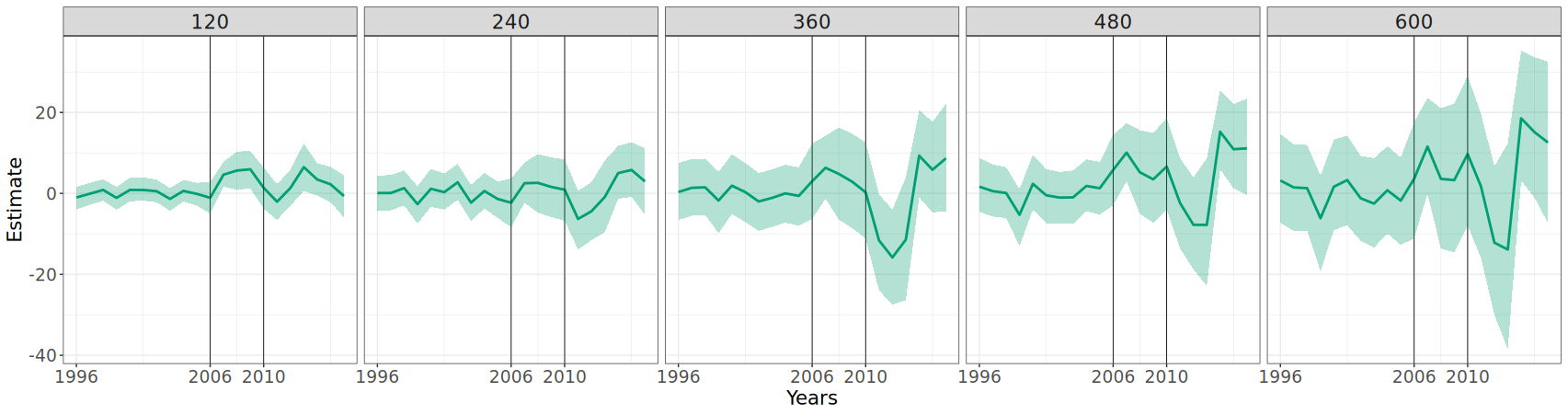}
    \caption{Posterior medians (lines) and 95\% posterior credible intervals (shaded areas) of the treatment effects     $\Delta_{it}$ for areas  within $d-$seconds of walking from a tramway stop, $d=120, 240, 360, 480,600$.
    The two vertical lines represent the beginning of the tramways construction and operational phases in 2006 and 2010, respectively.
    }
    \label{fig:results}
\end{figure}

In \cref{fig:smoothed_beta}, we report the estimated $\bm \beta_c$ coefficients of three control units across the treated units, from OLS, BVR and SVR, using the posterior mean for BVR and SVR. The estimates from OLS and BVR vary substantially across contiguous treated units, which may indicate overfitting, especially considering the short pre-treatment period.
In contrast, SVR borrows information across treated units and returns spatially coherent coefficients $\bm \beta_c$.


Figure \ref{fig:results} and Table \ref{tab:res} show the results using SVR. Figure \ref{fig:results} shows the posterior medians and 95\% posterior credible intervals for the effects $\Delta_{it}$ defined in Equation \ref{eq:effect}, where intervals are constructed using the 2.5 and 97.5 quantiles of the posterior distribution. 
Table \ref{tab:res} reports the same quantities for the intertemporal average causal effects $\Delta_{i}$ defined in Equation \ref{eq:ATE}, where the average is defined separately over the construction phase (2006--2009) and the operational phase (2010--2016). 
Results based on the alternative methods evaluated in Section \ref{sec:sim} are reported in Figure \ref{fig:application_comparison} in Supplement C.


\begin{table}

\caption{Posterior medians and 95\% posterior credible intervals of the intertemporal average causal effects $\Delta_{i}$ for areas within $d-$seconds walking distance from the tramway stop, $d=120, 240, 360, 480,600$, and posterior median of  the pre-treatment root mean squared prediction error (RMSPE). 
}
\label{tab:res}

\centering
\begin{tabular}{lccccc}
\toprule
  & 120 & 240 & 360 & 480 & 600\\
  \midrule
 &  \multicolumn{5}{c}{Construction period} \\
\cmidrule{2-6}
Estimate & 3.306 & 1.064 & 3.453 & 6.232 & 6.356\\
CI & (-1.003, 7.572) & (-5.596, 7.670) & (-6.772, 14.028) & (-3.239, 16.185) & (-9.463, 22.689)  \\
  \midrule
    &  \multicolumn{5}{c}{Operational period} \\
\cmidrule{2-6}
Estimate & 1.803 & 0.359 & -2.510 & 3.228 & 3.650\\
CI  & (-2.938, 6.485) & (-7.076, 7.839) & (-14.645, 10.008) & (-8.065, 15.378) & (-14.930, 23.361) \\
\midrule
RMSPE & 0.834 & 1.584 & 1.291 & 2.094 & 2.769 \\
\bottomrule
\end{tabular}
\end{table}

First of all, \cref{fig:results} suggests that SVR provides a good fit for the data, since the estimates and 95\% intervals in the pre-treatment period well-overlap with zero. Moreover, the overall RMSPE is ranging between 0.6\% and 2.2\% of the pre-treatment average outcome, suggesting a reasonable fit.
During the construction phase, our results show positive effects for the area nearest to the tramway stops ($\leq$2 minute-walk) which attenuate at larger distances, 
though the credible intervals always cover zero.
This illustrates that businesses might have invested in the tramway area during the construction phase, before the tramway became operational.
However, during the operational phase of the tramway line, we estimate
a sharp decline in the number of shops immediately following the opening of the line in 2010 and 2011, followed by a rebound and growth in the subsequent years. 
This pattern can be explained by the economic environment in Italy during those years. The 2008 economic crisis and the problematic state of Italian sovereign debt slowed the opening of new shops and commercial hubs everywhere in Florence. In the tramway area specifically, the effect might have been more pronounced than elsewhere, given its ongoing renewal of commercial offerings at precisely the time the crisis hit. As a result, the overall impact of the tramway’s construction was delayed, although it becomes noticeable again starting in 2013.
However, the credible intervals almost always cover zero, so we cannot draw firm conclusions.

We further study how the impact of the tramway on commercial vitality varies across types of stores, particularly regarding the purchase frequency of the products offered. 
Following \cite{grossi2025direct}, we distinguish between stores selling durable and non-durable goods. 
We find that 
stores selling durable goods exhibit limited sensitivity to the tramway, with positive effects only within a large area from the tram stops in the later post-treatment period. In contrast, non-durable goods stores experience a temporary contraction during construction, followed by a clear post-construction increase in activity, particularly within intermediate walking distance from the tram stops. We present the results in Supplement C.3.


\FloatBarrier

\section{Concluding Remarks}\label{sec:con}


In this work, we propose a novel approach, Spatial Vertical Regression (or SVR), for drawing causal inferences in panel data settings where units are spatial areas. 
Positioned within the framework of vertical regression for panel data,
SVR imputes the missing control potential outcomes  for treated units based on linear combinations of the observed control potential outcomes for the control units.
The core idea of SVR is that these linear combinations are expected to vary smoothly across contiguous areas.
This is encoded by employing a Gaussian process prior on the coefficients of the linear combinations, which imposes a correlation structure on the coefficients associated with the same control unit in the linear combinations across treated units.
In an extensive simulation study, 
SVR performs very well, outperforming existing alternative methods 
in terms of average bias, MSE and coverage,
even in challenging scenarios with short pre-treatment periods.
 
We deploy SVR to assess the effects on the commercial vitality (number of shops) arising from constructing the first line of the Florentine tramway. We consider the areas that are reachable within 120, 240, 360, 480, and 600 seconds of walking by each tramway stop as the treated areas,
and we design control areas based on hypothetical alternative tramway lines. 
We find small and generally statistically insignificant effects, with some indication for an increase in the area's commercial vitality in the later years.

Causal analysis of panel data with spatial units is challenging, and there are a number of extensions of our work that can be considered. Of particular interest to us is to develop methodology for panel data where the point pattern outcome is directly analyzed, rather than aggregating the point pattern data to an areal outcome. However, causal inference with point pattern treatment or outcome is challenging \citep{wang2020design, papadogeorgou2022causal}. Extending panel data methodology to spatio-temporal point pattern outcomes is an interesting line of future work, which could provide further insights on the localization of the impact of urban interventions such as the tramway line construction.

\newpage
\bibliographystyle{abbrvnat}
\bibliography{bibtex}
\newpage
\clearpage

\begin{appendices}
\setcounter{page}{1}
\setcounter{section}{0}    
\renewcommand{\theassumption}{S.\arabic{assumption}}
\renewcommand{\theequation}{S.\arabic{equation}}
\renewcommand{\thetable}{S.\arabic{table}}
\renewcommand{\thefigure}{S.\arabic{figure}}
\setcounter{equation}{0}    
\setcounter{table}{0}    
\setcounter{figure}{0}

\section{The choice of hyperparameters of the prior distributions}
\label{supp_sec:prior_justification}

\subsection{Prior distributions in SVR}

We choose the values of the hyperparameters such that the induced prior distribution is weakly-informative and covers all reasonable values, based on the following reasoning.

For standardized outcomes for treated and control units that have mean equal to zero and standard deviation equal to one, the coefficient of a control unit in the treated unit regression should remain within reasonable (and not too large) distance of zero. Furthermore, the choice of prior distributions and corresponding hyperparameters should allow for coefficients $\bm \beta_c$ that have anywhere from no to strong spatial correlation across treated areas. Towards this goal we specify:

\begin{itemize}
    \item For the parameters $b_c$ which represent the overall importance of a control unit across radii, we have specified $b_c \sim \hbox{Laplace}(\lambda_b)$ with $\lambda_b = 0.1$. Therefore, a priori the overall importance of a control unit is between $-1$ and 1 with probability 99.97\%. This is reasonable considering that all data are standardized. 
    Since outcomes are standardized to a unit standard deviation, the coefficients of the control units in the vertical regression model are expected to be limited in magnitude.
   \item The parameter $\sigma^2_\beta$ represents the variance of the coefficients in $\boldsymbol{\beta}_c$ around their mean $b_c$. 
   We specify a Truncated Normal prior in zero  to $\sigma^2_\beta$ with hyperparameters $(a_\beta, \eta_\beta) = (0,\, 0.35)$. 
   This choice corresponds to a commonly used weakly informative prior, \citep{gelman2006prior} which exerts minimal influence on the scale of $\boldsymbol{\beta}_c$ while still allowing the model to regularise extreme values of the variance. 
   Because the prior places substantial mass near zero and exhibits very heavy tails, it encourages the coefficients to remain concentrated around their centre unless the data provide strong evidence for a larger variance.
   Similarly, the error variance $\sigma^2_e$ is assigned an Truncated Normal prior in zero prior with hyperparameters $(a_e, \eta_e) = (0,\, 0.5)$, which again yields a weakly informative and scale–flexible prior.


    \item The spatial range of the coefficients $\bm \beta_c$ is dictated by $\rho^2_\beta$ on which we impose a Gamma prior with shape parameter $\alpha_{\rho_\beta} = 0.5$ and rate parameter $\eta_{\rho_\beta} = 2$. This prior specifies that a priori $P(\rho^2_\beta< 0.1) \approx 0.47, P(\rho^2_\beta < 0.3) \approx 0.73, P(\rho^2_\beta < 0.5) \approx 0.83$, and $P(\rho^2_\beta< 1) \approx 95\%$. Considering that all distances are standardized to cover $[0,1]$, this prior assigns sufficient weight to all reasonable values of $\rho^2_\beta$.
\end{itemize}

\begin{figure}[p]
\centering
\includegraphics[width=0.8\linewidth]{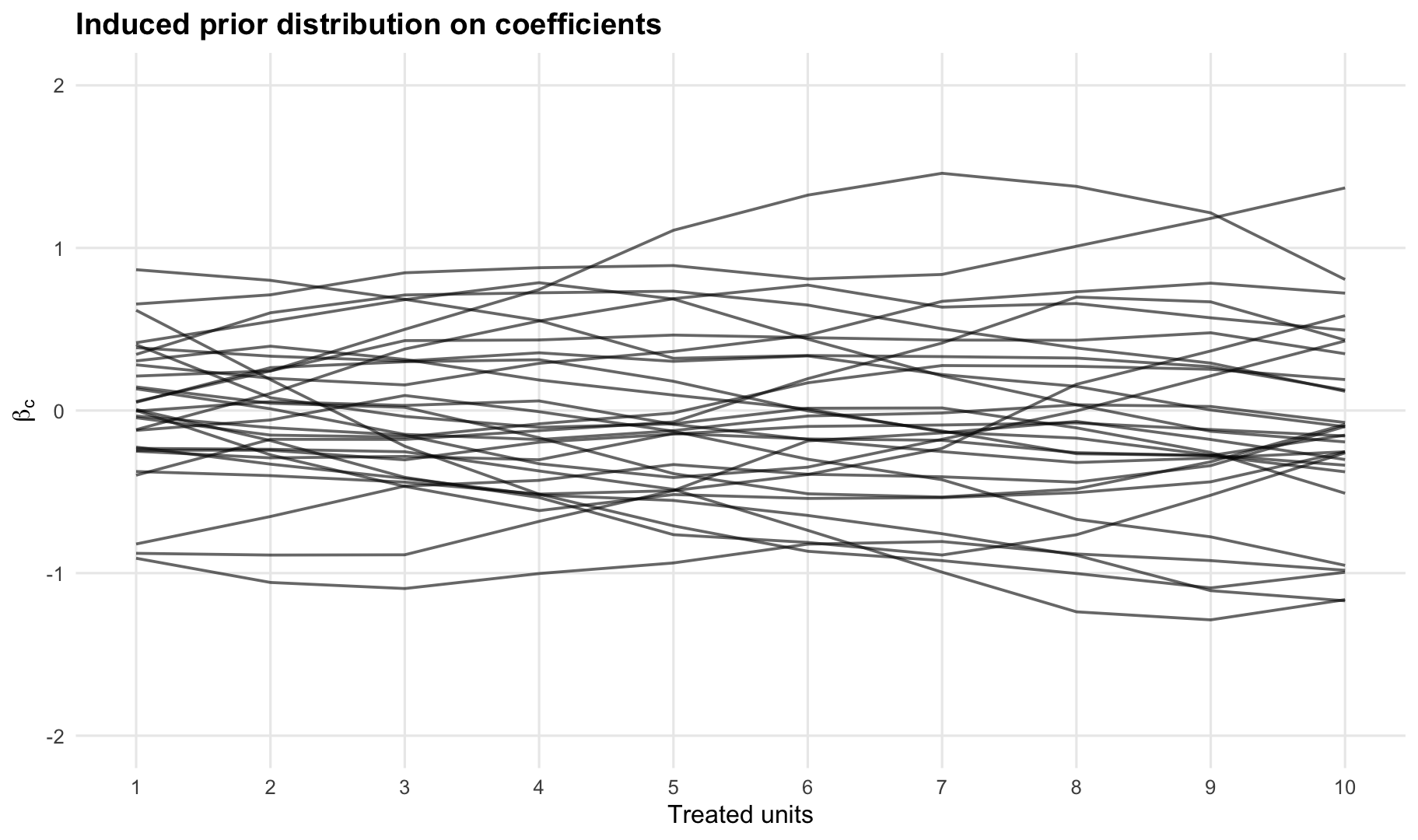}
    \caption{Draws of $\bm \beta_c$ from the induced prior distribution on these coefficients.}
    \label{fig:beta_from_prior}
\vspace{30pt}

\includegraphics[width=0.9\linewidth]{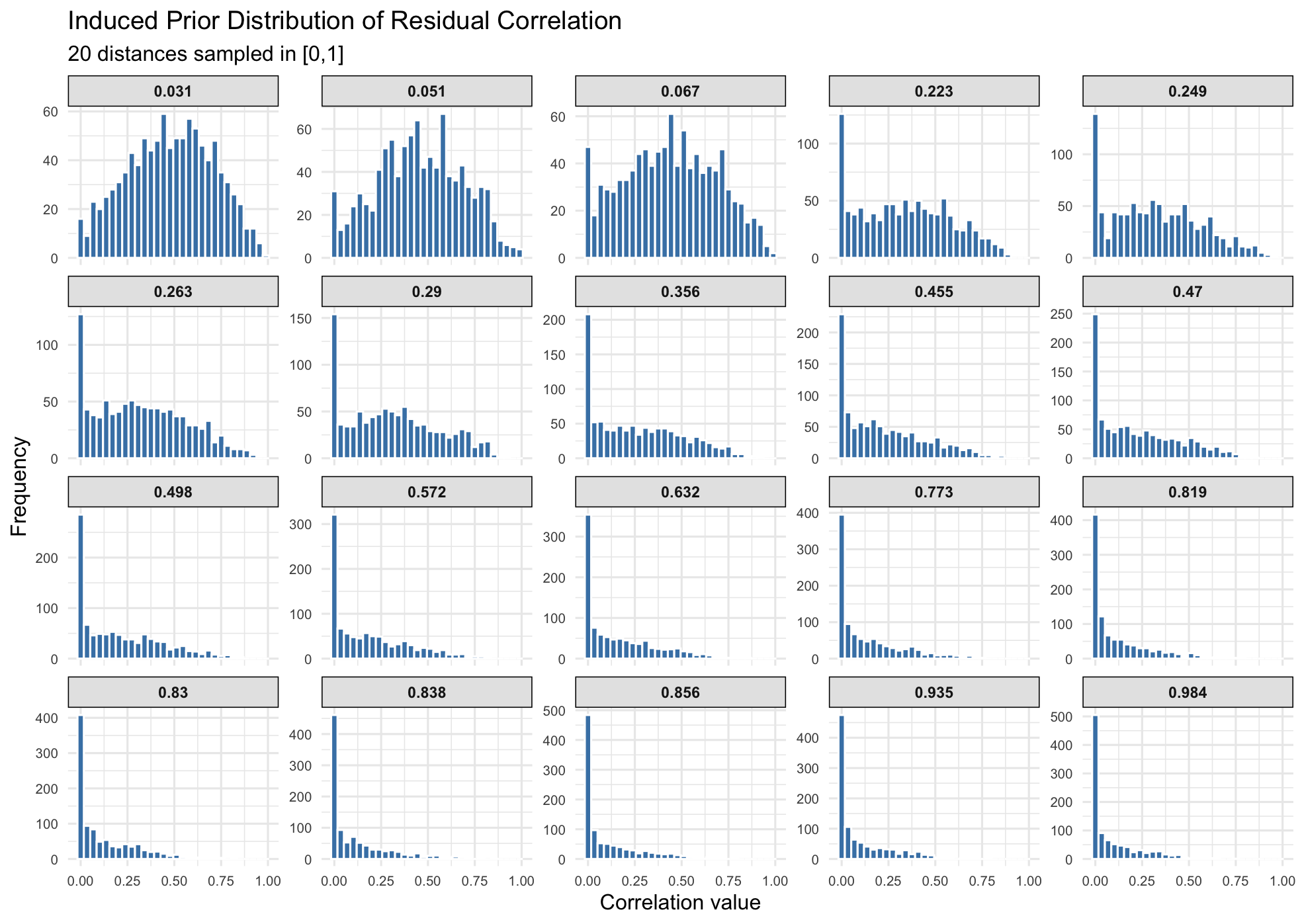}
\caption{Induced prior distribution for the correlation of the treated area with $d_i = 0$ with the remaining areas with $d_i\in[0,1]$.}

\label{fig:prior-residuals}
\end{figure}

In \cref{fig:beta_from_prior}, we include the curves for 30 independent draws from the induced prior distribution on $\bm \beta_c$. We see that the values cover a range from at least $-1$ to 1, and these draws include curves with strong or essentially absent spatial correlation.

Furthermore, we study the induced prior distribution on the treated units' residuals in the vertical regression. Again, we want our prior choice to allow for anywhere from weak to strong spatial correlation.

\begin{itemize}
    \item For the overall variance of the residuals we have specified the following shape and scale parameters in the Truncated Normal prior: $a_e=0$, and $\eta_e=0.5$. This distribution specifies that, a priori, $\sigma^2_e$ is less than 0.1, 0.3, 0.5, and 1 with probability that is approximately equal to 0.47, 0.73, 0.84, and 0.95, respectively. Since the outcome is standardized to have (unconditional) variance equal to 1, it is preferable that the prior distribution for the residual variance is concentrated on values below 1. Furthermore, this prior allows for a wide range of values for the residual variance, implying that the control units' outcomes can explain little or a lot of the variability in a treated unit's outcomes.
    
    \item For the lengthscale of the residuals we have specified $a_{\rho_e}=0.5 $, 
    and $\eta_{\rho_e}=1.5$ in the Gamma prior. This prior specifies that $\rho^2_\epsilon$ is less than 0.1, 0.3, 0.5 and 1 with probability 0.4, 0.67, 0.78 and 0.92. A similar argument to the prior on $\rho^2_\beta$ holds here.
\end{itemize}

For illustration, we consider 20 treated areas with values $d_i$ that are equally spread from 0 to 1, and we draw the matrix $\Sigma_e$ from its induced prior distribution 1,000 times. \cref{fig:prior-residuals} shows the histogram of the induced correlation of the treated area with $d_i = 0$ with the areas with the remaining values of $d_i$, across the 1,000 draws. We see that the induced prior distribution allows for spatial correlation at the smallest distances, while also allowing for a spatial correlation at moderate distances and declines at larger distances. Therefore, the induced prior covers a wide spectrum of spatial structures.

\subsection{Prior distributions in BVR and BSC}

For BVR, we specify a standard normal prior for the regression coefficients and a truncated normal distribution  at zero with parameters $0$ and $0.5$ for the residual variance.
Similarly to the justification of SVR, this choice of hyperparameters is made since it satisfies that the probability that the variance is below 1 is, a priori, approximately 85\%.
For BSC the prior specification is the same, while we constrain the regression coefficients to lie in the simplex.

\section{Illustration of simulation specifications}
\label{supp_sec:illustration_simulation}

Here, we illustrate the implied spatial structure imposed in our simulations with respect to the control units' coefficients and the temporal variation in the control units' outcomes.

\begin{figure}[p]
    \centering
    \includegraphics[width=0.8\linewidth]{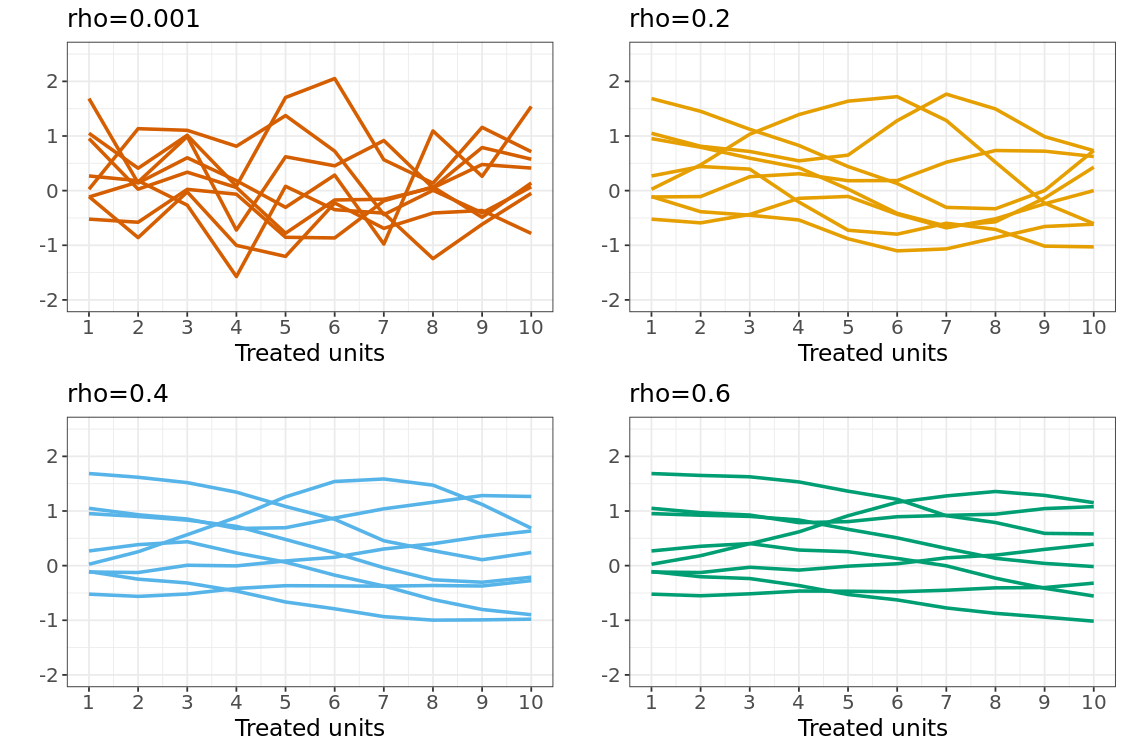}
    \caption{Realizations of $\bm \beta_c$ across the four different lengthscale values used in our simulations, $\rho^2_s \in \{0.001^2, 0.2^2, 0.4^2, 0.6^2\}$.}
    \label{fig:beta_example}

\vspace{30pt}

    \includegraphics[width=0.8\linewidth]{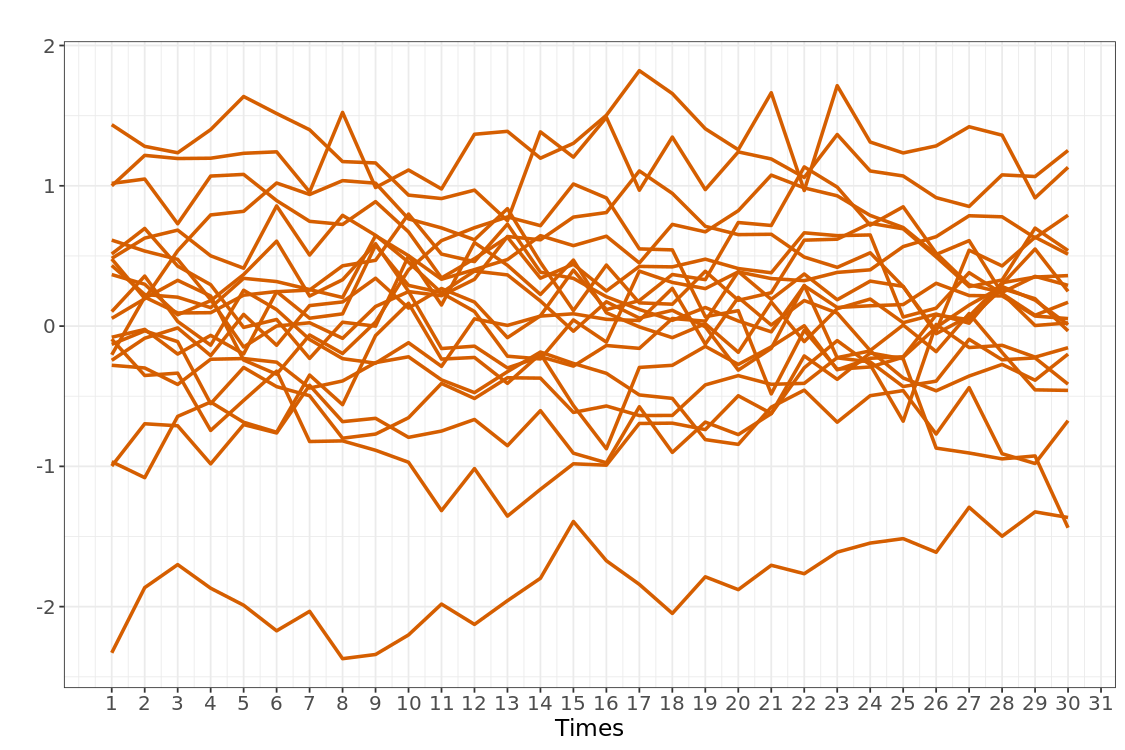}
    \caption{$Y_{ct}$ realizations across time under $\sigma^2_c=0.3^2$, $\rho_c=0.05^2$, $\eta_c=0.15^2$. }
    \label{fig:control_example}
\end{figure}

First, \cref{fig:beta_example} includes draws from the distribution of $\bm \beta_c$ based on the different values of $\rho^2_s$ used in our simulations. We see that indeed these choices of $\rho^2_s$ lead to increasing spatial structure in the coefficients, from independent to strongly spatially correlated.
Furthermore, in \cref{fig:control_example}, we visualize draws of the outcome values for the control units in our simulations, $Y_{ct}$, $c = N_1 + 1, \ldots, N$, across time, $t = 1, 2, \ldots, T$. As desired, the outcomes for the control units have a temporal structure with some independent variation, resembling the data in our analysis in \cref{sec:application}.
The temporal correlation in the control units' outcomes propagates to the treated units' control potential outcomes as well, which themselves will be temporally correlated.


\FloatBarrier

\section{Additional information on our study and results}
\label{supp_sec:additional_application}

\subsection{De-trended outcome time series}

\begin{figure}[p]
    \centering
    \includegraphics[width=\linewidth]{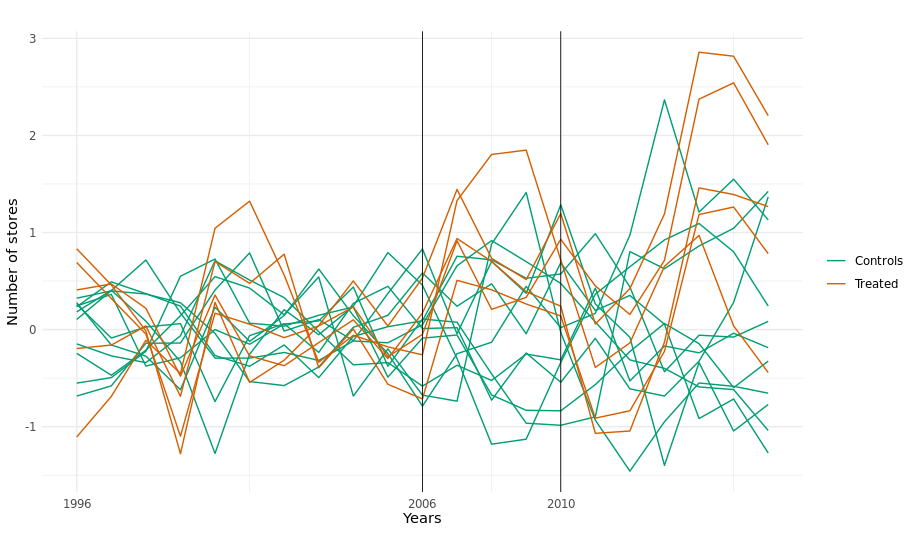}
    \caption{Normalized and de-trended time series plot for treated and control units over the observed period.}
    \label{fig:dataset_application}

\vspace{30pt}

    \includegraphics[width=.8\linewidth]{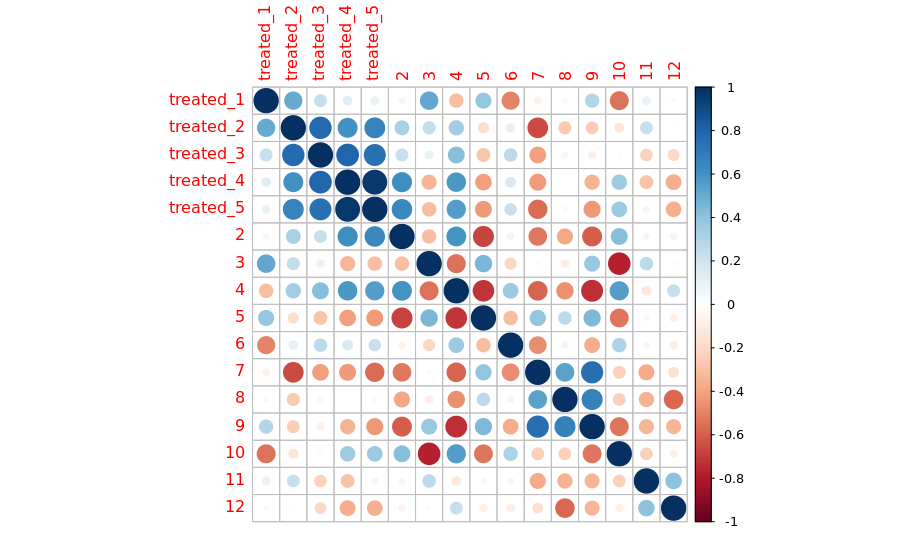}
    \caption{Correlation plot for the normalized and de-trended time series plot for treated and control units over the observed period.}
    \label{fig:correlation_dataset}
\end{figure}

Figure \ref{fig:dataset_application} reports the time series for the treated and control units once standardized for the pre-treatment mean and standard error, and detrended using the control units' average. We can notice that the treated units (in orange) shows a clear correlated trend, confirmed by the correlation plot in \cref{fig:correlation_dataset}. In the same figure we can notice that such correlation is not present for control units, stressing that the source for correlation is the spatial structure of treated units. 

\subsection{Point estimates across treated units}

\begin{figure}[htbp]
    \centering
    \includegraphics[width=\linewidth]{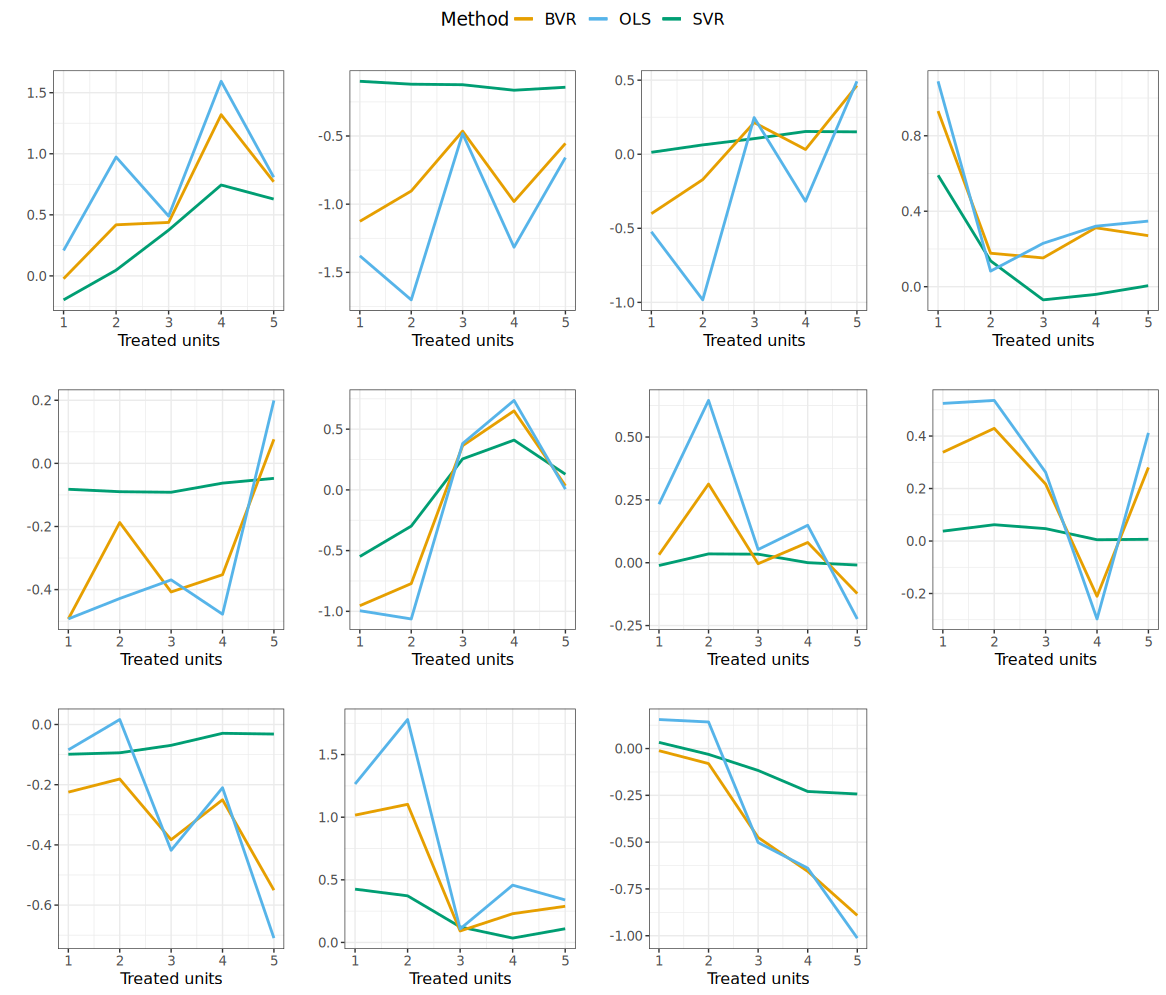}
    \caption{Estimated coefficients $\bm \beta_c$ for all the control units in our case study, and selected methods.}
    \label{fig:trace_all}
\end{figure}

\cref{fig:trace_all} reports the point estimation for $\bm \beta_c$ across the different control units employed in the dataset for three different estimation methods (BVR, OLS, SVR). 
As in \cref{fig:smoothed_beta} we notice that the estimation with SVR is the smoothest, with similar value for similar control units. OLS fails to achieve this, with sizeable differences across coefficients for correlated units. BVR has similar results to SVR. 

\subsection{Additional results for stores selling durable goods and non durable goods}

The impact of a tramway construction may have different effects if we consider different types of stores. \cite{grossi2025direct} distinguish two different types of stores, according to the nature of goods provided: 
\begin{itemize}
    \item Stores that sell \textbf{durable goods}, such as electronics, clothes, furniture. The purchase frequency for these types of goods is low, hereafter we can consider that the customers' choices may be less affected by the new tramway line, both in the construction phase, and in the operational period of the tramway. 
    \item Stores that are selling \textbf{non durable goods}: such as restaurant, bars, news kiosk, book shops, grocery stores, bakeries, etc . These stores have a high-frequency of purchase and thus, may be more affected by the construction of the tramway. 
\end{itemize}

\begin{figure}
    \centering
    \includegraphics[width=\linewidth]{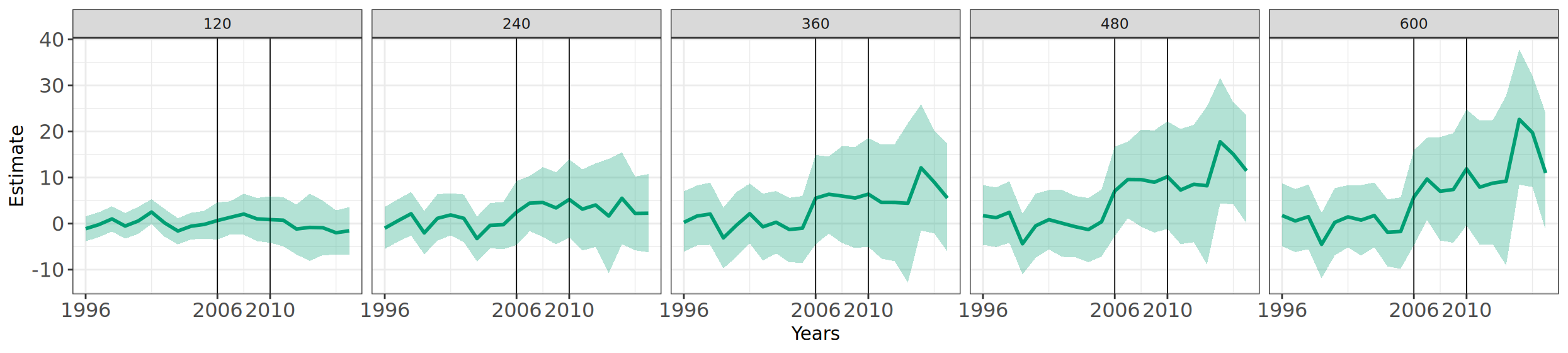}
    \caption{Posterior medians (lines) and 95\% posterior credible intervals (shaded areas) of the treatment effects     $\Delta_{it}$ for areas  within $d-$seconds of walking from a tramway stop, $d=120, 240, 360, 480,600$.
    The two vertical lines represent the beginning of the tramways construction and operational phases in 2006 and 2010, respectively.}
    \label{fig:application_durables}
\end{figure}

\begin{figure}
    \centering
    \includegraphics[width=\linewidth]{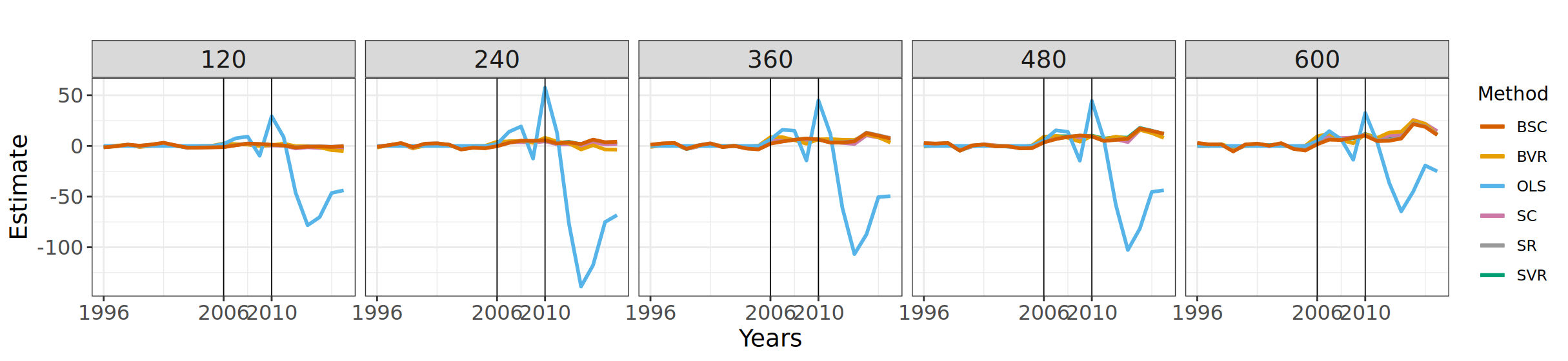}
    \caption{Comparison across alternative methods between posterior medians of the treatment effects $\Delta_{it}$ for areas  within $d-$seconds of walking from a tramway stop, $d=120, 240, 360, 480,600$. 
    The two vertical lines represent the beginning of the tramways construction and operational phases in 2006 and 2010, respectively.}
    \label{fig:comparison_durables}
\end{figure}

Figures \ref{fig:application_durables}, \ref{fig:comparison_durables} and table \cref{tab:res_dur} reports the estimation results at different distances from the closest tramway stops for stores that sells durable goods. 
For these stores we can see a positive effect for the areas located at higher distance from the tramway stops, with credibility intervals that do not comprise zero for the last time periods (2013-2016). 

\begin{figure}
    \centering
    \includegraphics[width=\linewidth]{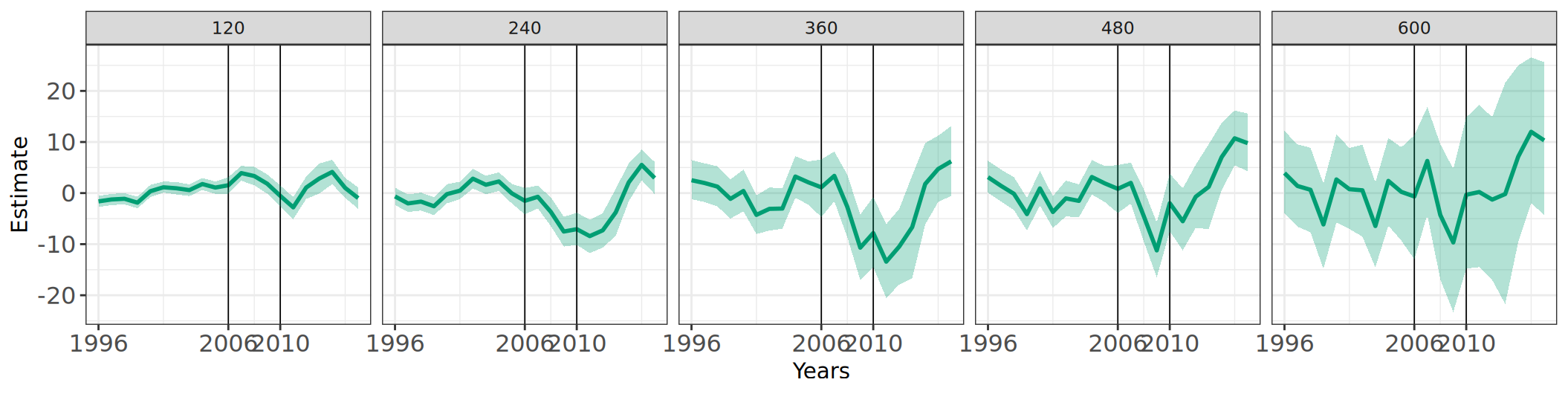}
    \caption{Posterior medians (lines) and 95\% posterior credible intervals (shaded areas) of the treatment effects     $\Delta_{it}$ for areas  within $d-$seconds of walking from a tramway stop, $d=120, 240, 360, 480,600$.
    The two vertical lines represent the beginning of the tramways construction and operational phases in 2006 and 2010, respectively.}
    \label{fig:application_nondurables}
\end{figure}

\begin{figure}
    \centering
    \includegraphics[width=\linewidth]{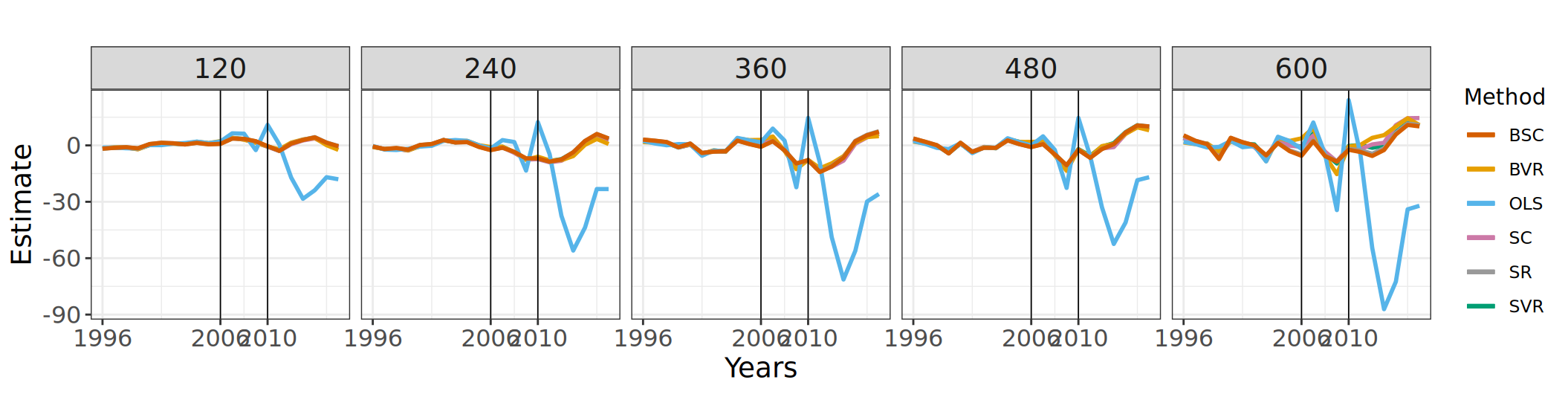}
    \caption{Comparison across alternative methods between posterior medians of the treatment effects $\Delta_{it}$ for areas  within $d-$seconds of walking from a tramway stop, $d=120, 240, 360, 480,600$. 
    The two vertical lines represent the beginning of the tramways construction and operational phases in 2006 and 2010, respectively.}
    \label{fig:comparison_nondur}
\end{figure}

Figures \ref{fig:application_nondurables}, \ref{fig:comparison_nondur} and table \cref{tab:res_nondur} reports the same results for stores that are selling non-durable goods. In this case we can underline a positive effect after the construction of the tramway, after a first contraction during the construction period, which is more evident for the areas between 120 and 480 seconds walking from the tramway. 

\subsection{MCMC diagnostics}

\begin{figure}[htbp]
    \centering
    \includegraphics[width=\linewidth]{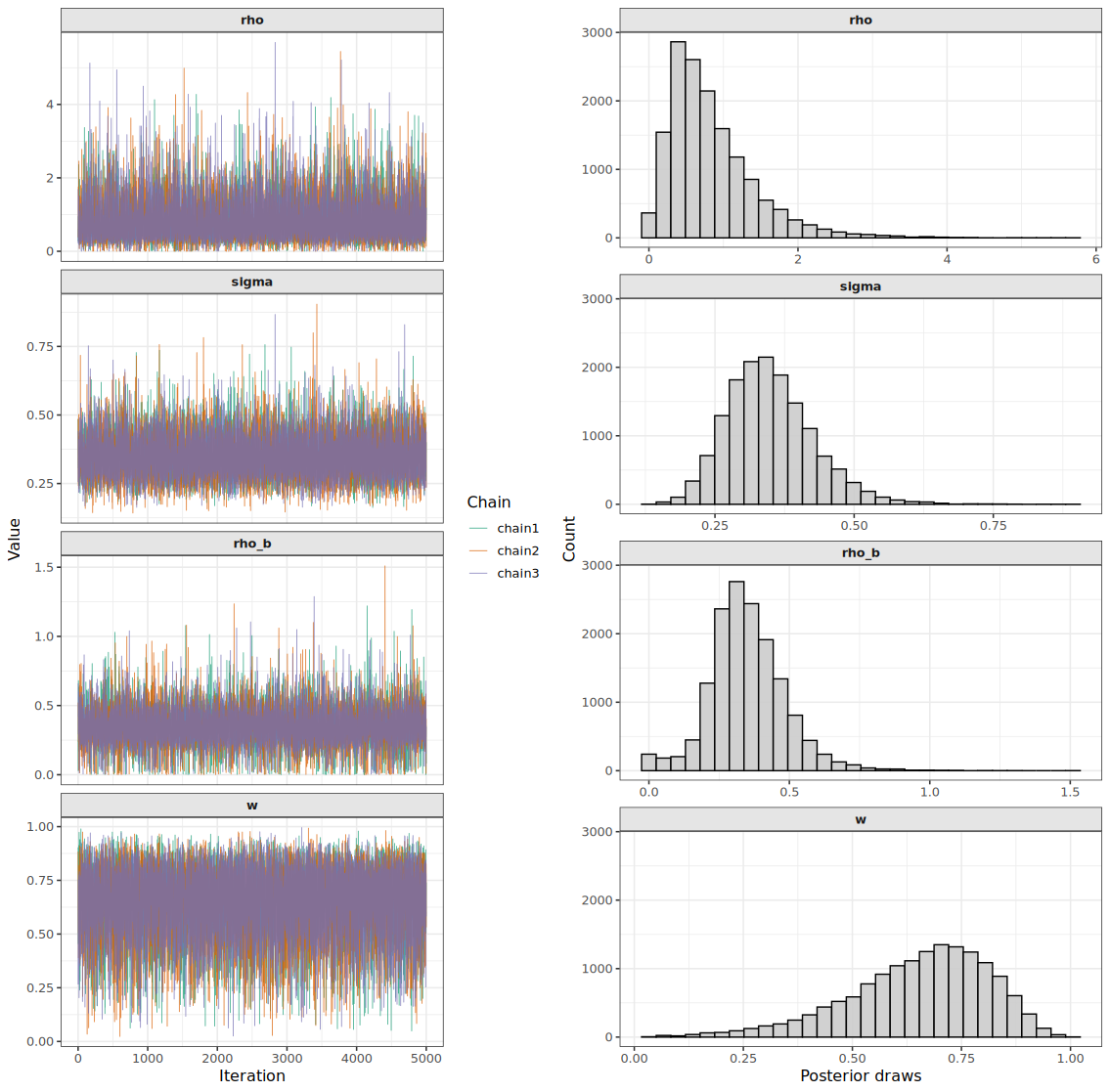}
    \caption{Traceplot and Posterior distribution for the hyperparaters of the model ($\rho^2_e, \rho^2_b, \sigma^2_e, w$). }
    \label{fig:diagnostics}
\end{figure}

Figure \ref{fig:diagnostics} presents the traceplots and the histograms of the posterior distributions for the parameters  $\rho^2_e$, $\rho^2_b$, $\sigma^2_e$ and $w$ estimated using 10,000 iterations (5,000 warmup) across three chains.
Figure \ref{fig:diagnostics} illustrates that the traceplots for all monitored parameters exhibit adequate mixing, with no visible trends or sticking behavior across chains.
Consistently $\widehat{R}$ values are close to 1 and Stan reports no warning for the model fit. These diagnostics provide 
evidence that the sampler has achieved satisfactory convergence and that posterior inference is reliable. Note that the posterior distribution of $w$, the $[0,1]$ parameter controlling the radio between spatial and IID errors is shifted toward 1, suggesting the presence of spatially correlated errors. 

\subsection{Point estimates from all methods compared in our simulations}

\begin{figure}
    \centering
    \includegraphics[width=\linewidth]{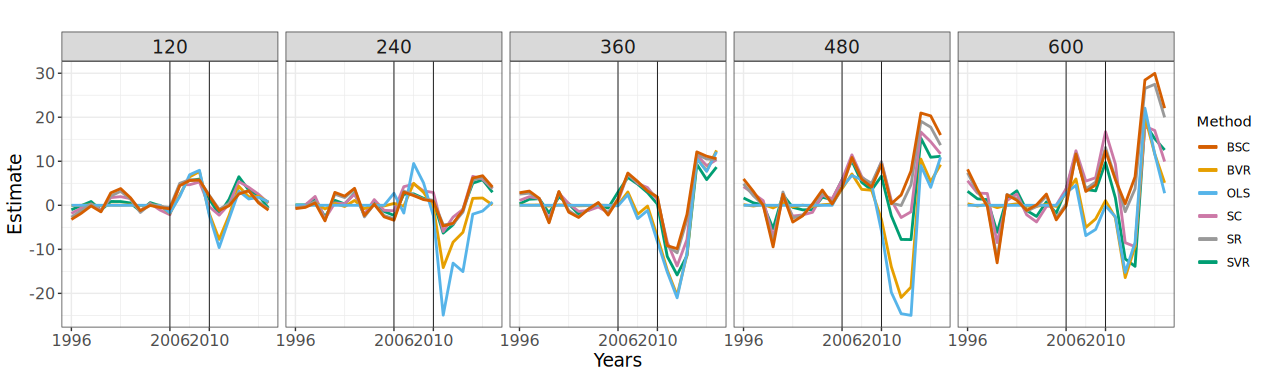}
    \caption{Comparison across alternative methods between posterior medians of the treatment effects $\Delta_{it}$ for areas  within $d-$seconds of walking from a tramway stop, $d=120, 240, 360, 480,600$. The two vertical lines represent the beginning of the tramways construction and operational phases in 2006 and 2010, respectively.}
    \label{fig:application_comparison}
\end{figure}

In \cref{fig:application_comparison} and Table \ref{tab:all_methods} we compare the point estimates for the different methods we propose in the simulations. We notice that all methods provide similar results, leading to a similar interpretation of results. Table \ref{tab:all_methods} reports RMSE, and average treatment effect for the served areas during the construction of the tramway and the operational period of the tramway.
Table~\ref{tab:all_methods} shows that differences across methods mainly concern pre-treatment fit and uncertainty quantification. Across most approaches, estimates point to moderate positive effects during the construction phase and more heterogeneous effects during the operational phase. Methods achieving very low pre-treatment RMSPE, such as OLS and BVR, tend to suffer from overfitting and inferential problems, as evidenced by degenerate or excessively wide confidence intervals. In contrast, SVR exhibits an intermediate pre-treatment RMSPE, higher than BVR but substantially lower than SC, BSC, and SR, while providing coherent and stable uncertainty quantification.


\begin{table}[ht]
\centering
\caption{Estimates, credible intervals, and pre-treatment RMSPE for all methods.}
\label{tab:all_methods}
\resizebox{\textwidth}{!}{
\begin{tabular}{lllccccc}
\toprule
Method & Period & Type & 120 & 240 & 360 & 480 & 600 \\
\midrule
SVR & Construction & Estimate & 3.306 & 1.064 & 3.453 & 6.232 & 6.356 \\
SVR & Construction & CI & (-1.003, 7.572) & (-5.596, 7.670) & (-6.772, 14.028) & (-3.239, 16.185) & (-9.463, 22.689) \\
SVR & Operational & Estimate & 1.803 & 0.359 & -2.510 & 3.228 & 3.650 \\
SVR & Operational & CI & (-2.938, 6.485) & (-7.076, 7.839) & (-14.645, 10.008) & (-8.065, 15.378) & (-14.930, 23.361) \\
SVR & Pre-treatment & RMSPE & 0.834 & 1.584 & 1.291 & 2.094 & 2.769 \\
\midrule
BVR & Construction & Estimate & 2.547 & 1.551 & -1.253 & 2.933 & 0.341 \\
BVR & Construction & CI & (-3.315, 8.491) & (-9.669, 12.214) & (-13.112, 12.205) & (-9.927, 17.441) & (-21.151, 25.106) \\
BVR & Operational & Estimate & -0.181 & -4.222 & -2.573 & -4.722 & 1.401 \\
BVR & Operational & CI & (-5.650, 7.504) & (-15.575, 10.434) & (-15.850, 11.561) & (-17.910, 13.760) & (-23.427, 26.346) \\
BVR & Pre-treatment & RMSPE & 0.066 & 0.569 & 0.085 & 0.312 & 0.288 \\
\midrule
SC & Construction & Estimate & 2.439 & 2.799 & 3.643 & 7.608 & 8.916 \\
SC & Construction & CI & -- & -- & -- & -- & -- \\
SC & Operational & Estimate & 1.744 & 1.124 & 0.147 & 6.606 & 6.058 \\
SC & Operational & CI & -- & -- & -- & -- & -- \\
SC & Pre-treatment & RMSPE & 1.314 & 1.673 & 1.767 & 3.178 & 3.798 \\
\midrule
BSC & Construction & Estimate & 3.435 & 0.803 & 3.843 & 6.851 & 6.329 \\
BSC & Construction & CI & (-1.094, 7.963) & (-4.751, 6.260) & (-2.617, 10.465) & (-2.477, 16.192) & (-6.479, 18.821) \\
BSC & Operational & Estimate & 0.669 & 1.186 & 2.126 & 11.291 & 15.525 \\
BSC & Operational & CI & (-3.976, 5.170) & (-4.559, 6.809) & (-4.848, 9.159) & (1.784, 21.111) & (2.575, 29.385) \\
BSC & Pre-treatment & RMSPE & 2.069 & 2.320 & 2.505 & 4.166 & 5.254 \\
\midrule
SR & Construction & Estimate & 3.654 & 1.012 & 3.778 & 7.173 & 6.833 \\
SR & Construction & CI & -- & -- & -- & -- & -- \\
SR & Operational & Estimate & 0.923 & 0.940 & 1.527 & 9.272 & 13.747 \\
SR & Operational & CI & -- & -- & -- & -- & -- \\
SR & Pre-treatment & RMSPE & 1.732 & 2.174 & 2.292 & 3.630 & 4.837 \\
\midrule
OLS & Construction & Estimate & 2.574 & 2.643 & -2.034 & 3.131 & -1.021 \\
OLS & Construction & CI & (2.574, 2.574) & (2.643, 2.643) & (-2.034, -2.034) & (3.131, 3.131) & (-1.021, -1.021) \\
OLS & Operational & Estimate & -0.910 & -9.288 & -2.758 & -7.560 & 1.696 \\
OLS & Operational & CI & (-0.910, -0.910) & (-9.288, -9.288) & (-2.758, -2.758) & (-7.560, -7.560) & (1.696, 1.696) \\
OLS & Pre-treatment & RMSPE & 0.000 & 0.000 & 0.000 & 0.000 & 0.000 \\
\bottomrule
\end{tabular}}
\end{table}

\begin{table}[!htbp]
\caption{Posterior medians and 95\% posterior credible intervals of the intertemporal average causal effects $\Delta_{i}$ for stores selling durable goods, for areas within $d-$seconds walking distance from the tramway stop, $d=120, 240, 360, 480,600$, and posterior median of  the pre-treatment root mean squared prediction error (RMSPE). 
}
\label{tab:res_dur}
\centering
\begin{tabular}{lccccc}
\toprule
  & 120 & 240 & 360 & 480 & 600\\
\midrule
  &  \multicolumn{5}{c}{Construction period} \\
\cmidrule(lr){2-6}
Estimate & 1.197 & 4.028 & 5.963 & 9.072 & 8.345\\
CI & (-3.239, 5.485) & (-3.348, 11.379) & (-4.245, 16.294) & (-1.070, 19.452) & (-2.431, 19.546)  \\
\midrule
  &  \multicolumn{5}{c}{Operational period} \\
\cmidrule(lr){2-6}
Estimate & -0.956 & 3.128 & 6.703 & 11.390 & 13.212\\
CI  & (-6.697, 4.627) & (-6.362, 12.553) & (-6.394, 19.948) & (-1.459, 24.805) & (-0.552, 27.751) \\
\midrule
RMSPE & 1.103 & 1.643 & 1.570 & 1.817 & 1.955 \\
\bottomrule
\end{tabular}%
\end{table}

\begin{table}[!htbp]
\caption{Posterior medians and 95\% posterior credible intervals of the intertemporal average causal effects $\Delta_{i}$ for stores selling non durable goods, for areas within $d-$seconds walking distance from the tramway stop, $d=120, 240, 360, 480,600$, and posterior median of  the pre-treatment root mean squared prediction error (RMSPE). 
}
\label{tab:res_nondur}
\centering
\begin{tabular}{lccccc}
\toprule
  & 120 & 240 & 360 & 480 & 600 \\
\midrule
  & \multicolumn{5}{c}{Construction period} \\
\cmidrule(lr){2-6}
Estimate & 2.012 & -4.101 & -3.341 & -2.979 & -1.729 \\
CI       & (0.222, 3.743) & (-6.841, -1.371) & (-9.281, 2.670) & (-7.884, 2.042) & (-14.420, 11.462) \\
\midrule
  & \multicolumn{5}{c}{Operational period} \\
\cmidrule(lr){2-6}
Estimate & 0.883 & -1.493 & -2.974 & 3.756 & 4.695 \\
CI       & (-1.434, 3.137) & (-5.018, 2.008) & (-10.595, 4.732) & (-2.459, 10.245) & (-11.515, 21.817) \\
\midrule
RMSPE    & 1.262 & 1.731 & 2.551 & 2.453 & 3.326 \\
\bottomrule
\end{tabular}%
\end{table}

\FloatBarrier

\section{Additional simulation results}
\label{supp_sec:additional_simulations}

\subsection{Figures of bias and MSE across all simulations}
\label{supp_subsec:figures_simulations}

\begin{figure}[p]
        \centering
            \includegraphics[width=\textwidth]{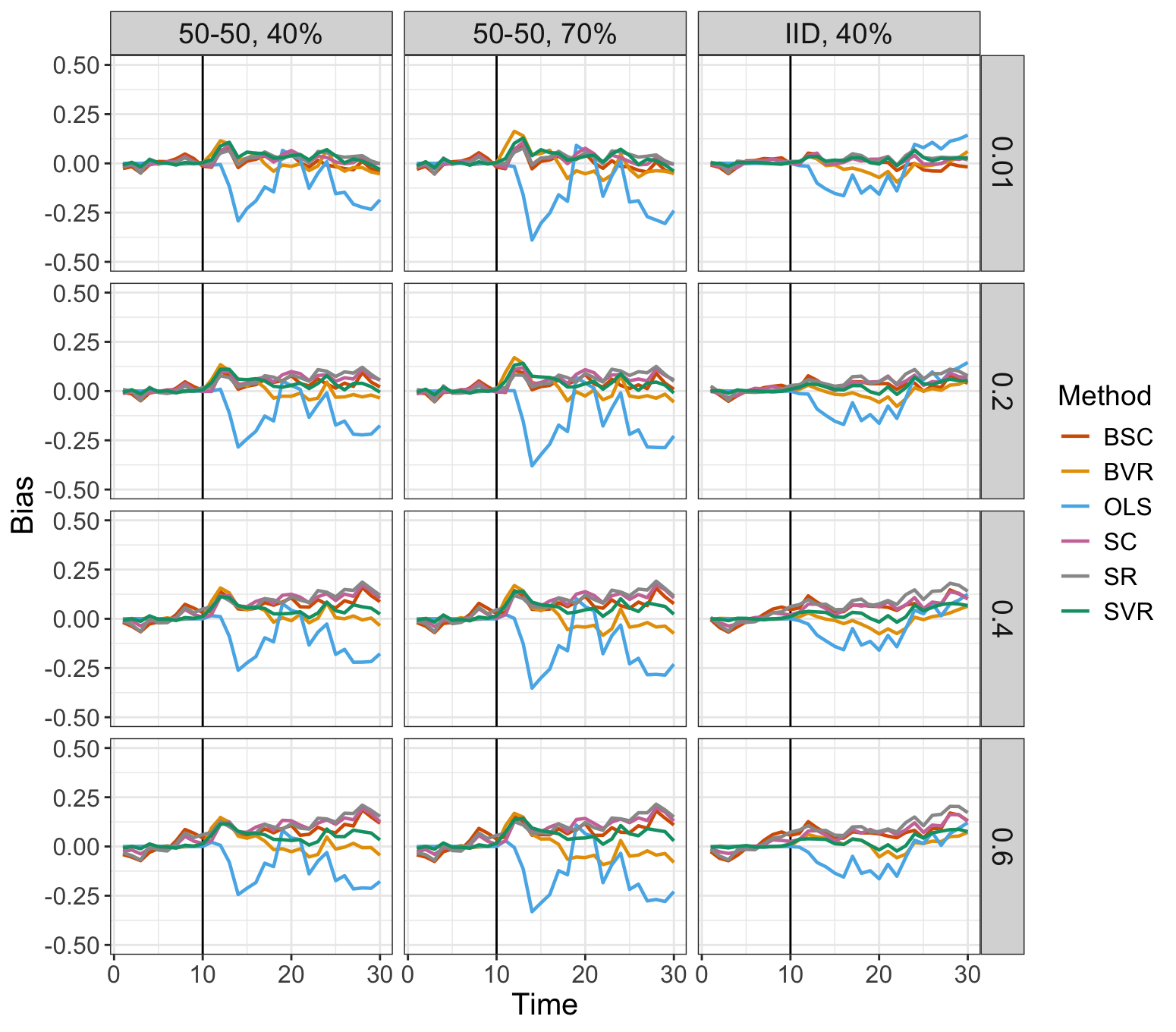}
            \caption{Average Bias with $T_0=10$. On rows: $\rho_s^2=\{0.001^2, 0.2^2, 0.4^2, 0.6^2\}$. On columns, type of errors: IID errors, noise to signal:40\%, 50\% spatial errors, 50\% IID with noise to signal:40\%, 50\% spatial errors, 50\% IID with noise to signal:70\%.}
            \label{fig:bias10}
        \end{figure}

\begin{figure}[p]
                \centering
            \includegraphics[width=\textwidth]{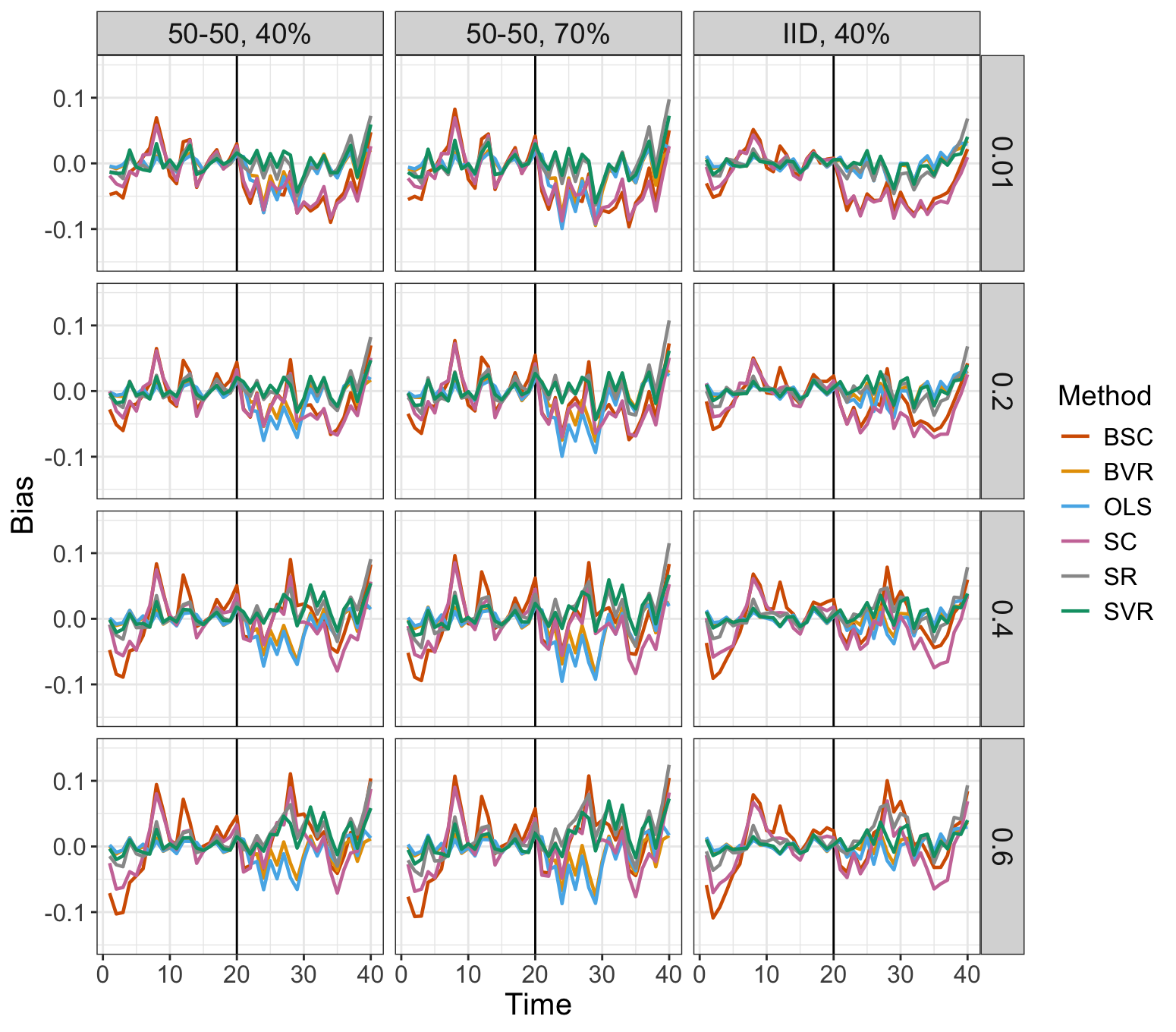}
            
                        \caption{Average Bias with $T_0=20$. On rows: $\rho_s^2=\{0.001^2, 0.2^2, 0.4^2, 0.6^2\}$. On columns, type of errors: IID errors, noise to signal:40\%, 50\% spatial errors, 50\% IID with noise to signal:40\%, 50\% spatial errors, 50\% IID with noise to signal:70\%.}
                        \label{fig:bias20}

        \end{figure}

\begin{figure}[p]
        \centering
            \includegraphics[width=\textwidth]{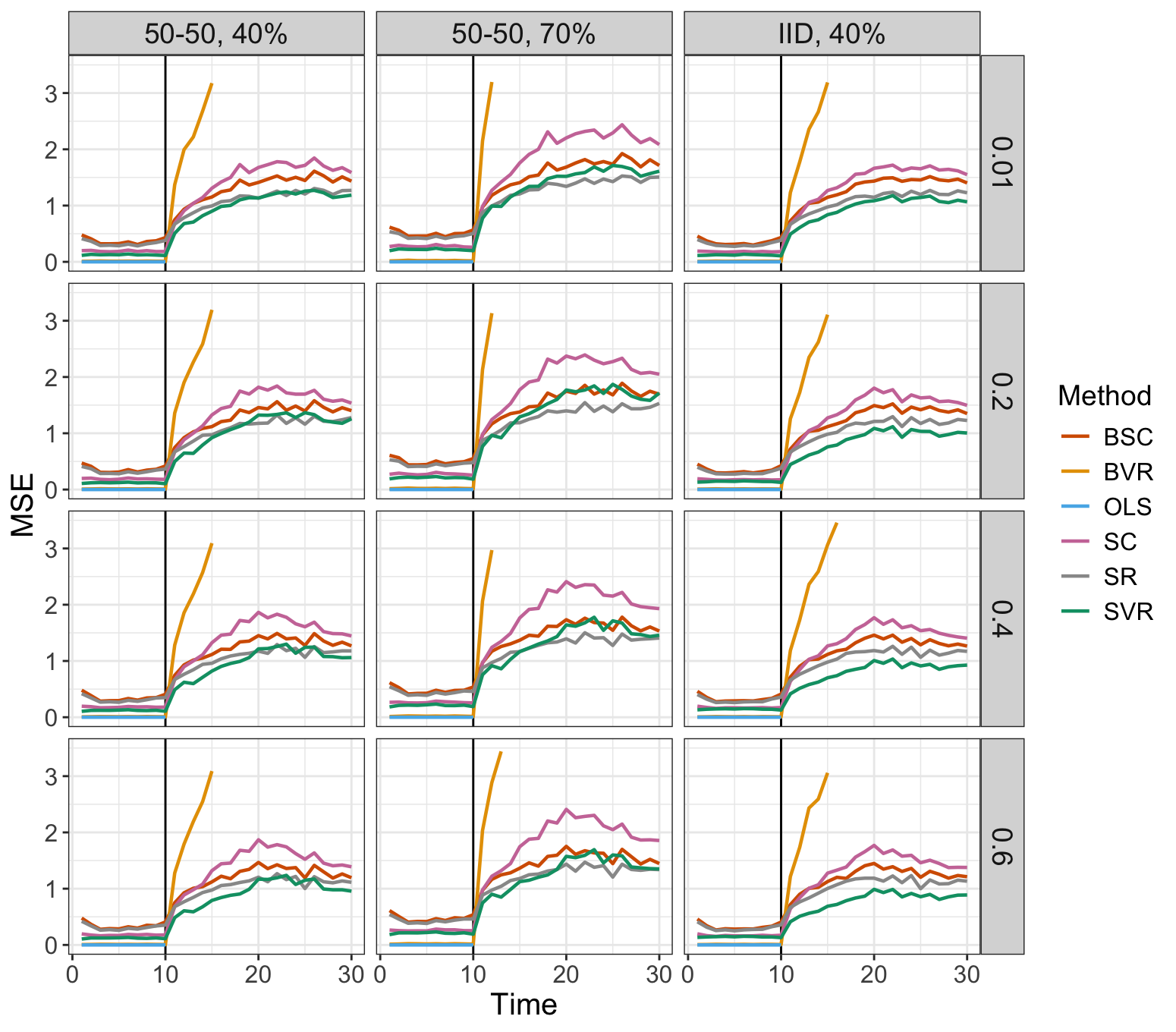}
            
            \caption{Average MSE with $T_0=10$. On rows: $\rho_s^2=\{0.001^2, 0.2^2, 0.4^2, 0.6^2\}$. On columns, type of errors: IID errors, noise to signal:40\%, 50\% spatial errors, 50\% IID with noise to signal:40\%, 50\% spatial errors, 50\% IID with noise to signal:70\%.}
            \label{fig:mse_10}
        \end{figure}

\begin{figure}[p]
                \centering
            \includegraphics[width=\textwidth]{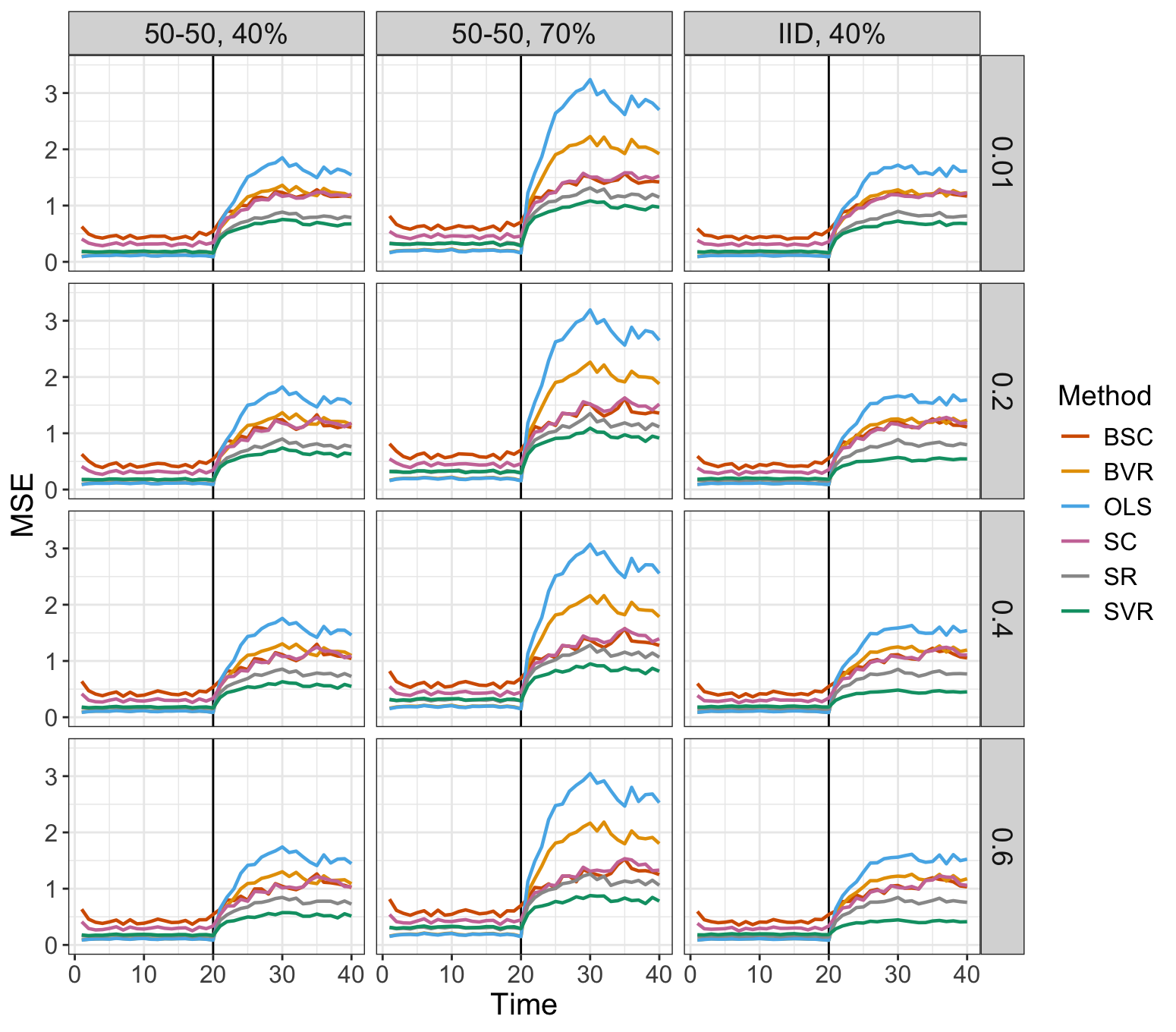}
            
                        \caption{Average MSE with $T_0=20$. On rows: $\rho_s^2=\{0.001^2, 0.2^2, 0.4^2, 0.6^2\}$. On columns, type of errors: IID errors, noise to signal:40\%, 50\% spatial errors, 50\% IID with noise to signal:40\%, 50\% spatial errors, 50\% IID with noise to signal:70\%.}
                        \label{fig:mse_20}

        \end{figure}

\begin{figure}[p]
        \centering
            \includegraphics[width=\textwidth]{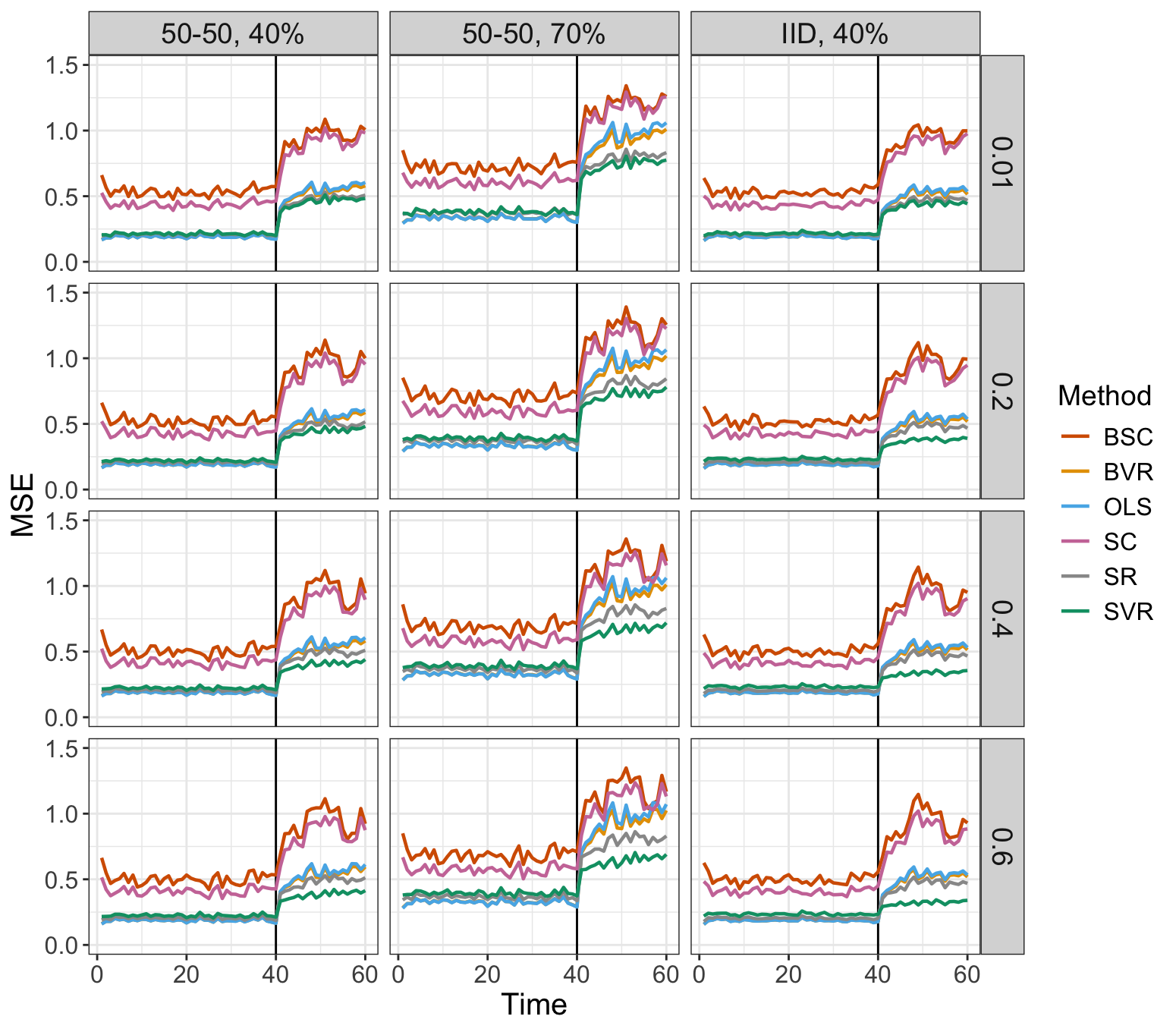}
           
            \caption{Average MSE with $T_0=40$. On rows: $\rho_s^2=\{0.001^2, 0.2^2, 0.4^2, 0.6^2\}$. On columns, type of errors: IID errors, noise to signal:40\%, 50\% spatial errors, 50\% IID with noise to signal:40\%, 50\% spatial errors, 50\% IID with noise to signal:70\%.}
             \label{fig:mse_40}
        \end{figure}

In this subsection we provide the graphs for bias and MSE under the several scenarios we tested. In particular Figures \ref{fig:bias10}, \ref{fig:bias20}, \ref{fig:bias40} reports the average bias, and Figures \ref{fig:mse_10}, \ref{fig:mse_20}, \ref{fig:mse_40} the MSE for $T_0 = 10$, $T_0 = 20$, and $T_0 = 40$, respectively. In all figures, the rows correspond to the different lengthscale parameter employed $\rho_s^2=\{0.001^2, 0.2^2, 0.4^2, 0.6^2\}$, while the columns represent the configurations for the error terms.

For the shortest pre-intervention time period ($T_0 = 10$), all estimation methods considered, except OLS, are still able to provide reasonably accurate predictions of the outcome (\cref{fig:bias10}). In contrast, OLS, which is known to be prone to overfitting when the pre-treatment period is short,  consistently underestimates the counterfactual. In terms of MSE (\cref{fig:mse_10}), OLS yields the highest post-treatment MSE, reflecting the same issues of overfitting. SVR achieves the lowest MSE across all methods when the errors are IID under all the scenarios considered, and also with spatially correlated residuals, when the time series of treated units are correlated, which is a reasonably common situation. 
SR has slightly lower MSE when the post treatment is very long, and the pre-treatment is short, but the estimates are comparable. 
Notably, SVR is able to leverage spatial information through the priors and reduce the MSE as the lengthscale increases, confirming the implicit regularization that we exploit with gaussian process priors.

When pre-treatment length increases ($T_0 \in \{20, 40\}$), the accuracy of the counterfactual estimates remains high across all methods (see Figures \ref{fig:bias20} and \ref{fig:bias40}). In scenarios characterized by low variance and uncorrelated errors, methods based on unconstrained models (BVR, SVR, OLS, SR) yield very similar and more stable results compared to those relying on constrained coefficients (BSC, SC).
In terms of MSE (see Figures \ref{fig:mse_20} and \ref{fig:mse_40}), OLS remains the noisiest estimator in terms of coefficient variability for $T_0 = 20$, although its performance becomes more comparable to that of other methods for $T_0 = 40$. SVR consistently delivers the lowest post-treatment MSE by effectively exploiting the spatial structure encoded in the priors. We notice that the MSE decreases with increasing lengthscale--see, for instance, column 3 in Figure \ref{fig:mse_40}.

        \begin{figure}[p]
        \centering
            \includegraphics[width=\textwidth]{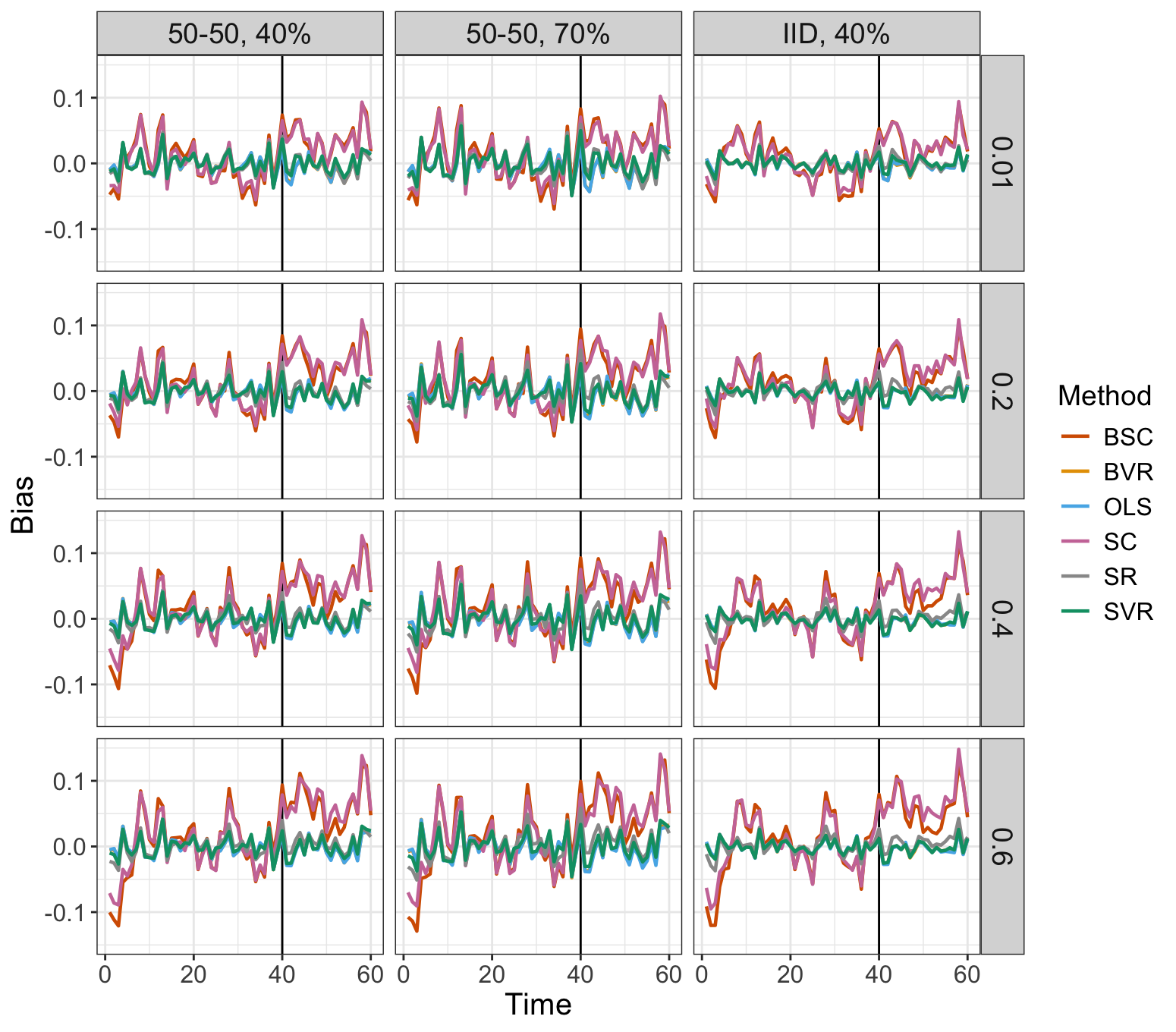}
            \label{fig:bias40}
            \caption{Average Bias with $T_0=40$. On rows: $\rho_s^2=\{0.001^2, 0.2^2, 0.4^2, 0.6^2\}$. On columns, type of errors: IID errors, noise to signal:40\%, 50\% spatial errors, 50\% IID with noise to signal:40\%, 50\% spatial errors, 50\% IID with noise to signal:70\%.}
            \label{fig:bias40}
        \end{figure}

\begin{figure}[p]
    \centering
    \includegraphics[width=\linewidth]{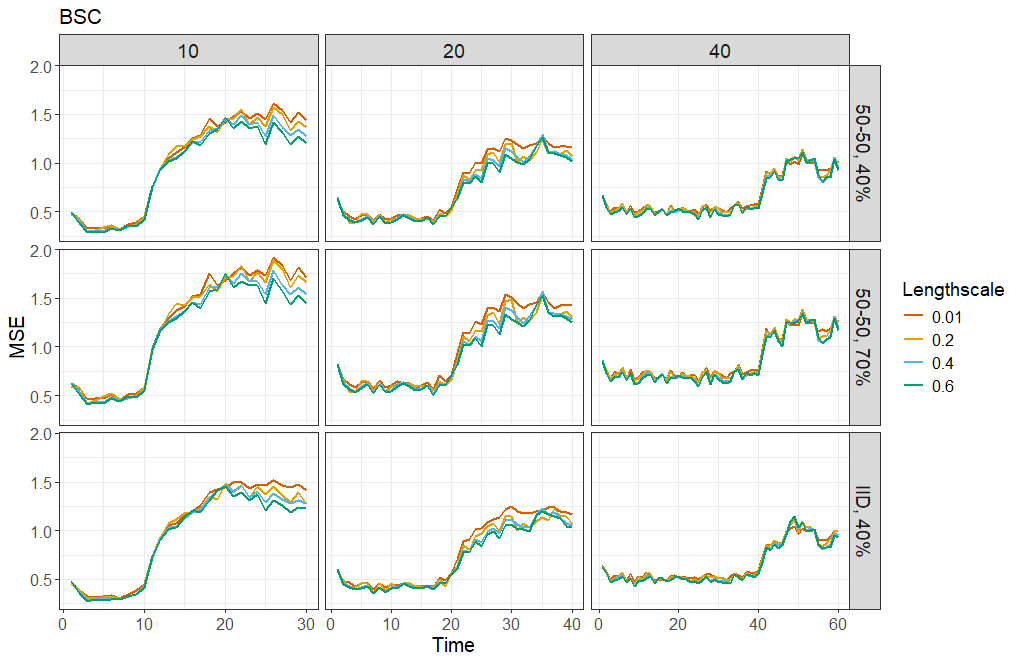}
    \caption{MSE for BSC estimates when $T_0 = 20$ by varying the spatial correlation $\rho_s$ (colors),  the length of the pre-treatment period (columns) and the error correlation structure (rows).}
    \label{fig:mse_bsc}

\vspace{30pt}
    
    \includegraphics[width=\linewidth]{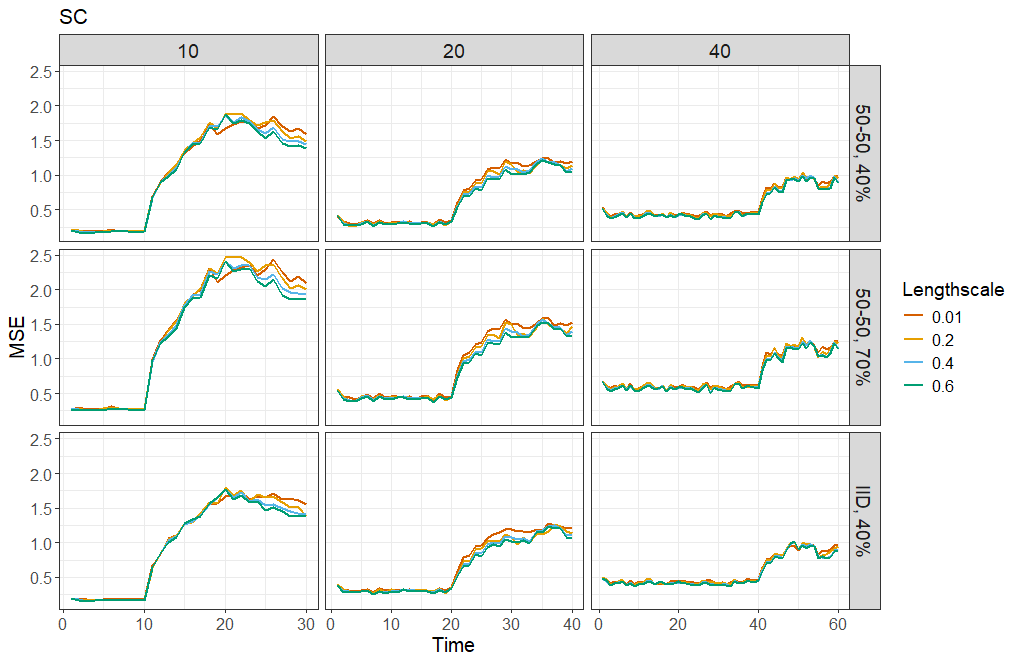}
    \caption{MSE for SC estimates when $T_0 = 20$ by varying the spatial correlation $\rho_s$ (colors),  the length of the pre-treatment period (columns) and the error correlation structure (rows).}
    \label{fig:mse_sc}
\end{figure}

\begin{figure}[p]
    \centering
    \includegraphics[width=\linewidth]{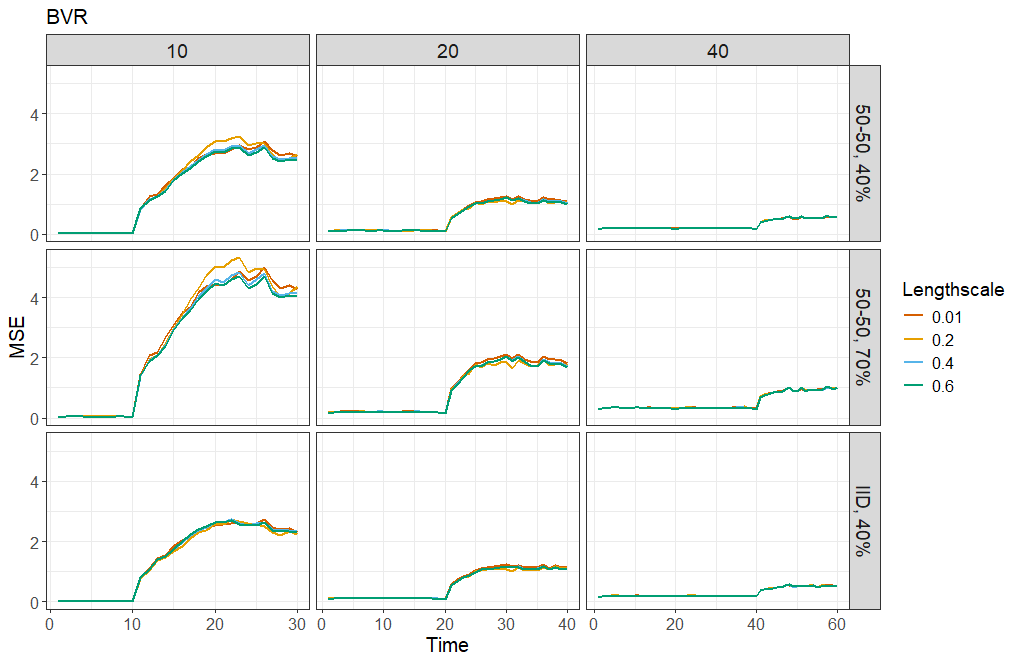}
    \caption{MSE for BVR estimates when $T_0 = 20$ by varying the spatial correlation $\rho_s$ (colors),  the length of the pre-treatment period (columns) and the error correlation structure (rows).}
    \label{fig:mse_bvr}
\end{figure}

Figures \ref{fig:mse_bsc}, \ref{fig:mse_sc}, and \ref{fig:mse_bvr} show the MSE of methods that do not explicitly leverage spatial information (SC, BVR, and BSC) as the spatial structure in the control units' coefficients increases, across different lengths of the pre-treatment period and model error configurations. The corresponding figure for SVR is given in \cref{fig:mse_smac_main}.
It is evident that SVR reduces the MSE as spatial correlation increases, whereas all other methods show no substantial change in performance under the same conditions. This highlights the source of the performance improvement in the model we propose—namely, its ability to incorporate and exploit spatial structure effectively.

\subsection{Additional tables of simulation results}

The tables presented provide a detailed analysis of the performance of several statistical methodologies for estimating counterfactual outcomes, focusing on bias, Mean Squared Error (MSE), and coverage intervals. 
Tables \ref{tab:bias001}, \ref{tab:bias02} in the Supplement, \cref{tab:bias_04} in \cref{sec:sim}, and \cref{tab:bias06} report the bias of all approaches under different strengths of spatial structure, $\rho^2_s \in \{0.001^2, 0.2^2, 0.4^2, 0.6^2\}$, respectively. Similarly, Tables \ref{tab:mse001} and \ref{tab:mse02} in the Supplement, \cref{tab:mse_04} in \cref{sec:sim}, and \cref{tab:mse06} in the Supplement include the MSE, and Tables \ref{tab:cov001}, \ref{tab:cov02} in the Supplement,  \cref{tab:cov_04} in \cref{sec:sim}, and \ref{tab:cov06} in the Supplement include the coverage across values of $\rho^2_s$.
These tables repeat some of the information discussed using the figures in Supplement \ref{supp_subsec:figures_simulations}. 

SVR stands out for its ability to reduce bias as spatial correlation increases, particularly for longer pre-treatment periods.
In terms of MSE, OLS performs poorly for low $T_0$, with very high error values. SVR, on the other hand, shows substantial improvements as spatial correlation increases, particularly when $T_0=40$. 
Coverage analysis further highlights OLS's limitations, with near-zero interval coverage for small $T_0$. BVR and SVR, however, achieve close-to-nominal 95\% coverage, even under spatial noise. SVR maintains reliable uncertainty estimates across varying conditions, underlining its robustness. BSC tends to undercover the 95\% nominal confidence interval. 
Overall, SVR consistently reduces bias and MSE while maintaining accurate uncertainty quantification, particularly in spatially dependent settings. This underscores the value of incorporating spatial structure into counterfactual estimation models.

\FloatBarrier

\begin{table}[p]
\centering
\begin{tabular}{cccccccccc}
\hline \\
& \multicolumn{3}{c}{$T_0=10$} & \multicolumn{3}{c}{$T_0=20$} & \multicolumn{3}{c}{$T_0=40$} \\
& IID & Sp--40\% & Sp--70\% & IID & Sp--40\% & Sp--70\% & IID & Sp--40\% & Sp--70\%  \\
\cmidrule(lr){2-4} \cmidrule(lr){5-7} \cmidrule(lr){8-10}
& \multicolumn{9}{c}{$T-T_0 = 5$} \\
\cmidrule(lr){2-10}
\textbf{True}& 1.70 & 1.70 & 1.70 & 1.68 & 1.68 & 1.67 & 1.68 & 1.68 & 1.68\\
\addlinespace
SC & 0.02 & 0.03 & 0.04 & -0.05 & -0.04 & -0.05 & 0.05 & 0.05 & 0.04\\
SR & 0.02 & 0.03 & 0.03 & -0.01 & -0.01 & -0.01 & 0.00 & 0.00 & 0.00\\
OLS & -0.08 & -0.13 & -0.18 & 0.00 & -0.03 & -0.04 & -0.01 & -0.01 & -0.02\\
BVR & 0.01 & 0.06 & 0.10 & 0.00 & -0.02 & -0.03 & -0.01 & -0.01 & -0.02\\
BSC & 0.02 & 0.02 & 0.02 & -0.06 & -0.05 & -0.05 & 0.05 & 0.05 & 0.05\\
\textbf{SVR} & 0.02 & 0.06 & 0.07 & 0.00 & 0.01 & 0.00 & 0.00 & 0.00 & 0.00\\
\hline \\
& \multicolumn{9}{c}{$T-T_0 = 10$} \\
\cmidrule(lr){2-10}
\textbf{True} & 1.70 & 1.68 & 1.68 & 1.69 & 1.70 & 1.70 & 1.70 & 1.70 & 1.71\\
\addlinespace
SC & 0.02 & 0.03 & 0.04 & -0.06 & -0.04 & -0.05 & 0.03 & 0.03 & 0.03\\
SR & 0.02 & 0.03 & 0.04 & -0.01 & -0.01 & -0.02 & 0.00 & 0.00 & 0.00\\
OLS & -0.11 & -0.10 & -0.13 & -0.01 & -0.04 & -0.05 & -0.01 & -0.01 & -0.01\\
BVR & -0.02 & 0.03 & 0.04 & -0.01 & -0.03 & -0.04 & -0.01 & -0.01 & -0.01\\
BSC & 0.02 & 0.02 & 0.02 & -0.05 & -0.05 & -0.06 & 0.03 & 0.03 & 0.03\\
\textbf{SVR} & 0.02 & 0.05 & 0.05 & 0.00 & 0.00 & -0.01 & 0.00 & 0.00 & 0.00\\
\hline \\
& \multicolumn{9}{c}{$T-T_0 = 0.5 T_0$} \\
\cmidrule(lr){2-10}
\textbf{True} &  &  &  &  &  &  & 1.71 & 1.72 & 1.72\\
\addlinespace
SC &  &  &  &  &  &  & 0.03 & 0.04 & 0.04\\
SR &  &  &  &  &  &  & 0.00 & 0.00 & 0.00\\
OLS &  \multicolumn{3}{c}{same as $T-T_0=5$}  & \multicolumn{3}{c}{same as $T-T_0=10$} & 0.00 & -0.01 & -0.01\\
BVR &  &  &  &  &  &  & 0.00 & -0.01 & -0.01\\
BSC &  &  &  &  &  &  & 0.03 & 0.04 & 0.04\\
{\textbf{SVR}} &  &  &  &  &  &  & 0.00 & 0.00 & 0.00\\
\hline
\end{tabular}
\caption{Average post-treatment bias of Synthetic Control (SC); Separate Ridge (SR) regression; OLS  vertical regression; Bayesian Vertical Regression (BVR); Bayesian Synthetic Control (BSC) and Spatial Vertical Regression (SVR) in scenarios with spatially non-correlated weights ($\rho^2_s = 0.001^2$). Columns correspond to the length of the pre-intervention period $T_0$ and the three cases of residual error structure, IID, Spatial with 40\% noise, and Spatial with 70\% noise. Rows correspond to the length of the post-intervention period and the different methods. Bias and outcome mean value are expressed in the same unit of measure. }
\label{tab:bias001}
\end{table}

\begin{table}[p]
\centering
\begin{tabular}{cccccccccc}
\hline \\
& \multicolumn{3}{c}{$T_0=10$} & \multicolumn{3}{c}{$T_0=20$} & \multicolumn{3}{c}{$T_0=40$} \\
& IID & Sp--40\% & Sp--70\% & IID & Sp--40\% & Sp--70\% & IID & Sp--40\% & Sp--70\%  \\
\cmidrule(lr){2-4} \cmidrule(lr){5-7} \cmidrule(lr){8-10}
& \multicolumn{9}{c}{$T-T_0 = 5$} \\
\cmidrule(lr){2-10}
\textbf{True} & 1.70 & 1.70 & 1.70 & 1.67 & 1.67 & 1.67 & 1.67 & 1.67 & 1.67\\
\addlinespace
SC & 0.03 & 0.05 & 0.06 & -0.04 & -0.03 & -0.04 & 0.06 & 0.06 & 0.06\\
SR & 0.04 & 0.04 & 0.05 & 0.00 & 0.00 & 0.00 & 0.00 & 0.00 & 0.00\\
OLS & -0.08 & -0.13 & -0.17 & -0.01 & -0.03 & -0.04 & -0.01 & -0.01 & -0.02\\
BVR & 0.01 & 0.07 & 0.10 & 0.00 & -0.02 & -0.03 & -0.01 & -0.01 & -0.02\\
BSC & 0.04 & 0.04 & 0.04 & -0.04 & -0.03 & -0.04 & 0.06 & 0.06 & 0.06\\
\textbf{SVR} & 0.02 & 0.08 & 0.09 & 0.01 & 0.00 & 0.00 & -0.01 & -0.01 & -0.01\\
\hline \\
& \multicolumn{9}{c}{$T-T_0 = 10$} \\
\cmidrule(lr){2-10}
\textbf{True} & 1.69 & 1.67 & 1.67 & 1.66 & 1.68 & 1.68 & 1.70 & 1.70 & 1.71\\
\addlinespace
SC & 0.03 & 0.06 & 0.07 & -0.04 & -0.02 & -0.03 & 0.04 & 0.05 & 0.04\\
SR & 0.05 & 0.05 & 0.06 & 0.00 & 0.00 & 0.00 & 0.00 & 0.00 & 0.00\\
OLS & -0.11 & -0.10 & -0.14 & -0.01 & -0.04 & -0.05 & -0.01 & -0.01 & -0.01\\
BVR & -0.01 & 0.03 & 0.05 & -0.01 & -0.03 & -0.03 & -0.01 & -0.01 & -0.01\\
BSC & 0.04 & 0.05 & 0.05 & -0.02 & -0.02 & -0.02 & 0.03 & 0.04 & 0.04\\
\textbf{SVR} & 0.02 & 0.06 & 0.07 & 0.01 & 0.00 & 0.00 & -0.01 & -0.01 & -0.01\\
\hline \\
& \multicolumn{9}{c}{$T-T_0 = 0.5 T_0$} \\
\cmidrule(lr){2-10}
\textbf{True} &  &  &  &  &  &  & 1.71 & 1.71 & 1.71\\
\addlinespace
SC &  &  &  &  &  &  & 0.04 & 0.05 & 0.05\\
SR &  &  &  &  &  &  & 0.00 & 0.00 & 0.00\\
OLS &  \multicolumn{3}{c}{same as $T-T_0=5$}  & \multicolumn{3}{c}{same as $T-T_0=10$} & 0.00 & -0.01 & -0.01\\
BVR &  &  &  &  &  &  & 0.00 & -0.01 & -0.01\\
BSC &  &  &  &  &  &  & 0.04 & 0.04 & 0.04\\
{\textbf{SVR}} &  &  &  &  &  &  & -0.01 & 00.00 & -0.01\\

\hline
\end{tabular}
\caption{Average post-treatment bias of Synthetic Control (SC); Separate Ridge (SR) regression; OLS  vertical regression; Bayesian Vertical Regression (BVR); Bayesian Synthetic Control (BSC) and Spatial Vertical Regression (SVR) in scenarios with spatially low-correlated weights ($\rho^2_s = 0.2^2$). Columns correspond to the length of the pre-intervention period $T_0$ and the three cases of residual error structure, IID, Spatial with 40\% noise, and Spatial with 70\% noise. Rows correspond to the length of the post-intervention period and the different methods. Bias and outcome mean value are expressed in the same unit of measure. }
\label{tab:bias02}
\end{table}

\begin{table}[p]
\centering
\begin{tabular}{cccccccccc}
\hline \\
& \multicolumn{3}{c}{$T_0=10$} & \multicolumn{3}{c}{$T_0=20$} & \multicolumn{3}{c}{$T_0=40$} \\
& IID & Sp--40\% & Sp--70\% & IID & Sp--40\% & Sp--70\% & IID & Sp--40\% & Sp--70\%  \\
\cmidrule(lr){2-4} \cmidrule(lr){5-7} \cmidrule(lr){8-10}
& \multicolumn{9}{c}{$T-T_0 = 5$} \\
\cmidrule(lr){2-10}
\textbf{True} & 1.71 & 1.71 & 1.71 & 1.69 & 1.68 & 1.68 & 1.66 & 1.66 & 1.66\\
\addlinespace
SC & 0.06 & 0.08 & 0.09 & -0.03 & -0.02 & -0.02 & 0.07 & 0.07 & 0.06\\
SR & 0.08 & 0.09 & 0.09 & 0.00 & 0.01 & 0.01 & 0.00 & 0.00 & 0.01\\
OLS & -0.07 & -0.10 & -0.14 & 0.00 & -0.03 & -0.03 & -0.01 & -0.01 & -0.02\\
BVR & 0.04 & 0.09 & 0.10 & 0.00 & -0.01 & -0.02 & -0.01 & -0.01 & -0.02\\
BSC & 0.08 & 0.09 & 0.09 & -0.03 & -0.02 & -0.02 & 0.07 & 0.08 & 0.07\\
\textbf{SVR} & 0.03 & 0.09 & 0.11 & 0.01 & 0.00 & 0.00 & -0.01 & -0.01 & -0.01\\
\hline \\
& \multicolumn{9}{c}{$T-T_0 = 10$} \\
\cmidrule(lr){2-10}
\textbf{True} & 1.71 & 1.69 & 1.69 & 1.65 & 1.66 & 1.66 & 1.71 & 1.72 & 1.72\\
\addlinespace
SC & 0.06 & 0.10 & 0.10 & 0.00 & 0.01 & 0.00 & 0.07 & 0.07 & 0.07\\
SR & 0.08 & 0.09 & 0.09 & 0.03 & 0.02 & 0.03 & 0.00 & 0.01 & 0.01\\
OLS & -0.10 & -0.08 & -0.10 & -0.01 & -0.03 & -0.04 & -0.01 & -0.01 & -0.01\\
BVR & 0.02 & 0.05 & 0.04 & 0.00 & -0.02 & -0.03 & -0.01 & 0.00 & -0.01\\
BSC & 0.07 & 0.09 & 0.09 & 0.01 & 0.02 & 0.01 & 0.06 & 0.06 & 0.06\\
\textbf{SVR} & 0.02 & 0.07 & 0.08 & 0.01 & 0.01 & 0.01 & -0.01 & 0.00 & 0.00\\
\hline\\
& \multicolumn{9}{c}{$T-T_0 = 0.5 T_0$} \\
\cmidrule(lr){2-10}
\textbf{True}&  &  &  &  &  &  & 1.71 & 1.71 & 1.72\\
\addlinespace
SC &  &  &  &  &  &  & 0.07 & 0.07 & 0.07\\
SR &  &  &  &  &  &  & 0.01 & 0.01 & 0.01\\
OLS &  \multicolumn{3}{c}{same as $T-T_0=5$}  & \multicolumn{3}{c}{same as $T-T_0=10$} & 0.00 & 0.00 & -0.01\\
BVR &  &  &  &  &  &  & 0.00 & 0.00 & 0.00\\
BSC &  &  &  &  &  &  & 0.06 & 0.06 & 0.06\\
\hline \\
\end{tabular}
\caption{Average post-treatment bias of Synthetic Control (SC); Separate Ridge (SR) regression; OLS  vertical regression; Bayesian Vertical Regression (BVR); Bayesian Synthetic Control (BSC) and Spatial Vertical Regression (SVR) in scenarios with spatially high-correlated weights ($\rho^2_s = 0.6^2$). Columns correspond to the length of the pre-intervention period $T_0$ and the three cases of residual error structure, IID, Spatial with 40\% noise, and Spatial with 70\% noise. Rows correspond to the length of the post-intervention period and the different methods. Bias and outcome mean value are expressed in the same unit of measure. }
\label{tab:bias06}
\end{table}

\begin{table}[p]
\centering
\begin{tabular}{cccccccccc}
\hline \\
& \multicolumn{3}{c}{$T_0=10$} & \multicolumn{3}{c}{$T_0=20$} & \multicolumn{3}{c}{$T_0=40$} \\
& IID & Sp--40\% & Sp--70\% & IID & Sp--40\% & Sp--70\% & IID & Sp--40\% & Sp--70\%  \\
\cmidrule(lr){2-4} \cmidrule(lr){5-7} \cmidrule(lr){8-10}
& \multicolumn{9}{c}{$T-T_0 = 5$} \\
\cmidrule(lr){2-10}
SC & 0.99 & 1.02 & 1.40 & 0.82 & 0.81 & 1.08 & 0.77 & 0.80 & 1.05\\
SR & 0.84 & 0.86 & 1.07 & 0.61 & 0.62 & 0.95 & 0.41 & 0.43 & 0.72\\
OLS & 13.28 & 17.93 & 31.35 & 1.06 & 1.10 & 1.92 & 0.45 & 0.48 & 0.83\\
BVR & 2.24 & 2.29 & 3.71 & 0.84 & 0.88 & 1.47 & 0.44 & 0.47 & 0.81\\
BSC & 0.98 & 1.00 & 1.24 & 0.90 & 0.90 & 1.14 & 0.84 & 0.86 & 1.11\\
\textbf{SVR} & 0.68 & 0.72 & 1.03 & 0.53 & 0.54 & 0.82 & 0.39 & 0.41 & 0.68\\
\hline \\
& \multicolumn{9}{c}{$T-T_0 = 10$} \\
\cmidrule(lr){2-10}
SC & 1.25 & 1.30 & 1.75 & 0.98 & 0.97 & 1.28 & 0.84 & 0.86 & 1.10\\
SR & 0.98 & 0.99 & 1.20 & 0.71 & 0.72 & 1.08 & 0.44 & 0.46 & 0.75\\
OLS & 21.07 & 25.76 & 45.02 & 1.35 & 1.41 & 2.46 & 0.50 & 0.51 & 0.90\\
BVR & 3.23 & 3.28 & 5.30 & 1.03 & 1.08 & 1.78 & 0.48 & 0.50 & 0.86\\
BSC & 1.16 & 1.17 & 1.43 & 1.03 & 1.04 & 1.29 & 0.91 & 0.92 & 1.16\\
\textbf{SVR} & 0.84 & 0.90 & 1.24 & 0.60 & 0.63 & 0.92 & 0.42 & 0.44 & 0.71\\
\hline\\
& \multicolumn{9}{c}{$T-T_0 = 0.5 T_0$} \\
\cmidrule(lr){2-10}
SC &  &  &  &  &  &  & 0.88 & 0.90 & 1.15\\
SR &  &  &  &  &  &  & 0.46 & 0.48 & 0.78\\
OLS &  \multicolumn{3}{c}{same as $T-T_0=5$}  & \multicolumn{3}{c}{same as $T-T_0=10$} & 0.52 & 0.55 & 0.96\\
BVR &  &  &  &  &  &  & 0.50 & 0.52 & 0.91\\
BSC &  &  &  &  &  &  & 0.94 & 0.95 & 1.19\\
\textbf{SVR}  &  &  &  &  &  &  &  0.43 & 0.46 & 0.74\\

\hline \\
\end{tabular}
\caption{Average post-treatment MSE of Synthetic Control (SC); Separate Ridge (SR) regression; OLS  vertical regression; Bayesian Vertical Regression (BVR); Bayesian Synthetic Control (BSC) and Spatial Vertical Regression (SVR) in scenarios with spatially non-correlated weights ($\rho^2_s = 0.001^2$). Columns correspond to the length of the pre-intervention period $T_0$ and the three cases of residual error structure, IID, Spatial with 40\% noise, and Spatial with 70\% noise. Rows correspond to the length of the post-intervention period and the different methods.}
\label{tab:mse001}
\end{table}

\begin{table}[p]
\centering
\begin{tabular}{cccccccccc}
\hline \\
& \multicolumn{3}{c}{$T_0=10$} & \multicolumn{3}{c}{$T_0=20$} & \multicolumn{3}{c}{$T_0=40$} \\
& IID & Sp--40\% & Sp--70\% & IID & Sp--40\% & Sp--70\% & IID & Sp--40\% & Sp--70\%  \\
\cmidrule(lr){2-4} \cmidrule(lr){5-7} \cmidrule(lr){8-10}
& \multicolumn{9}{c}{$T-T_0 = 5$} \\
\cmidrule(lr){2-10}
SC & 0.99 & 1.00 & 1.38 & 0.77 & 0.77 & 1.04 & 0.76 & 0.78 & 1.03\\
SR & 0.84 & 0.85 & 1.05 & 0.60 & 0.60 & 0.92 & 0.40 & 0.43 & 0.72\\
OLS & 12.90 & 17.40 & 30.42 & 1.04 & 1.09 & 1.90 & 0.45 & 0.47 & 0.83\\
BVR & 2.21 & 2.26 & 3.60 & 0.84 & 0.86 & 1.45 & 0.44 & 0.46 & 0.80\\
BSC & 0.97 & 0.98 & 1.23 & 0.85 & 0.85 & 1.08 & 0.83 & 0.86 & 1.11\\
\textbf{SVR} & 0.60 & 0.70 & 1.01 & 0.44 & 0.52 & 0.80 & 0.34 & 0.40 & 0.66\\
\hline \\
& \multicolumn{9}{c}{$T-T_0 = 10$} \\
\cmidrule(lr){2-10}
SC & 1.28 & 1.32 & 1.77 & 0.93 & 0.94 & 1.23 & 0.84 & 0.85 & 1.10\\
SR & 0.99 & 0.99 & 1.20 & 0.70 & 0.70 & 1.06 & 0.44 & 0.46 & 0.76\\
OLS & 20.93 & 25.46 & 44.49 & 1.32 & 1.38 & 2.42 & 0.50 & 0.51 & 0.90\\
BVR & 3.28 & 3.38 & 5.33 & 1.02 & 1.06 & 1.77 & 0.48 & 0.50 & 0.86\\
BSC & 1.15 & 1.16 & 1.42 & 0.99 & 1.00 & 1.24 & 0.92 & 0.93 & 1.17\\
\textbf{SVR} & 0.77 & 0.92 & 1.27 & 0.49 & 0.59 & 0.89 & 0.36 & 0.42 & 0.69\\
\hline\\
& \multicolumn{9}{c}{$T-T_0 = 0.5 T_0$} \\
\cmidrule(lr){2-10}
SC &  &  &  &  &  &  & 0.87 & 0.88 & 1.14\\
SR &  &  &  &  &  &  & 0.46 & 0.48 & 0.79\\
OLS & \multicolumn{3}{c}{same as $T-T_0=5$}  & \multicolumn{3}{c}{same as $T-T_0=10$} & 0.52 & 0.55 & 0.96\\
BVR &  &  &  &  &  &  & 0.50 & 0.52 & 0.91\\
BSC &  &  &  &  &  &  & 0.94 & 0.96 & 1.19\\
\textbf{SVR} &  &  &  &  &  &  & 0.37 & 0.44 & 0.72\\

\hline \\
\end{tabular}
\caption{Average post-treatment MSE of Synthetic Control (SC); Separate Ridge (SR) regression; OLS  vertical regression; Bayesian Vertical Regression (BVR); Bayesian Synthetic Control (BSC) and Spatial Vertical Regression (SVR) in scenarios with spatially low-correlated weights ($\rho^2_s = 0.2^2$). Columns correspond to the length of the pre-intervention period $T_0$ and the three cases of residual error structure, IID, Spatial with 40\% noise, and Spatial with 70\% noise. Rows correspond to the length of the post-intervention period and the different methods.}
\label{tab:mse02}
\end{table}

\begin{table}[p]
\centering
\begin{tabular}{cccccccccc}
\hline \\
& \multicolumn{3}{c}{$T_0=10$} & \multicolumn{3}{c}{$T_0=20$} & \multicolumn{3}{c}{$T_0=40$} \\
& IID & Sp--40\% & Sp--70\% & IID & Sp--40\% & Sp--70\% & IID & Sp--40\% & Sp--70\%  \\
\cmidrule(lr){2-4} \cmidrule(lr){5-7} \cmidrule(lr){8-10}
& \multicolumn{9}{c}{$T-T_0 = 5$} \\
\cmidrule(lr){2-10}
SC & 0.97 & 0.99 & 1.34 & 0.69 & 0.70 & 0.95 & 0.71 & 0.73 & 0.97\\
SR & 0.84 & 0.84 & 1.05 & 0.57 & 0.58 & 0.89 & 0.40 & 0.42 & 0.70\\
OLS & 13.04 & 16.31 & 28.50 & 0.99 & 1.04 & 1.81 & 0.44 & 0.47 & 0.82\\
BVR & 2.21 & 2.18 & 3.44 & 0.82 & 0.83 & 1.40 & 0.43 & 0.46 & 0.80\\
BSC & 0.95 & 0.97 & 1.21 & 0.77 & 0.78 & 0.99 & 0.80 & 0.82 & 1.07\\
\textbf{SVR} & 0.55 & 0.63 & 0.92 & 0.36 & 0.43 & 0.68 & 0.31 & 0.35 & 0.60\\
\hline \\
& \multicolumn{9}{c}{$T-T_0 = 10$} \\
\cmidrule(lr){2-10}
SC & 1.25 & 1.31 & 1.73 & 0.84 & 0.84 & 1.11 & 0.81 & 0.81 & 1.04\\
SR & 0.99 & 0.98 & 1.18 & 0.68 & 0.69 & 1.03 & 0.44 & 0.45 & 0.74\\
OLS & 21.94 & 24.06 & 42.01 & 1.25 & 1.32 & 2.31 & 0.49 & 0.51 & 0.89\\
BVR & 3.21 & 3.19 & 5.06 & 1.00 & 1.02 & 1.71 & 0.48 & 0.50 & 0.86\\
BSC & 1.13 & 1.14 & 1.38 & 0.88 & 0.89 & 1.11 & 0.91 & 0.91 & 1.14\\
\textbf{SVR} & 0.70 & 0.79 & 1.11 & 0.39 & 0.48 & 0.75 & 0.31 & 0.36 & 0.61\\
\hline\\
& \multicolumn{9}{c}{$T-T_0 = 0.5 T_0$} \\
\cmidrule(lr){2-10}
SC &  &  &  &  &  &  & 0.84 & 0.85 & 1.09\\
SR &  &  &  &  &  &  & 0.46 & 0.48 & 0.78\\
OLS &  \multicolumn{3}{c}{same as $T-T_0=5$} & \multicolumn{3}{c}{same as $T-T_0=10$} & 0.52 & 0.55 & 0.95\\
BVR &  &  &  &  &  &  & 0.50 & 0.52 & 0.91\\
BSC &  &  &  &  &  &  & 0.92 & 0.93 & 1.16\\
\textbf{SVR}  &  &  &  &  &  &  &  0.32 & 0.38 & 0.64\\

\hline \\
\end{tabular}
\caption{Average post-treatment MSE of Synthetic Control (SC); Separate Ridge (SR) regression; OLS  vertical regression; Bayesian Vertical Regression (BVR); Bayesian Synthetic Control (BSC) and Spatial Vertical Regression (SVR) in scenarios with spatially high-correlated weights ($\rho^2_s = 0.6^2$). Columns correspond to the length of the pre-intervention period $T_0$ and the three cases of residual error structure, IID, Spatial with 40\% noise, and Spatial with 70\% noise. Rows correspond to the length of the post-intervention period and the different methods.}
\label{tab:mse06}
\end{table}

\begin{table}[p]
\centering
\begin{tabular}{cccccccccc}
\hline \\
& \multicolumn{3}{c}{$T_0=10$} & \multicolumn{3}{c}{$T_0=20$} & \multicolumn{3}{c}{$T_0=40$} \\
& IID & Sp--40\% & Sp--70\% & IID & Sp--40\% & Sp--70\% & IID & Sp--40\% & Sp--70\%  \\
\cmidrule(lr){2-4} \cmidrule(lr){5-7} \cmidrule(lr){8-10}
& \multicolumn{9}{c}{$T-T_0 = 5$} \\
\cmidrule(lr){2-10}
OLS & 0.00 & 0.00 & 0.00 & 0.84 & 0.83 & 0.83 & 0.74 & 0.74 & 0.74\\
BVR & 0.93 & 0.92 & 0.90 & 0.95 & 0.93 & 0.93 & 0.95 & 0.95 & 0.94\\
BSC & 0.81 & 0.81 & 0.83 & 0.86 & 0.87 & 0.87 & 0.90 & 0.90 & 0.91\\
\textbf{SVR} & 0.94 & 0.93 & 0.93 & 0.93 & 0.94 & 0.94 & 0.94 & 0.94 & 0.94\\
\hline \\
& \multicolumn{9}{c}{$T-T_0 = 10$} \\
\cmidrule(lr){2-10}
OLS & 0.00 & 0.00 & 0.00 & 0.85 & 0.84 & 0.84 & 0.77 & 0.77 & 0.77\\
BVR & 0.94 & 0.93 & 0.91 & 0.95 & 0.94 & 0.93 & 0.95 & 0.95 & 0.94\\
BSC & 0.79 & 0.79 & 0.81 & 0.84 & 0.85 & 0.86 & 0.89 & 0.89 & 0.90\\
\textbf{SVR} & 0.93 & 0.93 & 0.93 & 0.93 & 0.93 & 0.94 & 0.94 & 0.94 & 0.94\\
\hline\\
& \multicolumn{9}{c}{$T-T_0 = 0.5 T_0$} \\
\cmidrule(lr){2-10}
OLS &  &  &  &  &  &  & 0.78 & 0.77 & 0.77\\
BVR &  \multicolumn{3}{c}{same as $T-T_0=5$}  & \multicolumn{3}{c}{same as $T-T_0=10$} & 0.97 & 0.96 & 0.95\\
BSC&  &  &  &  &  &  & 0.89 & 0.89 & 0.90\\
\textbf{SVR}&  &  &  &  &  &  & 0.94 & 0.94 & 0.94\\

\hline \\
\end{tabular}
\caption{Average post-treatment   coverage of  95\% confidence / posterior credible intervals for the imputed post-treatement control  potential outcomes for the treated units based on  OLS vertical regression; Bayesian Vertical Regression (BVR); Bayesian Synthetic Control (BSC) and Spatial Vertical Regression (SVR) in scenarios with spatially non-correlated weights ($\rho^2_s = 0.001^2$). Columns correspond to the length of the pre-intervention period $T_0$ and the three cases of residual error structure, IID, Spatial with 40\% noise, and Spatial with 70\% noise. Rows correspond to the length of the post-intervention period and the different methods.}
\label{tab:cov001}
\end{table}

\begin{table}[p]
\centering
\begin{tabular}{cccccccccc}
\hline \\
& \multicolumn{3}{c}{$T_0=10$} & \multicolumn{3}{c}{$T_0=20$} & \multicolumn{3}{c}{$T_0=40$} \\
& IID & Sp--40\% & Sp--70\% & IID & Sp--40\% & Sp--70\% & IID & Sp--40\% & Sp--70\%  \\
\cmidrule(lr){2-4} \cmidrule(lr){5-7} \cmidrule(lr){8-10}
& \multicolumn{9}{c}{$T-T_0 = 5$} \\
\cmidrule(lr){2-10}
OLS & 0.00 & 0.00 & 0.00 & 0.84 & 0.83 & 0.83 & 0.74 & 0.74 & 0.74\\
BVR & 0.93 & 0.92 & 0.91 & 0.94 & 0.94 & 0.93 & 0.95 & 0.94 & 0.94\\
BSC & 0.80 & 0.80 & 0.82 & 0.87 & 0.87 & 0.88 & 0.90 & 0.90 & 0.91\\
\textbf{SVR} & 0.93 & 0.92 & 0.93 & 0.94 & 0.94 & 0.94 & 0.94 & 0.94 & 0.94\\
\hline \\
& \multicolumn{9}{c}{$T-T_0 = 10$} \\
\cmidrule(lr){2-10}
OLS & 0.00 & 0.00 & 0.00 & 0.85 & 0.84 & 0.84 & 0.77 & 0.77 & 0.77\\
BVR & 0.93 & 0.93 & 0.91 & 0.94 & 0.94 & 0.93 & 0.95 & 0.94 & 0.94\\
BSC & 0.78 & 0.78 & 0.81 & 0.85 & 0.85 & 0.87 & 0.89 & 0.89 & 0.90\\
\textbf{SVR} & 0.93 & 0.93 & 0.93 & 0.94 & 0.93 & 0.93 & 0.94 & 0.94 & 0.94\\
\hline\\
& \multicolumn{9}{c}{$T-T_0 = 0.5 T_0$} \\
\cmidrule(lr){2-10}
OLS &  &  &  &  &  &  & 0.78 & 0.77 & 0.77\\
BVR &  \multicolumn{3}{c}{$T-T_0=5$}  & \multicolumn{3}{c}{$T-T_0=10$} & 0.97 & 0.96 & 0.95\\
BSC &  &  &  &  &  &  & 0.88 & 0.88 & 0.89\\
\textbf{SVR}  &  &  &  &  &  &  & 0.94 & 0.94 & 0.94\\
\hline \\
\end{tabular}
\caption{Average post-treatment   coverage of  95\% confidence / posterior credible intervals for the imputed post-treatement control  potential outcomes for the treated units based on  OLS vertical regression; Bayesian Vertical Regression (BVR); Bayesian Synthetic Control (BSC) and Spatial Vertical Regression (SVR) in scenarios with spatially high-correlated weights ($\rho^2_s = 0.2^2$). Columns correspond to the length of the pre-intervention period $T_0$ and the three cases of residual error structure, IID, Spatial with 40\% noise, and Spatial with 70\% noise. Rows correspond to the length of the post-intervention period and the different methods.}
\label{tab:cov02}
\end{table}

\begin{table}[p]
\centering
\begin{tabular}{cccccccccc}
\hline \\
& \multicolumn{3}{c}{$T_0=10$} & \multicolumn{3}{c}{$T_0=20$} & \multicolumn{3}{c}{$T_0=40$} \\
& IID & Sp--40\% & Sp--70\% & IID & Sp--40\% & Sp--70\% & IID & Sp--40\% & Sp--70\%  \\
\cmidrule(lr){2-4} \cmidrule(lr){5-7} \cmidrule(lr){8-10}
& \multicolumn{9}{c}{$T-T_0 = 5$} \\
\cmidrule(lr){2-10}
OLS & 0.00 & 0.00 & 0.00 & 0.84 & 0.83 & 0.83 & 0.74 & 0.74 & 0.74\\
BVR & 0.93 & 0.93 & 0.91 & 0.94 & 0.94 & 0.93 & 0.95 & 0.94 & 0.94\\
BSC & 0.80 & 0.80 & 0.81 & 0.88 & 0.88 & 0.89 & 0.90 & 0.90 & 0.90\\
\textbf{SVR} & 0.93 & 0.92 & 0.93 & 0.94 & 0.94 & 0.94 & 0.94 & 0.94 & 0.94\\
\hline \\
& \multicolumn{9}{c}{$T-T_0 = 10$} \\
\cmidrule(lr){2-10}
OLS & 0.00 & 0.00 & 0.00 & 0.85 & 0.84 & 0.84 & 0.77 & 0.77 & 0.77\\
BVR & 0.93 & 0.93 & 0.91 & 0.94 & 0.94 & 0.94 & 0.95 & 0.94 & 0.94\\
BSC & 0.78 & 0.78 & 0.80 & 0.86 & 0.86 & 0.88 & 0.88 & 0.89 & 0.90\\
\textbf{SVR} & 0.94 & 0.93 & 0.93 & 0.94 & 0.94 & 0.94 & 0.94 & 0.94 & 0.94\\
\hline\\
& \multicolumn{9}{c}{$T-T_0 = 0.5 T_0$} \\
\cmidrule(lr){2-10}
OLS &  &  &  &  &  &  & 0.78 & 0.77 & 0.77\\
BVR &  \multicolumn{3}{c}{same as $T-T_0=5$}  & \multicolumn{3}{c}{same as $T-T_0=10$} & 0.97 & 0.96 & 0.95\\
BSC &  &  &  &  &  &  & 0.88 & 0.88 & 0.90\\
\textbf{SVR}  &  &  &  &  &  &  & 0.94 & 0.94 & 0.94\\

\hline \\
\end{tabular}
\caption{Average post-treatment   coverage of  95\% confidence / posterior credible intervals for the imputed post-treatement control  potential outcomes for the treated units based on  OLS vertical regression; Bayesian Vertical Regression (BVR); Bayesian Synthetic Control (BSC) and Spatial Vertical Regression (SVR) in scenarios with spatially high-correlated weights ($\rho^2_s = 0.6^2$). Columns correspond to the length of the pre-intervention period $T_0$ and the three cases of residual error structure, IID, Spatial with 40\% noise, and Spatial with 70\% noise. Rows correspond to the length of the post-intervention period and the different methods.}
\label{tab:cov06}
\end{table}

\end{appendices}

\end{document}